\documentclass[aps,prx,twocolumn,superscriptaddress,amsmath,nofootinbib,10pt,floatfix]{revtex4-1}
 
\usepackage{graphicx}
\usepackage[normalem]{ulem}
\usepackage{xcolor}
\usepackage[utf8]{inputenc}
\usepackage{comment}
\DeclareMathSizes{8}{7}{7}{7}

\usepackage{textgreek}
\usepackage[colorlinks, linkcolor=blue, citecolor=blue]{hyperref}
\usepackage{gensymb}
\usepackage[compat=1.1.0]{tikz-feynman}
\usepackage{tikz}
\tikzset{
pattern size/.store in=\mcSize, 
pattern size = 5pt,
pattern thickness/.store in=\mcThickness, 
pattern thickness = 0.3pt,
pattern radius/.store in=\mcRadius, 
pattern radius = 1pt}

\usepackage{amssymb}
\newcommand{\JH}{J_\tn{H}}
\newcommand{\JA}{J_\tn{A}}
\newcommand{\be}{\begin{equation}}
\newcommand{\ee}{\end{equation}}
\newcommand{\beq}{\begin{eqnarray}}
\newcommand{\eeq}{\end{eqnarray}}
\newcommand{\ba}{\[\begin{aligned}}
\newcommand{\ea}{\end{aligned}\]}
\newcommand{\bal}{\begin{aligned}}
\newcommand{\eal}{\end{aligned}}

\newcommand{\la}{\langle}
\newcommand{\ra}{\rangle}

\newcommand{\Tr}{{\rm Tr\,}}

\renewcommand{\vec}[1]{{\bf #1}}
\renewcommand{\epsilon}{\varepsilon}

\def\nn{\nonumber}

\newcommand{\avg}[1]{\left\langle #1 \right\rangle}

\renewcommand{\vec}[1]{\boldsymbol{#1}}
\colorlet{green}{teal}

\def \up{\uparrow}
\def \down{\downarrow}

\def \w{{\omega}}

\def \k{{\vec{k}}}
\def \q{{\vec{q}}}

\def \r{{\bf {r}}}

\def \ket#1{{\,|\,#1\,\rangle\,}}
\def \bra#1{{\,\langle\,#1\,|\,}}
\def \braket#1#2{{\,\langle\,#1\,|\,#2\,\rangle\,}}

\def \ra{{\rangle}}
\def \la{{\langle}}
\def \tn{\textnormal}

\def \ba{\begin{align*}}
\def \ea{\end{align*}}

\newcounter{indice}

\begin{document}
\title{Intertwined Orders, Quantum Criticality and Skyrmions  in Tunable Topological Bands}
\author{Xuepeng Wang} 
\affiliation{Department of Physics, Cornell University, Ithaca, New York 14853, USA.}
\author{Johannes S. Hofmann} \email{jhofmann@pks.mpg.de}
\affiliation{Max-Planck-Institut f\"ur Physik Komplexer Systeme, N\"othnitzer Strasse 38, 01187 Dresden, Germany}
\author{Debanjan Chowdhury}\email{debanjanchowdhury@cornell.edu}
\affiliation{Department of Physics, Cornell University, Ithaca, New York 14853, USA.}
\date{\today}

\begin{abstract}
Skyrmions are emergent many-body excitations that lie at the heart of both multi-component quantum Hall-like systems and deconfined quantum criticality. In a companion article (X. Wang et al., arXiv:2507.22971), we studied a microscopic time-reversal symmetric model of tunable interacting Chern bands using numerically exact determinant quantum Monte Carlo calculations, and presented evidence for the emergence of robust skyrmion excitations. These charged excitations emerge in the vicinity of a many-body insulator at a commensurate filling of the Chern bands, and lead to the onset of superconductivity when doped away from the insulating phase. Here, we present quantum Monte-Carlo results and a complementary field-theoretical analysis for the quantum phase transition(s) that arise between the intertwined phases as a function of two distinct tuning parameters. Our numerical results are consistent with a single continuous quantum phase transition between an insulating Chern antiferromagnet and a fully gapped superconductor, with an emergent $SO(5)$ symmetry at the putative critical point, highly suggestive of deconfined quantum (pseudo-)criticality. We also present a detailed comparison between the momentum-resolved spectral functions associated with the neutral collective modes, single electron and composite spin-polaron excitations obtained using a combination of Monte-Carlo computations and a Bethe-Salpeter analysis built on top of the self-consistent Hartree-Fock calculation. We end with a brief outlook on some of the interesting open problems.
\end{abstract}

\maketitle

%\tableofcontents

\section{Introduction}
The advent of moir\'e materials has unleashed a new experimental era in the field of correlated electron physics, where the effects of Mott-localization, frustration, and topology can be studied in a highly tunable setting \cite{Andrei2021,Mak2022,young21}. Specifically, the ability to vary the ratio of effective interactions and single-particle bandwidth, as well as electronic band-filling, provides access to a wide range of competing many-body phases and phase-transitions \cite{Li2021,Ghiotto_2021} within a single device. Such tunability together with the interplay between topology and interaction offers an ideal platform not only for hosting intertwined orders \cite{SSRMP,LNW,intertwined} including correlated insulators, magnetism and unconventional superconductivity \cite{yc_tbg1, yc_tbg2, yc_tbg3,Lu19,Sharpe19,dean19_tbg,Jiang19,YC_nematic,Nuckolls20,Young20,AV18,Stepanov20,Serlin2020Science,AV20_prx,MacDonald20,SS20,Bultinck20_prl,Park21_tbg,Pierce21,park21,Efetov21,MacDonald21,Wu21,PhysRevB.103.195127,PhysRevB.98.081102,PhysRevLett.127.247703,PhysRevB.102.035136,PhysRevB.102.045107,KangVafekPRL,TBG4,dai_prb21, peri2021fragile, Bernevig21,Hao21,Yves_21prx,dc21,Oh21,Yves22,Zeldov22, zaletel_dmrg,YahuiTS,BABHF,LedwithPRX,jfmv_prl, tmdsc25,zhang2025,tmdsc_fai,tmdsc_dean}, but also for non-electronic charged composite quasiparticles, such as skyrmions (or spin-polarons, corresponding to electrons dressed by spin-flip excitations) \cite{ashvin_skrymion, baby_skyrmion, chatterjee20, chatterjee22,wagner22, Schindler22, prembabu2025spinpolaronsflatband}. From the viewpoint of quantum Hall physics, properties of these emergent excitations have been studied in both Landau-level like system \cite{sondhi,barrett,girvin1999quantum, girvin2000} and beyond \cite{skyr_short}. On the other hand, skyrmions as topological textures can also emerge near putative deconfined quantum critical points (DQCP) in a variety of setups, such as the N\'eel to valence bond solid (VBS) transition \cite{dqcp1, dqcp_prb_04} and quantum spin Hall insulator to superconductor transition \cite{assaad19, assaad21, assaad23}. Recent numerical studies based on fuzzy-sphere regularization have further revealed new insights into the compatibility between DQCP and emergent symmetries \cite{sphere1, zhou_fuzzy2, hofmann_fuzzy}. Studying explicit models using controlled theoretical methods that host all of the features introduced above in a setting that shares varying degrees of microscopic similarity with moir\'e systems remains an exciting topic of research.

In a companion article \cite{skyr_short}, we have studied the connection between band-topology and a specific class of strong interactions in an exactly solvable model. We have observed stable skyrmionic excitations away from any Landau-level-like limit in a wide regime of parameter space using determinant quantum Monte Carlo (DQMC) computations. We provided numerical evidence for the emergence of an insulating many-body Chern antiferromagnet (ChAF) at a commensurate filling of the Chern bands, which when doped leads to a gapped superconductor (SC). From a technical perspective, although we demonstrated explicitly the existence of composite charged excitations (spin-polarons or skyrmions) via tracking the spin quantum number associated with the doped charges, we did not discuss their momentum-resolved spectra. 
In the present manuscript, we will present new results for the dispersion of the collective neutral excitations (i.e. the magnons associated with the ChAF insulator) and the composite charged excitations, and provide insights into the technical differences that arise between the fully non-perturbative DQMC computation vs. a combined Hartree-Fock and Bethe-Salpeter calculation.

We have also demonstrated via the DQMC computations that tuning the electronic bandwidth or the ratio of the distinct interaction parameters at the commensurate filling drives a phase transition between the ChAF and the gapped SC. However, the nature of these phase transition(s) have not been investigated previously in any detail. The focus of the current manuscript will be on a number of complementary and detailed aspects of this rich theoretical problem. One of our motivations is to investigate the possibility of realizing a deconfined quantum critical point (DQCP) between these two phases. We note that a large body of work studying DQCP has focused predominantly on either microscopic lattice models of interacting spins \cite{dqcp_prb_04,dqcp1,sandvik07, sandvik09, sandvik10}, electronic models with Dirac points but where topology is not built in \cite{so5_3,assaad19,assaad21,assaad23}, or in the explicit projected limit of the lowest Landau level or active Chern bands \cite{fierz, sphere2, sphere3}. In contrast, our focus will be on models of interacting Chern bands which are not necessarily in the fully projected limit and far away from any Landau-level-like limit (i.e. realizing ideal quantum geometry), but where the low-energy physics is governed primarily by the partially filled Chern bands with strong interactions. The promising route towards realizing DQCP in such interacting topological nearly flat-band models is exciting due to its potential connection to experimental advances with tunable moir\'e materials, and furthermore can help reveal new avenues for engineering novel DQCPs with a variety of emergent symmetries. 

\section{Overview of Results}\label{sec::overview}

\begin{figure}[htb]
\includegraphics[width=90mm,scale=1]{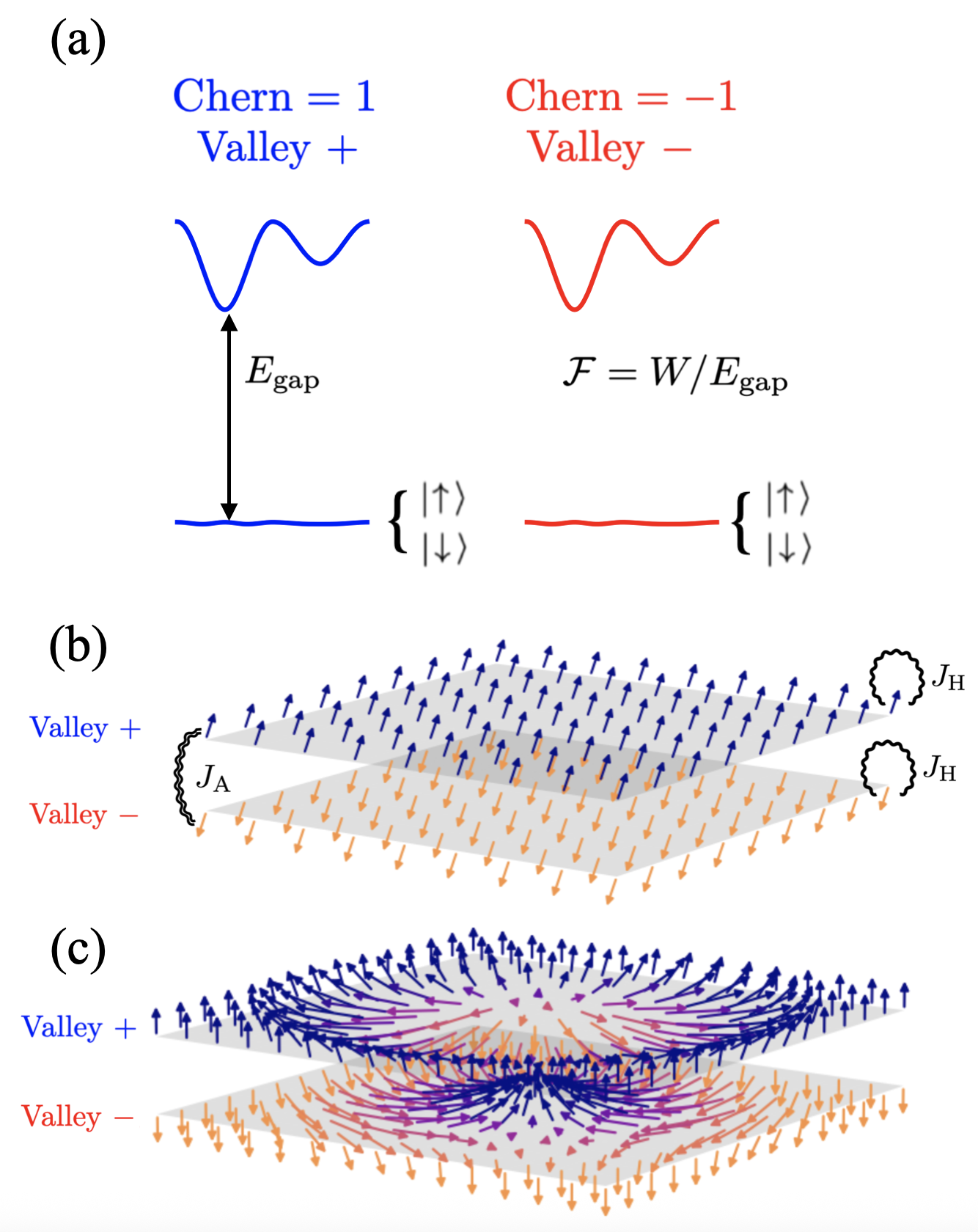}
\caption{(a) Schematics of band structures for a specific high-symmetry cut in the Brillouin zone for $\mathcal{F}\equiv W/E_{\tn{gap}} = 0.01$, where blue (red) curve denotes spinful bands with $\tau=+$ ($\tau=-$). The Chern number $C=\tau$. (b) Schematics of the correlated ChAF insulator. Color scheme and notations for interaction vertices are the same as in panel (a). (Double) Wavy line denotes microscopic (anti-)ferromagnetic interaction in the microscopic real space basis used in the QMC simulation. (c) Schematics of the doped skyrmions on top of the ChAF insulator. }\label{fig::schematics}
\end{figure}

Let us now briefly review the model introduced in the companion article \cite{skyr_short}, and discuss the key results obtained previously through DQMC computations. We studied an interacting time-reversal symmetric model of tunable isolated topological bands with Chern number $C=\pm1$ that does not suffer from the sign-problem. The two-dimensional model of spinful electrons ($\sigma=\uparrow,\downarrow$) carrying an additional ``valley" ($\tau=\pm$) and sublattice ($\eta = \tn{A},\tn{B}$) degree of freedom (d.o.f.) is governed by a non-interacting Hamiltonian per valley, $H^{(\tau)}_{\tn{kin}}$  \cite{hkin,single_chern,svh},
\beq\label{h_kin0}
H^{(\tau)}_{\tn{kin}} = \sum_\k\psi^{\dagger}_{\k,\tau}\bigg[B^{0}_{\k,\tau}\eta_0 + \vec{B_{\k,\tau}\cdot\vec{\eta}}\bigg]\sigma_0\psi_{\k,\tau},
\eeq
where $\psi^{\dagger}_{\k,\tau} = (c^{\dagger}_{\k,A,\tau,\uparrow}, c^{\dagger}_{\k,B,\tau,\uparrow}, c^{\dagger}_{\k,A,\tau,\downarrow}, c^{\dagger}_{\k,B,\tau,\downarrow})$ and $c^{\dagger}_{\k,\eta,\tau,\sigma}$ denotes the electron creation operator on sublattice $\eta$ with spin $\sigma$ and valley $\tau$; similarly, $\eta_\alpha$, $\sigma_{\alpha}$ and $\tau_{\alpha}$ ($\alpha =0,x,y,z$) are Pauli matrices in sublattice, spin and valley subspace, respectively, where $\alpha=0$ denote identity matrices. The explicit form of the matrices $B^{0}_{\k,\tau}$ and $\vec{B}_{\k,\tau}$ for a square lattice model with a first-neighbor ($t$) and staggered second-neighbor ($t_2=t/\sqrt{2}$) hopping with a $\pi-$flux per square plaquette, and a fifth ($t_5$) neighbor hopping that helps tune the flatness ratio ${\cal{F}}=W/E_{\tn{gap}}$ ($W\equiv$bandwidth, $E_{\tn{gap}}\equiv$remote bandgap) are adapted from \cite{skyr_short} as
\begin{equation}
\begin{aligned}
&
B^{x}_{\k,\tau=+} + iB^{y}_{\k,\tau=+} = -2t \bigg[e^{-i\frac{\pi}{4}}\cos(k_y) + e^{i\frac{\pi}{4}}\cos(k_x)\bigg] \\&
B^{z}_{\k,\tau=+} = -2t_2 \bigg[ \cos(k_x + k_y) - \cos(k_x - k_y) \bigg] \\&
B^{0}_{\k,\tau=+} = -2t_5 \bigg[ \cos(2k_x + 2k_y) + \cos(2k_x - 2k_y) \bigg]
\end{aligned},
\end{equation}
.By including two copies of $H^{(\tau)}_{\tn{kin}}$ in a time-reversal-invariant fashion under $\mathcal{T}=i\tau_y\mathcal{K}$ where $\mathcal{K}$ denotes complex conjugation, we obtain a set of degenerate Chern bands carrying spin and valley with $C=\tau$. A schematic of the band structures are shown in Fig.~\ref{fig::schematics}(a). 

 We include an interaction Hamiltonian, $H_{\tn{interaction}} = H_{\tn{intra-valley}} + H_{\tn{inter-valley}}$, 
\begin{subequations}\label{eq::ham_int}
\beq
H_{\tn{intra-valley}} &=& 
J_{\tn{H}}\sum_{\vec{R},\eta,\tau=\pm} \vec{S}^{\tau}_{\vec{R},\eta}\cdot\vec{S}^{\tau}_{\vec{R},\eta}, \label{intra_valley_int}\\
H_{\tn{inter-valley}} &=& J_{\tn{A}}\sum_{\vec{R},\eta}\vec{S}^{+}_{\vec{R},\eta}\cdot\vec{S}^{-}_{\vec{R},\eta},~~\tn{where}\label{inter_valley_int}\\
\vec{S}^{\tau}_{\vec{R},\eta} &=& \sum_{\alpha,\beta = \uparrow,\downarrow}c^{\dagger}_{\vec{R},\tau,\eta,\alpha}\vec{\sigma}_{\alpha\beta} c_{\vec{R},\tau,\eta,\beta}.
\eeq
\end{subequations}
% Note that we combine the sublattice index ($\eta$) into the two-dimensional spatial coordinate $\r$. 
with a ferromagnetic intra-valley Hund's-type exchange with $J_{\tn{H}}<0$ and an antiferromagnetic inter-valley exchange with $J_{\tn{A}}>0$. We denote $\vec{R}$ as unit-cell coordinates and $\r\equiv\vec{r}_{\vec{R}\eta}=\vec{R}+\vec{\delta}_\eta$ as the coordinate of $\eta$-sublattices in $\vec{R}$-th unit-cell, where $\vec{\delta}_\eta$ is the coordinate of sublattice $\eta$ within a unit-cell. We will also use $\vec{r'}$ as shorthand for $\vec{r}_{\vec{R'}\eta'}$. The above model is sign-problem-free as long as $J_{\rm{A}}\geq 2|J_{\rm{H}}|$. Our primary focus in the companion study was on the emergence of an interaction-induced ChAF phase at half-filling ($\nu=2$) of the topological bands, which when doped with electrons yields charged skrymions. A schematic of the correlated ChAF insulator and a skyrmionic excitation are shown in Fig.~\ref{fig::schematics}(b) and Fig.~\ref{fig::schematics}(c), respectively. Furthermore, the skyrmions across time-reversed valleys pair up into spin-singlet Cooper pairs. 

In Section \ref{sec::hf}, we will perform a Hartree-Fock (HF) calculation to analyze the collective magnon excitations and the composite charged spin-polaron excitations \cite{baby_skyrmion, Schindler22}
on top of the ChAF phase, which serves as a complementary exercise to the full non-perturbative DQMC computations presented in Ref.~\cite{skyr_short}. We focus on two different interaction anisotropy parameters at a fixed $\mathcal{F}=0.01$: $|\JH|/\JA=0.5$, where we have  performed zero temperature projective quantum Monte-Carlo (PQMC) simulations; and $\JA=0$ which lies beyond the sign-problem-free regime of the QMC computations. Interestingly, we find that within the HF computation, the charge$-e$, spin$-3/2$ spin-polaron is \textit{not} energetically favored compared to the electronic quasi-particles in both cases. This suggests that the fluctuation effects captured using the DQMC computations in Ref.~\cite{skyr_short} are important for stabilizing the {\it skyrmionic} excitations with $S\geq 3/2$.

Based on our DQMC results reported in Ref.~\cite{skyr_short}, we had already noticed the possibility of an interaction anisotropy (i.e. $|J_{\rm{H}}|/J_{\rm{A}}$) tuned quantum phase transition from the ChAF to a superconductor at $\nu=2$. Our primary focus here will be on the field-theoretic aspects of the problem of fluctuating intertwined orders  near $\nu=2$.  We will examine critically the tendency towards deconfined quantum criticality as a function of interaction anisotropies and study the role played by topology in this non-perturbative setting. Since the fermionic degrees of freedom remain gapped in both phases, in Section \ref{sec::so5} we derive a low-energy effective field theory starting from the microscopic Hamiltonian with the interactions projected to the isolated topological bands. The resulting theory obtained by integrating out the gapped fermions takes the form of an $SO(5)$ non-linear sigma model (NL$\sigma$M) with a level$-1$ Wess-Zumino-Witten (WZW) term.

In Section \ref{sec::qpt}, we present our numerical PQMC results and perform a scaling analysis near the quantum phase transition(s). We provide suggestive numerical evidence for a single interaction-anisotropy tuned transition that is reminiscent of a deconfined quantum phase transition with an emergent approximate $SO(5)$ symmetry. The bandwidth tuned transition(s) associated with onset of superconductivity and loss of ChAF appear to show broadly similar behavior, though one of the critical exponents is further removed from those associated with interaction-anisotropy tuned transition. Some of these results are expected in light of recent numerical results obtained using fuzzy-sphere regularization of a problem with the same underlying symmetries \cite{sphere1, sphere2, sphere3}. We end with a summary of proposed future directions in Section \ref{sec::outlook}.

\section{Hartree-Fock Analysis: Quasiparticle(s) \& Collective Excitations}\label{sec::hf}

As per our DQMC computations \cite{skyr_short}, the model introduced above hosts a competition between an intertwined ChAF order and superconductivity in the vicinity of $\nu=2$ as a function of $|\JH|/\JA$. Moreover, in the vicinity of the phase transition between the two orders, superconductivity is driven by the condensation of skyrmion-antiskyrmion pairs \cite{ashvin_skrymion}. In the classic spinful quantum Hall setting, the skyrmion excitations on top of a quantum Hall ferromagnet can be obtained within a mean-field computation \cite{Fertig94, Fertig97, baby_skyrmion}. In this section, we analyze the same problem within mean-field theory based on a HF approximation, and investigate the extent of agreement with the non-perturbative DQMC computations. Interestingly, we observe that within the above approximation and including a single electron dressed by one spin-flip (i.e. in the ``baby-skyrmion" limit \cite{baby_skyrmion, Schindler22}), the composite excitation is \textit{not} the cheapest charged excitation both at the level of the single Chern sector ($\JA=0$) or both Chern sectors ($\JA=2|\JH|$), respectively.

Starting from the full Hamiltonian in Eqs.~\eqref{h_kin0},~\eqref{intra_valley_int} and \eqref{inter_valley_int}, we consider the following mean-field decomposition within the HF approximation,
\begin{equation}\label{ham_hf}
\begin{aligned}
H_{\tn{HF}}[P,Q]= \sum_{\tau}H^{\tau}_{\tn{kin}} +& \sum_{i;\alpha\beta} \bigg[H_{\tn{ph}}[P_{mn}]_{i} c^\dagger_{im} c_{in}^{\phantom\dagger} \\&
+ H_{\tn{SC}}[Q_{mn}]_{i} c_{im}c_{in} + \tn{h.c.}\bigg],
\end{aligned}
\end{equation}
where $m,n$ are flavor indices including spin, orbital and sublattice, and $P \equiv \avg{c^\dagger_{im}c_{in}^{\phantom\dagger}} $ and $Q \equiv \avg{c_{im} c_{in}}$ represent the single-particle density matrix in particle-hole and superconducting channels, respectively. A few representative mean-field channels along with their density matrices are summarized in Table~\ref{mfchannel}.
\begin{table}[h!]
    \centering
    \begin{tabular}{|c|c|}
    \hline
        \bf{Order Parameter} & \bf{Density Matrix}  \\
        \hline \hline
         Spin order & $c^{\dagger}_{\tau\alpha}\,\sigma^{\mu}_{\alpha\beta}\,c_{\tau\beta}^{\phantom\dagger}$ \\
         Valley order & $c^{\dagger}_{\alpha\sigma}\,\tau^{\mu}_{\alpha\beta}\,c_{\beta\sigma}^{\phantom\dagger}$ \\
         Spin-polarized inter-valley coherence~~~ & $c^\dagger_{+\sigma}\,c_{-\sigma}^{\phantom\dagger} \pm c^\dagger_{-\sigma}\,c_{+\sigma}^{\phantom\dagger}$ \\
         Spin-mixed inter-valley coherence & $c^\dagger_{+\sigma}\,c_{-\bar{\sigma}} \pm c^\dagger_{-\sigma}\,c_{+\bar{\sigma}}$ \\
         Spin-singlet SC ($\tau_z = 0$) & $c_{-\alpha}\,(i\sigma_y)_{\alpha\beta}\,c_{+\beta}$  \\
         Spin-triplet SC  & $c_{-\alpha}\,(i\sigma_y\sigma_\mu)_{\alpha\beta}\,c_{+\beta}$ \\
         \hline
    \end{tabular}
    \caption{Representative mean-field channels for decomposing the interactions in Eq.~\eqref{intra_valley_int}-\eqref{inter_valley_int} along with their quadratic density-matrices (see Eq.~\eqref{ham_hf}); repeated indices are summed over.}\label{mfchannel}
\end{table}

\begin{figure}
\includegraphics[width=90mm,scale=1]{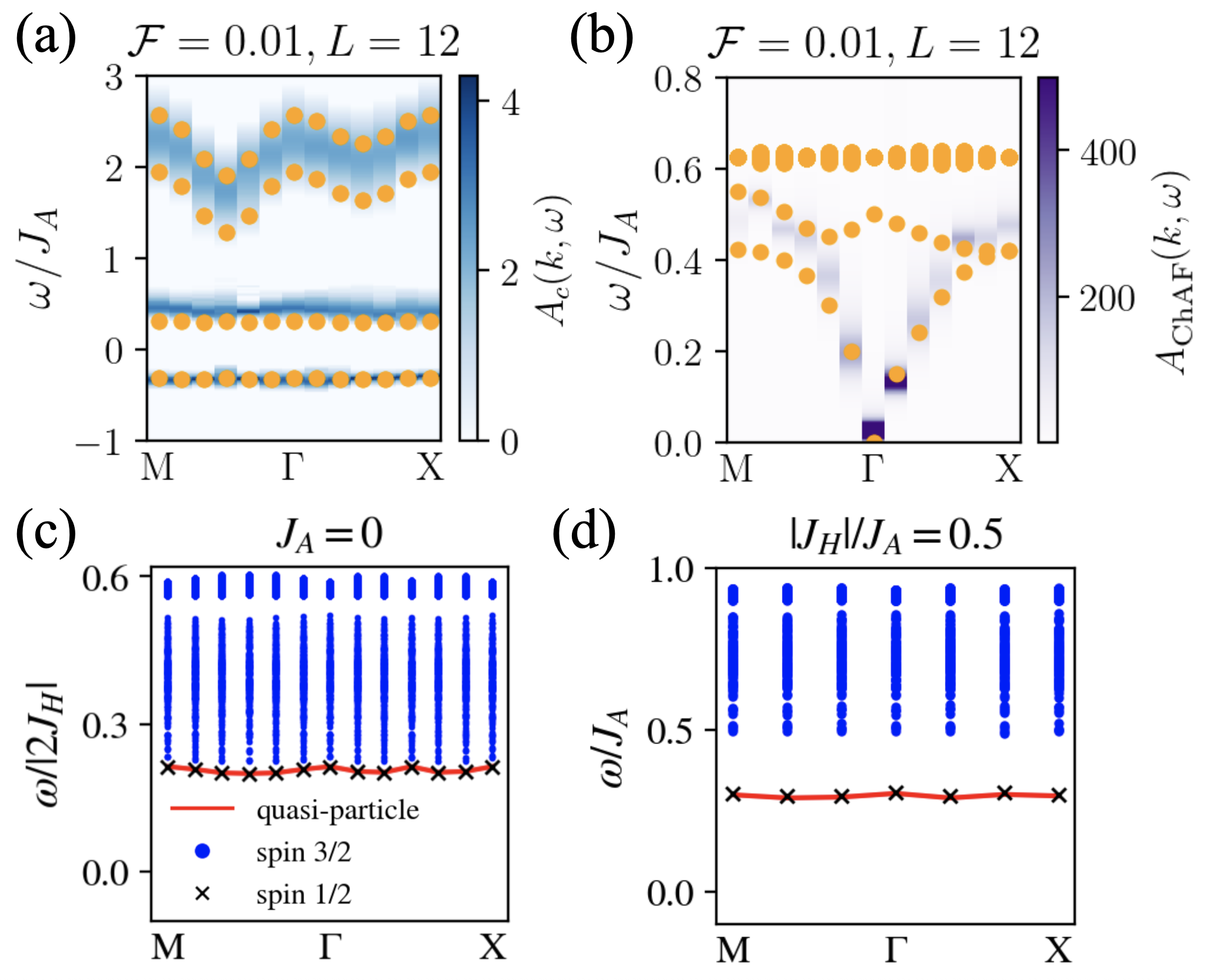}
\caption{(a)-(b) Excitation spectrum obtained from Hartree-Fock calculation (orange dots) at $|\JH|/\JA=0.5$ for (a) quasi-particle and for (b) magnon in the ChAF phase. Diffuse blue/violet curves denote the spectral function $A_\lambda(\k,\omega)$ obtained from stochastic analytical continuation in PQMC simulation, where $\lambda\equiv \tn{electron, ChAF}$. (c)-(d) Excitation spectrum for composite charged spin-polaron obtained from Hartree-Fock calculation at (c) $\JA=0$ and at (d) $|\JH|/\JA=0.5$. Blue dots denote spin-3/2 spin-polaron continuum; Black crosses denote the spin-1/2 spin-polaron excitation; red lines denote the electronic quasi-particle excitation spectrum, serving as a guide to eye for comparison. 
}
\label{fig::hf}
\end{figure}

Our HF computation for the ground-state lead to results that are qualitatively in agreement with the DQMC results. At $\nu=2$ the ground-state is a correlated ChAF with a density matrix, $P^{\mu}_{\mathrm{ChAF}}\equiv c^{\dagger}_{\gamma\alpha}\,\sigma^{\mu}_{\alpha\beta}\tau^z_{\gamma\gamma'}\,c_{\gamma'\beta}^{\phantom\dagger}$, which when doped with additional electrons or holes yields a spin-singlet valley-triplet superconductor with density matrix, $Q_{\mathrm{SC}}\equiv c_{-\alpha}\,(i\sigma_y)_{\alpha\beta}\,c_{+\beta}$. In the symmetry broken ground state, the single quasi-particle excitation spectrum along a specific high-symmetry cut is plotted in Fig.~\ref{fig::hf}(a); see orange dots. For a direct quantitative comparison with the DQMC results, we also show the electronic spectral function, $A_c(\k,\w)$, obtained from stochastic analytical continuation in the same figure obtained for the same set of microscopic parameters; see blue curves. We have found that the interaction-induced quasi-particle gaps broadly agree with each other. We will next turn to the collective mode excitations obtained within the mean-field setup. Additional details can be found in Appendix \ref{App::eig}.

\subsection{Collective mode excitations}
\begin{figure}
\includegraphics[width=60mm,scale=1]{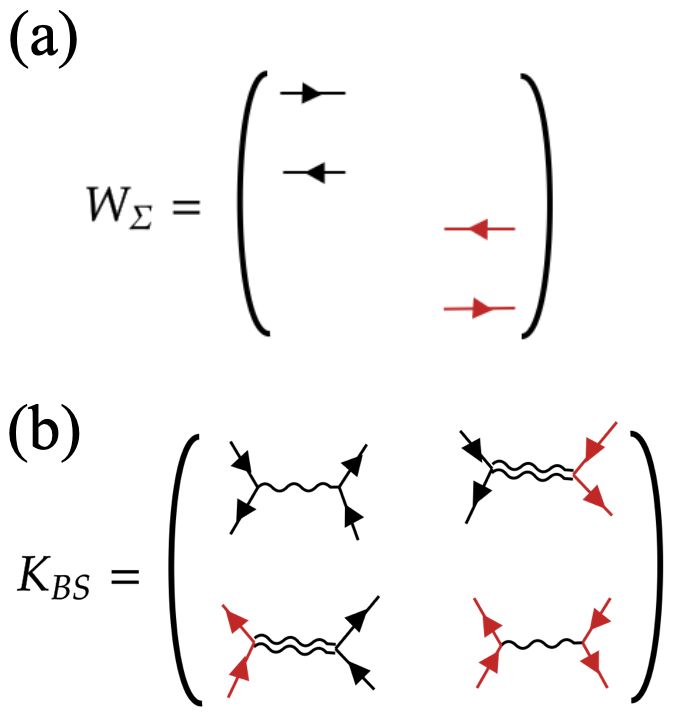}
\caption{\label{fig::feyn1}
Feynman diagrams for ChAF collective mode (a) propagator, (b) Bethe-Salpeter Kernel. Black (red) solid lines with arrow denote the Green's function of electronic quasi-particle. Wavy lines and double wavy lines denote the interaction vertex from intra-valley Hunds coupling ($\sim \JH$) and inter-valley anti-ferromagnetic coupling ($\sim\JA$), respectively.
}
\end{figure}

In this section, we will focus on the collective excitations, e.g. the Goldstone mode in the ChAF phase at $\nu=2$, within the mean-field computation. Consider the classical symmetry-broken ChAF ground state in the valleys $\tau=\pm$, characterized by $\ket{+\down}$ and $\ket{-\up}$. The magnon excitations are created by,
\begin{subequations}
\beq
&
b^\dagger_{\k,\q;+} \equiv d^{\dagger}_{\k+\q,+\up} d_{\k,+\down}^{\phantom\dagger}, \\&
b^\dagger_{\k,\q;-} \equiv d^{\dagger}_{\k+\q,-\down} d_{\k,-\up}^{\phantom\dagger}.
\eeq
\end{subequations}
Here $d_{\tau\alpha}(\k)$ denotes the electronic operator projected onto the Hilbert space of active bands with Bloch wavefunctions, $u_{\eta\tau}(\k)$, which is related to the microscopic electronic operators, $c_{\eta\tau\alpha}(\k)$, 
\beq
c_{\eta\tau\alpha}(\k) = u_{\eta\tau}(\k)d_{\tau\alpha}(\k).\label{eq::basis_change}
\eeq
It is natural to group $(b^\dagger_{\k,\q;+}, b_{\k,\q;-}^{\phantom\dagger})$, as is done within the usual Holstein-Primakoff transformation for the antiferromagnetic problem. The propagator for this collective excitation can be expressed diagrammatically in Fig.~\ref{fig::feyn1}(a), where we represent the electronic propagator for the $\tau=+~(-)$ valley in black (red); the wavy and double wavy lines denote the interaction vertex from intra-valley Hunds coupling ($\sim \JH$) and inter-valley anti-ferromagnetic coupling ($\sim\JA$), respectively. By setting up the Bethe-Salpeter kernel in Fig.~\ref{fig::feyn1}(b), the two-particle excitation spectrum in the ChAF phase can be obtained by solving for the poles of the fully dressed magnon propogator, which is governed by Eq.~\eqref{eq::bse_caf} below. 

\beq\label{eq::bse_caf}
&
\det(W_{\Sigma}^{-1} - K_{\tn{BS}}) = 0.
\eeq
We evaluate the particle-hole propagator in Eq.~\eqref{eq::bse_caf} with respect to the HF ground state at $\nu=2$, i.e. assuming that the filled states are $\ket{+\down}$ and $\ket{-\up}$. The entries are given by,
\begin{equation}\label{eq::w00}
\begin{aligned}
W_{\Sigma}^{(0,0)} 
&=
\sum_{k_0}\frac{1}{ik_0 - E^{+\up}_{\k}} \frac{1}{i(k_0-\Omega) - E^{+\down}_{\k-\q}} 
\\&
= 
\frac{1}{i\Omega - (E^{+\up}_{\k} - E^{+\down}_{\k-\q})},
\end{aligned}
\end{equation}
\begin{equation}\label{eq::w11}
\begin{aligned}
W_{\Sigma}^{(1,1)} 
&=
\sum_{k_0}\frac{1}{ik_0 + E^{-\down}_{-\k}} \frac{1}{i(k_0-\Omega) + E^{-\up}_{-\k+\q}} 
\\&
= 
\frac{1}{-i\Omega - (E^{-\down}_{-\k} - E^{-\up}_{-\k+\q})}.
\end{aligned}
\end{equation}
Here, $ik_0$ ($i\Omega$) denotes the fermionic (bosonic) Matsubara frequency. $E^{\tau\sigma}_{\k}$ is the Hartree-Fock renormalized band dispersion for valley $\tau$ and spin $\sigma$. 
When arranging the momentum (frequency) labels in Fig.~\ref{fig::feyn1}, we use the convention that the incoming momentum (frequency) is given by $\q$ ($\Omega$). The internal momentum, $\k$, can be treated as matrix indices when diagonalizing the spectrum. We notice that the external frequency $\Omega$ in Eq.~\eqref{eq::w00}-\eqref{eq::w11} has contrasting sign, which is reminiscent of the Holstein-Primakoff treatment of the antiferromagnetic spin-wave problem on a bipartite lattice, where the effective Hamiltonian of spin-wave excitation is a bosonic Bogoliubov Hamiltonian. 

The solution to Eq.~\eqref{eq::bse_caf} can be treated as a generalized eigenvalue problem. The excitation spectrum given by the eigenvalues of this equation are shown in Fig.~\ref{fig::hf}b along a high symmetry cut. At the $\Gamma$-point, the spectrum is gapless, as expected based on Goldstone's theorem. For a direct quantitative comparison, we also show the ChAF spectral function, $A_{\rm{ChAF}}(\k,\omega)=\tn{Im}\la P^\dagger_{\rm{ChAF}}(\k,\omega) P_{\rm{ChAF}}(\k,\omega)\ra |_{\omega\rightarrow\omega+i0^{+}}$, obtained from stochastic analytical continuation using our DQMC data in Fig.~\ref{fig::hf}b. In spite of the usual numerical uncertainties associated with analytical continuation, we find a qualitative agreement between the ChAF spectrum obtained from QMC and the momentum-resolved spectrum from the HF calculation. We note that the velocity of the Goldstone mode is slightly overestimated in the HF computation compared to the QMC results.

\subsection{Composite charged excitations}

\begin{figure}[htb]
\includegraphics[width=90mm,scale=1]{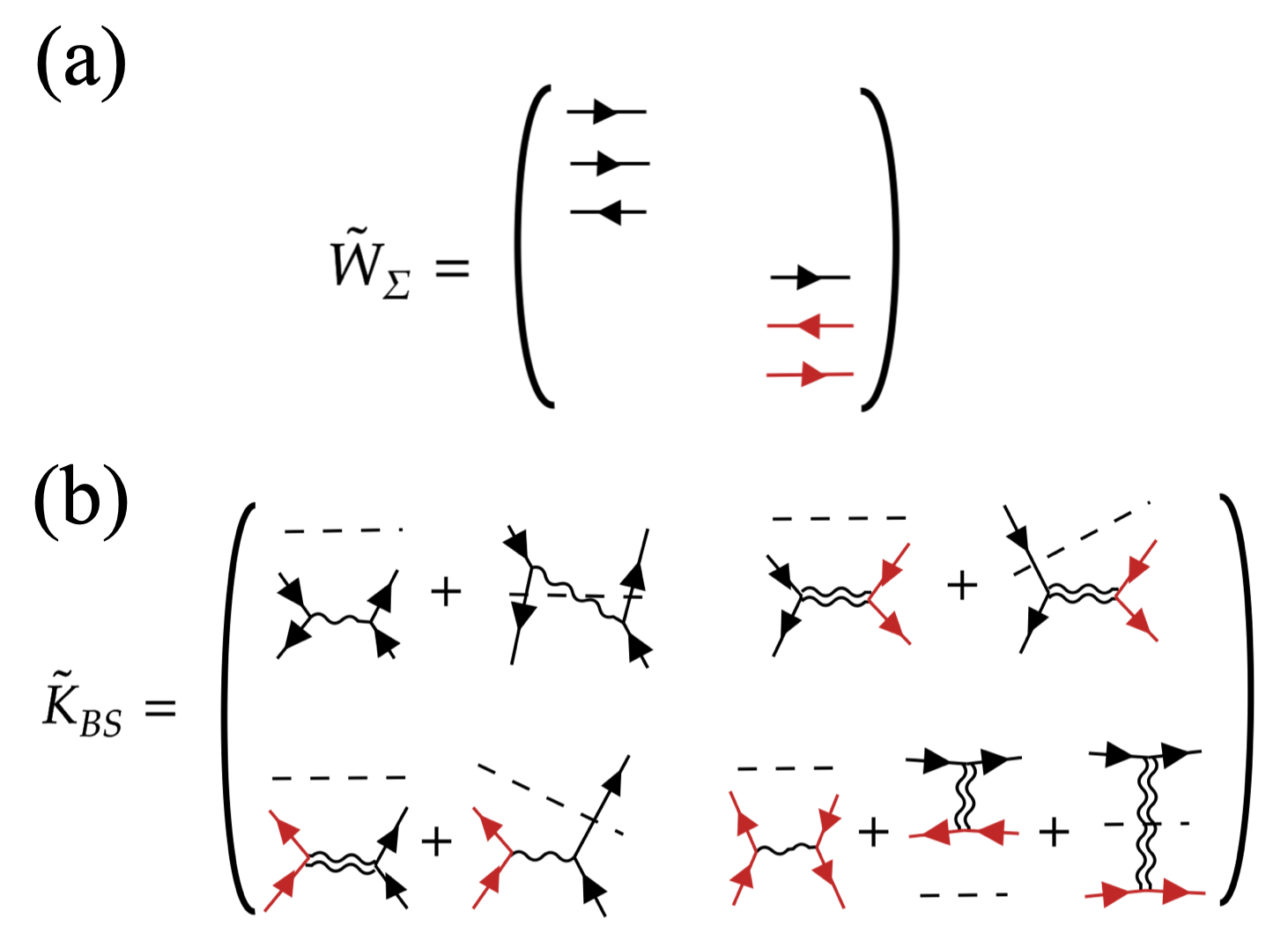}
\caption{\label{fig::feyn2}
Feynman diagrams for composite charged excitations showing the (a) propagator, and (b) Bethe-Salpeter Kernel. Dashed lines denote identity in the corresponding matrix entries. Color convention is the same as in Fig.~\ref{fig::feyn1}.
}
\end{figure}

Starting from the ChAF ground state obtained within HF approximation at $\nu=2$, we now examine the energetics of the cheapest charge$-e$ excitation. Based on the DQMC results which find evidence of spin-polaron excitations \cite{skyr_short}, we pay special attention to the energy of the composite excitation formed by binding a single electron with one spin-flip \cite{baby_skyrmion}. Due to the underlying $SU(2)_{\rm{spin}}\otimes U(1)_{\rm{charge}}\otimes U(1)_{\rm{valley}}$ symmetry at generic $\JA$ and $\JH$, the excitations of key interest have $S_z=3/2$, $\delta N_{\tau=+}=1$ and $\delta N_{\tau=-}=0$. Specifically, two distinct forms of such composite excitations can be formed: electron with $\tau=+,~\sigma=\up$ bound to a magnon creation operator in $\tau=+$ valley, whose creation operator is defined as,
\beq\label{eq::s_polaron_dag}
s^{\dagger}_{+}\equiv 
d^{\dagger}_{\k_{e1};+\up}
b^\dagger_{\k_{e2},\q;+},
\eeq
and an electron with $\tau=+, \sigma=\up$ bound to a magnon annihilation operator in $\tau=-$ valley, whose creation operator is defined as
\beq\label{eq::s_polaron}
s^{\dagger}_{-}\equiv 
d^{\dagger}_{\k_{e1};+\up}
b_{\k_{e2},\q;-}.
\eeq
The propagators of $s^{\dagger}_{+}$ and $s^{\dagger}_{-}$ are depicted diagrammatically by $\tilde{W}_{\Sigma}^{(0,0)}$ and $\tilde{W}_{\Sigma}^{(1,1)}$ in Fig.~\ref{fig::feyn2}(a). As in the case of the previous Bethe-Salpeter computation, the internal momentum $\k_{e1}$ and $\k_{e2}$ are treated as indices of the matrices in Fig.~\ref{fig::feyn2}(a)-~\ref{fig::feyn2}(b). We use the same notation for interaction vertices and electronic Green's functions in Fig.~\ref{fig::feyn2} as Fig.~\ref{fig::feyn1}; the dashed lines in Fig.~\ref{fig::feyn2}(b) denote identity in the corresponding matrix entries.
The interaction in this composite excitation basis is depicted in Fig.~\ref{fig::feyn2}(b). The excitation spectrum can be obtained by solving the generalized eigenvalue problem in Eq.~\eqref{eq::tw_eig} below, 
\beq
\det(\tilde{W}_{\Sigma}^{-1} - \tilde{K}_{\tn{BS}}) = 0.\label{eq::tw_eig}
\eeq
The propagators for these composite excitations can be evaluated as,
\begin{equation}\label{eq::tw00}
\begin{aligned}
\tilde{W}_{\Sigma}^{(0,0)} 
&=
\sum_{\Omega} \frac{1}{i(\Omega_{\tn{tot}}-\Omega) - E^{+\up}_{\k_{e1}}} \frac{1}{i\Omega - ( E^{+\up}_{\k_{e2}} - E^{+\down}_{\k_{e1} + \k_{e2} - \q} )} 
\\&
= 
\frac{-1}{i\Omega_{\tn{tot}} - (E^{+\up}_{\k_{e1}} +  E^{+\up}_{\k_{e2}} - E^{+\down}_{\k_{e1} + \k_{e2} - \q} )},
\end{aligned}
\end{equation}
and similarly,
\begin{equation}\label{eq::tw11}
\begin{aligned}
\tilde{W}_{\Sigma}^{(1,1)} 
&
= 
\frac{1}{i\Omega_{\tn{tot}} - (E^{+\up}_{\k_{e1}}  - E^{-\down}_{-\k_{e2}} + E^{-\up}_{\q -\k_{e1} - \k_{e2}} )}.
\end{aligned}
\end{equation}

Let us start by discussing the results for $\JA=0$, when the two valleys decouple. Note that this problem lies outside the reach of the DQMC computations. 
Due to the SU(2)$^{\tau=+}_{\tn{spin}}\otimes$SU(2)$^{\tau=-}_{\tn{spin}}$ spin symmetry, the spin quantum numbers per valley are separately conserved. Let us evaluate the total (i.e. macroscopic) spin quantum number $S^{\tau}_{\tn{tot}}$ of a given Hartree-Fock state $\ket{\Psi_{\mathrm{HF}}^{\delta\nu}}$ at doping $\delta\nu$ (defined below) via
\begin{subequations}\label{eq::s_tot_def}
\beq
\chi^{\tau}(\q) &=& \frac{1}{L^2}\sum_{\r,\r'}e^{-i(\r-\r')\cdot\q}\avg{\vec{S}^{\tau}_{\r'}\cdot\vec{S}^{\tau}_{\r}},~~~\\
\chi^{\tau}(\q=0) &=& S_{\tn{tot}}^{\tau}(S_{\tn{tot}}^{\tau}+1)/L^2, 
\eeq
\end{subequations}
where the average in Eq.
\eqref{eq::s_tot_def}a is evaluated w.r.t. $\ket{\Psi_{\mathrm{HF}}^{\delta\nu}}$ and $\vec{S}^{\tau}_{\r} \equiv \sum_{\alpha,\beta = \uparrow,\downarrow}c^{\dagger}_{\r,\tau,\alpha}\vec{\sigma}_{\alpha\beta}c_{\r,\tau,\beta}
$. Here, given $\ket{\Psi_{\mathrm{HF}}^{\delta\nu=0}}$ is a fully polarized state, the macroscopic spin quantum number associated with it is  $S_{\tn{tot}}^{\tau}(\delta\nu=0) = N/2$ ($N=L^2$ where $L$ is the linear system size). The excited state with a single composite excitation in $\tau=+$ valley is defined as $\ket{\Psi_{\mathrm{HF}}^{\delta\nu=1}}\equiv s^{\dagger}_{+} \ket{\Psi_{\mathrm{HF}}^{\delta\nu=0}}$ (where $s^{\dagger}_{+}$ is defined in Eq.~\eqref{eq::s_polaron_dag}), whose spin polarization in $z-$direction is $S^{\tau=+}_{z} = (N-3)/2$. 
The macroscopic spin quantum number of $\ket{\Psi_{\mathrm{HF}}^{\delta\nu=1}}$ can be $S_{\tn{tot}}^{\tau=+}(\delta\nu=1) = (N-1)/2$ or $(N-3)/2$.
The spin quantum number of the composite excitation $s^{\dagger}_{+}$ is defined as the change of macroscopic spin quantum number $\Delta S_{\tau=+} \equiv S_{\tn{tot}}^{\tau=+}(\delta\nu=0) - S_{\tn{tot}}^{\tau=+}(\delta\nu=1)$, which can take the value of $\Delta S_{\tau=+} =3/2$ or $\Delta S_{\tau=+} = 1/2$. For simplicity, let us drop the $\Delta$ in denoting the spin quantum number of the composite excitation.
 In Fig.~\ref{fig::hf}(c), we plot the excitation spectrum for the $S_{\tau=+}=3/2$ spin-polarons (blue dots) and the $S_{\tau=+}=1/2$ electronic quasi-particles (black crosses) for $\JA=0$. We also compare the dispersion of these composite excitations with the renormalized HF single-particle spectrum (red line). These results suggest that for $\JA=0$ the cheapest charged excitation is the $S_{\tau=+}=1/2$ electronic quasi-particle. Moreover, there is no energy gap between the $S_{\tau=+}=1/2$ and $S_{\tau=+}=3/2$ charge$-e$ excitation, highlighting that within the single-Chern sector, the charged spin-polaron excitations are nearly degenerate with electronic quasi-particles.

We now turn to $|\JH|/\JA=0.5$, where a vast majority of our QMC computations have been performed. In Fig.~\ref{fig::hf}(d), we show the spectrum of composite and single-particle excitations. Interestingly, the electronic quasi-particle excitations continue to remain the cheapest charged excitation and are now well-separated by an energy gap from the $S_{\tau=+}=3/2$ spin-polaron excitations. Note that for $\JA\neq0$, the spin quantum number in $\tau=+$ valley is no longer separately conserved; the composite excitations can be in a coherent superposition among spin$-1/2$, spin$-3/2$ etc. However, the quantitative agreement between the dispersion of the HF renormalized single-particle excitation and the lowest energy composite excitation suggests that the cheapest charged excitation has a large overlap with the spin$-1/2$ and {\it not} the spin$-3/2$ charged spin-polaron.

To summarize, we have found that while the single electron-like excitations and the neutral collective modes show broad agreement between the HF computations and DQMC, the charged spin-polarons do not. In particular, while the cheapest charged excitations near $\nu=2$ are skyrmionic in nature with $S>1/2$ for a range of $\JA/\JH$ as determined by our DQMC computations \cite{skyr_short}, these are not the lowest energy excitations within our Bethe-Salpeter treatment of the problem on top of the HF renormalized ground state. 

\section{Low Energy Effective Field Theory}\label{sec::so5}

In this section starting from the model introduced above we will analyze the underlying field-theoretic structure that governs the possible transition(s) between a superconductor and insulating ChAF at $\nu=2$ and $T=0$ as a function of increasing $|\JH|/\JA$. We analyze the salient aspects of the low-energy effective theory after integrating out the gapped fermionic degrees of freedom on either side of the purported transition. In particular, we show that the effective theory is described in terms of an $SO(5)$ NL$\sigma$M with a level$-1$ WZW term for a bosonic ``superspin". We will analyze quantitatively the nature of the critical point(s) obtained within our DQMC computations in Section \ref{sec::qpt} below. 

Given the ordering tendencies towards a ChAF and spin-singlet valley-triplet ($\tau_z=0$) SC near $\nu=2$, we introduce the order-parameters, 
\begin{subequations}
\beq
\label{so5:1}
O^{\mu}_{\tn{ChAF}}(\vec{r}) &\equiv& \sum_{\tau;\alpha\beta=\up\down}\tau c^{\dagger}_{\vec{r}\tau\alpha}\sigma^{\mu}_{\alpha\beta}c_{\vec{r}\tau\beta} ,~ \mu=1,2,3, ~~\\
\label{so5:2}
O_{\tn{SC}}(\vec{r}) &\equiv& c^{\dagger}_{\vec{r}+\up} c^{\dagger}_{\vec{r}-\down} - c^{\dagger}_{\vec{r}+\down} c^{\dagger}_{\vec{r}-\up}
\eeq
\end{subequations}
Here, $\vec{r}\equiv \vec{r}_{\vec{R}\eta}$ denotes the coordinates of the sublattice $\eta$ in the unit cell $\vec{R}$. Let us now introduce the Nambu-spinor projected onto the basis of the active low-energy bands, $\Psi_{\vec{k}}^\dagger\equiv [d^\dagger_{+\up}(\vec{k})~, d^\dagger_{+\down}(\vec{k})~, d_{-\up}(-\vec{k})~, d_{-\down}(-\vec{k})]$, and use $\mu_\alpha$ as Pauli matrices in particle-hole space. Here $d_{\tau\alpha}(\k)$ denotes the electronic operator in the Hilbert space of active bands defined in Eq.\eqref{eq::basis_change}. 

It is natural to combine the order parameters in Eq.~\eqref{so5:1}-\eqref{so5:2} into a five-dimensional super-spin \cite{so5_1, so5_2, so5_3, so5_demler, so5_param1, auerbach1998quantum,TH05,SF06}. After summing over sublattice indices $\eta$, the super-spin can be expressed as a product between a common form-factor which factorize\cite{si} and bilinears of the projected operator $\Psi$ as follows:
\beq
\begin{aligned}
    &
    N_{1} = \tn{Re} [\Delta_{\tn{S}}] = \frac{1}{2}\Psi^\dagger\sigma_y\mu_y\Psi \\&
    N_{2} = S_x \tau_z = \frac{1}{2}\Psi^\dagger\sigma_x\Psi \\&
    N_{3} = S_y \tau_z = \frac{1}{2}\Psi^\dagger\sigma_y \mu_z\Psi \\&
    N_{4} = S_z \tau_z = \frac{1}{2}\Psi^\dagger\sigma_z\Psi \\&
    N_{5} = -\tn{Im} [\Delta_{\tn{S}}] = \frac{1}{2}\Psi^\dagger\sigma_y \mu_x\Psi.
\end{aligned}\label{eq::rotor_main}
\eeq
The low-energy effective theory in terms of the $SO(5)$ rotor $N_{\alpha}$ and the corresponding angular momentum $iL_{\alpha\beta}\equiv[N_\alpha, N_\beta] $ acting on $N_{\alpha}$ can be formulated as, 
\beq
\begin{aligned}
    &
    L_{15} = -(n_{\tn{tot}}-2)/2 = -\mu_z/2 \\&
    L_{23} = S_z = \sigma_z\mu_z/2 \\&
    L_{24} = -S_y = -\sigma_y/2 \\&
    L_{34} = S_x = \sigma_x\mu_z/2 \\&
    L_{45} = -\tn{Re}[\Delta_{\tn{T}z}] = -\sigma_x\mu_x/2, ~~ L_{14} = \tn{Im}[\Delta_{\tn{T}z}] = \sigma_x\mu_y/2 \\&
    L_{25} = \tn{Re}[\Delta_{\tn{T}x}] = -\sigma_z\mu_x/2, ~~ L_{12} = \tn{Im}[\Delta_{\tn{T}x}] = \sigma_z\mu_y/2 \\&
    L_{35} = \tn{Re}[\Delta_{\tn{T}y}] = \mu_y/2, ~~ L_{13} = \tn{Im}[\Delta_{\tn{T}y}] = \mu_x/2.
\end{aligned}\label{eq::Angular_momentum}
\eeq
Additional details can be found in Appendix \ref{app::so5}. Let us now rewrite the microscopic theory with interactions projected to the low-energy bands in terms of the rotor variables introduced above, which yields
% \begin{widetext}
\beq\label{eq::ham_rotor_k_main}
\begin{aligned}
H_{\tn{rotor}} = &\sum_{\vec{k},\vec{p},\vec{q}} 
F_{\vec{k},\vec{p},\vec{q}}
\bigg[
U_0 \Psi^{\dagger}_{\vec{k}+\vec{q}}  \Psi_{\vec{k}} 
\Psi^{\dagger}_{\vec{p}-\vec{q}}  \Psi_{\vec{p}}\\&
+
\sum_{\alpha\beta} U_{\alpha\beta}
\Psi^{\dagger}_{\vec{k}+\vec{q}} L_{\alpha\beta} \Psi_{\vec{k}} 
\Psi^{\dagger}_{\vec{p}-\vec{q}} L_{\alpha\beta} \Psi_{\vec{p}}\\&
+
\sum_{\mu}
V_{\mu} \Psi^{\dagger}_{\vec{k}+\vec{q}} N_{\mu} \Psi_{\vec{k}} 
\Psi^{\dagger}_{\vec{p}-\vec{q}} N_{\mu} \Psi_{\vec{p}}
\bigg] \\&
+ \frac{3|\JH|}{16}\sum_{\alpha\eta; \vec{k}\vec{p} }~\bigg[
|u_{\eta +}(\vec{k})|^2|u_{\eta +}(\vec{p})|^2 d^{\dagger}_{\vec{p};+\alpha} 
d_{\vec{p};+\alpha} 
\bigg],
\end{aligned}
\eeq
% \end{widetext}
where $F_{\vec{k},\vec{p},\vec{q}} \equiv \sum_{\eta} u
_{\eta +}^{*}(\vec{k}+\vec{q}) u_{\eta +}(\vec{k}) u_{\eta +}^{*}(\vec{p}-\vec{q}) u_{\eta +}(\vec{p})$ is the form-factor from the flat-band projection and the coupling strengths are given by,
%\begin{subequations}
\beq
% \begin{aligned}
&U_0 = U_{15} = \frac{3|\JH|}{8}, U_{23}=U_{34}=U_{24} = \frac{\JA-3|\JH|}{8},\nn\\&
U_{25}=U_{35}=U_{45} =U_{12} = U_{13} = U_{14} = \frac{\JA}{16},\nn \\&
V_1=V_5 = -\frac{3\JA}{16}, ~~~~ V_2=V_3=V_4 = -\frac{\JA+3|\JH|}{8}.\,
\label{eq::param}
% \end{aligned}
\eeq
%\end{subequations}
Clearly, the ratio, $|\JH|/\JA$, tunes the anisotropy between the ChAF and SC orders. Remarkably, at $\JA=6|\JH|$, an $SO(5)$ symmetry emerges in the projected low-energy effective theory in the flat-band limit. It is important to point out that we have {\it not} included the kinetic energy contribution from the bare bandwidth term in the above Hamiltonian. Thus any low-energy emergent symmetry associated with the projected Hamiltonian and possible critical points will likely arise at a different ratio of $\JA/|\JH|$. Note that the last term in Eq.~\eqref{eq::ham_rotor_k_main} appears as a valley-dependent potential. However, assuming that the projector to the flat-band $u^*_{\eta+}(\k)u_{\eta+}(\k)$ is site-independent (a.k.a the more commonly invoked uniform pairing condition \cite{uniform-pairing}), this term becomes a conserved quantity due to the underlying microscopic U(1)$_\tn{charge}\otimes$ U(1)$_\tn{valley}$ symmetry and thus does not affect the dynamics. Furthermore, the time-reversal symmetry $\mathcal{T}\equiv i\tau_y\mathcal{K}$ dictates that the electron density in different valleys should be equal, which enables us to drop the terms in the fourth line in Eq.~\eqref{eq::ham_rotor_k_main} and focus only on the quartic contributions.

When the interaction-induced single-particle gap is finite, a WZW term can be naturally generated from the underlying topological band \cite{lee_sachdev1, lee_sachdev2, qptbook}, which encodes the fluctuations in the intertwined order-parameters included in the super-spin.
By integrating out the gapped fermionic degrees of freedom in the vicinity of the transition at $\nu=2$ and at $T=0$, and performing a standard gradient expansion, we obtain (see Appendix \ref{app::so5})
\beq\label{eq::wzw_main}
\begin{aligned}
&
S_{\tn{WZW}} = 2\pi i  ~W[N_a],\\&
W[N_a] = \frac{3}{8\pi^2}\int_u \int_{\r,\tau} ~\varepsilon_{abcde} N^{a}\partial_x N^b \partial_y N^c \partial_\tau N^d \partial_u N^e.
\end{aligned}
\eeq
The quantized coefficient of the WZW term in Eq.~\eqref{eq::wzw_main} reflects the topology of the Goldstone manifold $\pi_4(\frac{\rm{SO(5)}}{\rm{SO(4)}})=\mathcal{Z}$. More importantly, Eq.~\eqref{eq::wzw_main} assigns electric charge $2e$ to a topological skyrmion texture \cite{lee_sachdev2} formed by the ChAF order parameter in real-space, which can be physically viewed as a bound state between charge $e$ skyrmion in $C=1$ valley and charge $e$ anti-skyrmion in $C=-1$ valley \cite{ashvin_skrymion}. Thus, the above low-energy effective theory serves as an ideal platform to study a possibly deconfined quantum phase transition.  

\section{Quantum Phase Transitions at $\nu=2$}\label{sec::qpt}

\begin{figure}[htb]
\includegraphics[width=70mm,scale=1]{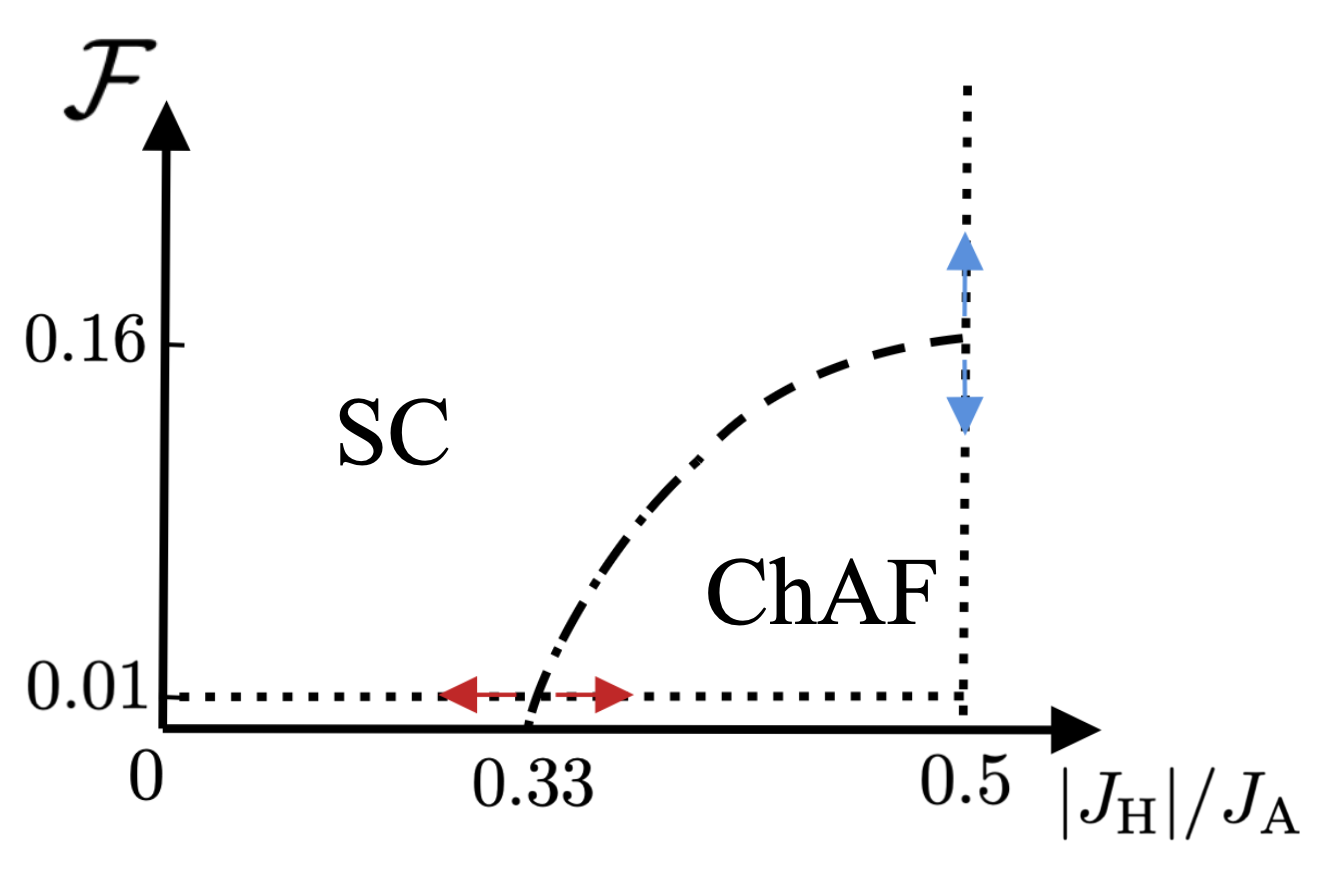}
\caption{\label{fig::qpt_schematic}
A schematic representation of the flatness ratio and interaction-anisotropy tuned quantum phase transition(s). Dotted line denotes the parameter set scanned in PQMC. Red and blue arrows denote the vicinity of the transitions observed in PQMC. 
}
\end{figure}

In this section, we present our zero temperature PQMC results studying the distinct phase transitions at a fixed filling $\nu=2$ as a function of $|\JH|/\JA$ and bandwidth, which highlights the role of the emergent $SO(5)$ symmetry discussed in Section \ref{sec::so5} above and the proximity to a deconfined critical point. Within PQMC, for a system defined on a lattice with $L\times L$ unit cells, the ground state is obtained via a projection, $\ket{\Psi^{\Theta}_{\rm{G}}}\equiv e^{-\Theta H}\ket{\Psi_{\rm{trial}}}$, along the imaginary time axis with respect to a trial state. Here $\ket{\Psi_{\rm{trial}}}$ is a Slater determinant state with a fixed filling fraction, $\nu\equiv\sum_{\r}\avg{n_{\r}}/L^2=2$, carefully chosen to preserve the microscopic $U(1)_{\rm{charge}}\times U(1)_{\rm{valley}}\times SU(2)_{\rm{spin}}$ symmetry. The implementation of the imaginary time evolution is performed via a Trotter decomposition with $\Delta \tau = \Theta/N_{\tn{Trotter}}$; the interaction terms are factorized through a discrete Hubbard-Stratonovich transformation, with auxiliary fields sampled stochastically via single spin-flip updates. Again, we stress that we have use the short hand $\vec{r}\equiv \vec{r}_{\vec{R}\eta}$ to denote the coordinates of the sublattice $\eta$ of the $\vec{R}$-th unit cell, and use the shorthand $\r'\equiv \r_{\vec{R'}\eta'}$.

In order to diagnose the quantum phase transition(s) and the associated critical exponents, we introduce the correlation length $\xi_O$ and equal-time correlation function $S_O(\q)$,
\begin{subequations}\label{eq::obs_main}
\beq
\xi_O &\equiv& \frac{1}{2 \sin(\pi/L)}\sqrt{\frac{S_O(\vec{q} = \vec{0})}{S_O(\vec{q} = (2\pi/L,0))} - 1},\quad\\
S_O(\q) &=& \frac{1}{L^2}\sum_{\r,\r'}e^{-i(\r-\r')\cdot\q}\avg{O^{\dagger}(\r')O(\r)},
\eeq
\end{subequations}
where $O\equiv$ChAF, SC are the order parameters of interest; see Eq.~\eqref{so5:1}-\eqref{so5:2}. A finite-size-scaling analysis will be performed based on Eq.~\eqref{eq::obs_main} for varying linear system size $L$. In PQMC computations, the imaginary-time projection length $\Theta$ is also chosen to scale with linear system size $L$, assuming a dynamical exponent $z=1$ \cite{fakher_scaling}.

In the following, we first present the numerical results for the interaction-anisotropy and bandwidth-tuned phase transitions (see Fig.~\ref{fig::qpt_schematic} for a schematic) in Sections \ref{sub::aniso-phase-transition} and \ref{sub::bandwidth-phase-transition}, respectively, and then provide a possible interpretation in Section \ref{sub::interpretation}.

\subsection{Interaction anisotropy-tuned transition}\label{sub::aniso-phase-transition}
\begin{figure*}[htb]
\includegraphics[width=180mm,scale=1]{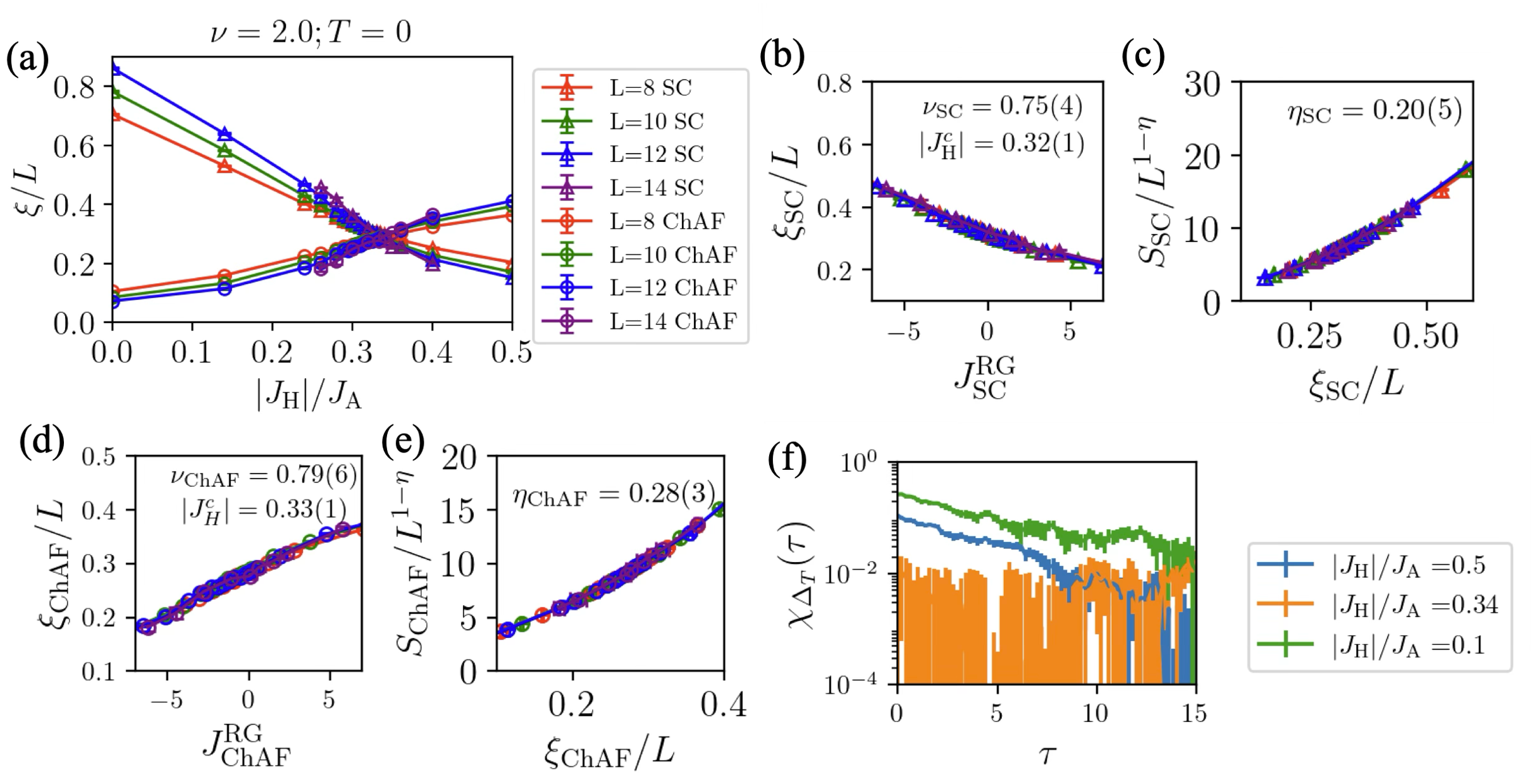}
\caption{\label{Fig::aniso} Interaction anisotropy tuned transition at $\nu=2.0, T=0$ and $\mathcal{F}=0.01$. (a) Correlation length $\xi_O$ for $O=\rm{ChAF, SC}$ for different system sizes. Scaling collapse analysis for (b) SC correlation length, where $J^{\tn{RG}}_{\tn{O}}$ is defined in Eq.~\eqref{eq::collect_fit1}b. Scaling collapse analysis for (c) static SC susceptibility, (d) ChAF correlation length, and (e) static ChAF susceptibility. The critical exponents in (b)-(e) are obtained from collective fitting using Eq.~\eqref{eq::collect_fit1}a and Eq.~\eqref{eq::collect_fit2}. (f) Dynamical correlation function $\chi_{\Delta_{\tn{T}}}(\tau)$ defined in Eq.~\eqref{eq::dyn_so5} evaluated at $L=8$. For $|\JH|/\JA=0.1$ and $|\JH|/\JA=0.5$ situated far away from the quantum critical point, $\chi_{\Delta_{\tn{T}}}(\tau)$ decays exponentially; for $|\JH|/\JA=0.34$, $\chi_{\Delta_{\tn{T}}}(\tau)$ is approximately constant.
}
\end{figure*}

Let us begin by studying the problem at $\nu=2$ and $T=0$, keeping $\mathcal{F}=0.01$ fixed. To tune the interaction anisotropy, we modify $|\JH|$ while keeping $\JA$ fixed. Since $\JA\geq2|\JH|$ to preserve the sign-problem-free nature of the model, the interaction strengths remain smaller than the gap to the remote bands, $E_{\tn{gap}}$, as we vary $|\JH|$. Finite-size-scaling results for the RG-invariant correlation length, $\xi_{O}/L$, are shown in Fig.~\ref{Fig::aniso}(a). Starting from the limiting case of $|\JH|/\JA=0.5$, where the ground state at $\nu=2$ is an insulating ChAF, decreasing $|\JH|$ drives a tendency towards SC while destroying the ChAF order. In order to determine the location of the quantum critical point(s) and their associated universality, we use the following fitting analysis by utilizing a scaling ansatz \cite{fakher_scaling},
\begin{subequations}\label{eq::collect_fit1}
\beq
\frac{\xi_O(J_\tn{H}, L)}{L} &=& f_\xi(J_{\tn{O}}) + L^{-\omega} g_{\xi}(J_{\tn{O}}),\\
J^{\tn{RG}}_{\tn{O}} &=& [(J_{\tn{H}}-J_{\tn{H}}^{\tn{c}})/J_{\tn{H}}^{\tn{c}}]L^{1/\nu_{\tn{O}}},\label{eq::Jrg}
\eeq
\end{subequations}
where the equal-time correlation functions,
\beq\label{eq::collect_fit2}
S_{O}(\q=0;\JH,L) = L^{1-\eta_O}~h_{S,\xi}(\xi_O/L).
\eeq
We group the data together from the different (linear) system-sizes for fitting and assume a polynomial form for the scaling functions, $f_\xi(J_{\tn{O}})$, $g_{\xi}(J_{\tn{O}})$ and $h_{S,\xi}(\xi_O/L)$ in Eq.~\eqref{eq::collect_fit1}-\eqref{eq::collect_fit2}, respectively. The results show scaling-collapse for both superconductivity (Fig.~\ref{Fig::aniso}b-c) and ChAF order (Fig.~\ref{Fig::aniso}d-e). Interestingly, as a function of decreasing $|\JH|$, we determine the quantum critical point associated with onset of superconductivity, $|\JH^{c1}|/\JA=0.32(1)$ to almost coincide with the quantum critical point for loss of ChAF order,  $|\JH^{c2}|/\JA=0.33(1)$.  
Relatedly, the correlation length critical exponent, $\nu_{\tn{SC}} = 0.75(4)$, and anomalous dimension, $\eta_{\tn{SC}}=0.20(5)$ deviate from the usual 3D-XY universality class. Similarly, $\nu_{\tn{ChAF}} = 0.79(6)$ and $\eta_{\tn{ChAF}}=0.28(3)$ also deviate from the usual 3D-O(3) universality class. In light of these results, and signatures of an emergent $SO(5)$ symmetry in the numerical data which we turn to next (see also discussion in Section \ref{sec::so5}), there are clear indications of a proximity to a deconfined quantum critical point (DQCP) \cite{dqcp1, dqcp2}. Note that we have also performed an independent crossing-point analysis which yields similar critical exponents but with larger uncertainties due to the limited range of available system sizes (Appendix \ref{App::cross}).

Next, we turn to one of the most interesting questions vis-\`a-vis the low-energy field theory, namely whether an emergent $SO(5)$ symmetry appears near the above quantum critical point(s). To probe this possible symmetry, we focus on the associated emergent conserved charge \cite{noether1,noether2}, namely the generator that rotates between the ChAF and SC order parameters $\Delta_T^{(\alpha)} \equiv L_{\alpha 5} + iL_{1\alpha}$, for $\alpha=2,3,4$. Note that $\Delta_T^{(\alpha)}$ physically represents the superconducting order-parameters in the spin-triplet, valley-singlet channel, 
\begin{subequations}
\beq
\label{eq:Deltax}
    \Delta_{\tn{T}x} &=& c_{\r +\up}c_{\r -\up}-c_{\r +\down}c_{\r -\down} \\
    \Delta_{\tn{T}y} &=& -i(c_{\r+\up}c_{\r-\up}+c_{\r+\down}c_{\r-\down}) \label{eq:Deltay}\\
    \Delta_{\tn{T}z} &=& c_{\r+\up}c_{\r-\down}+c_{\r+\down}c_{\r -\up}. \label{eq:Deltaz}
\eeq
\end{subequations}
We have computed the SU(2)-symmetrized imaginary time-displaced correlation function, $\chi_{\Delta_{\tn{T}}}(\tau)$, at $\q=0$, 
\beq\label{eq::dyn_so5}
\chi_{\Delta_{\tn{T}}}(\tau)\equiv \sum_{\alpha}\frac{\bra{\Psi_{\tn{G}}^{\Theta}} e^{\tau H} \Delta^{\dagger}_{T\alpha} e^{-\tau H} \Delta_{T\alpha} \ket{\Psi_{\tn{G}}^{\Theta}}}{\braket{\Psi_{\tn{G}}^{\Theta}}{\Psi_{\tn{G}}^{\Theta}}},
\eeq
as a function of $\tau$ for three different values of $|\JH|/\JA$. Note that an emergent $SO(5)$ symmetry with an approximately conserved $\Delta_{\tn{T}}$ (i.e., $[H, \Delta_{\tn{T}}] \approx 0$), should be reflected in a $\tau-$independent $\chi_{\Delta_{\tn{T}}}(\tau)$. Otherwise, $\chi_{\Delta_{\tn{T}}}(\tau)$ should exhibit an exponential decay with a rate that is roughly determined by the energy to rotate between the ChAF and SC orders. The numerical data for $\chi_{\Delta_{\tn{T}}}(\tau)$ at $\nu=2$ are shown in Fig.~\ref{Fig::aniso}(f) on a semi-log scale. For $|\JH|/\JA = 0.1$ and $|\JH|/\JA = 0.5$, located sufficiently away from the critical point, $\chi_{\Delta_{\tn{T}}}(\tau)$ decays exponentially. On the other hand, for $|\JH|/\JA = 0.34$, which lies much closer to the critical point(s), $\chi_{\Delta_{\tn{T}}}(\tau)$ 
does not exhibit clear indications of exponential decay and  behaves as noisy fluctuations near zero\footnote{If the SO(5) symmetry does emerge at the critical point, the noisy fluctuations are expected due to the approximately vanishing static susceptibility, see Appendix. \ref{App:emer_symm}} 
at the range of studied values of $\tau$. This suggests strong evidence for $\Delta_{\tn{T}}$ being an emergent nearly conserved quantity---lending further credence to the possibility of realizing a deconfined quantum critical point in the above model separating the insulating ChAF and a gapped SC.

\subsection{Bandwidth-tuned transition}\label{sub::bandwidth-phase-transition}
\begin{figure}[htb]
\includegraphics[width=80mm,scale=1]{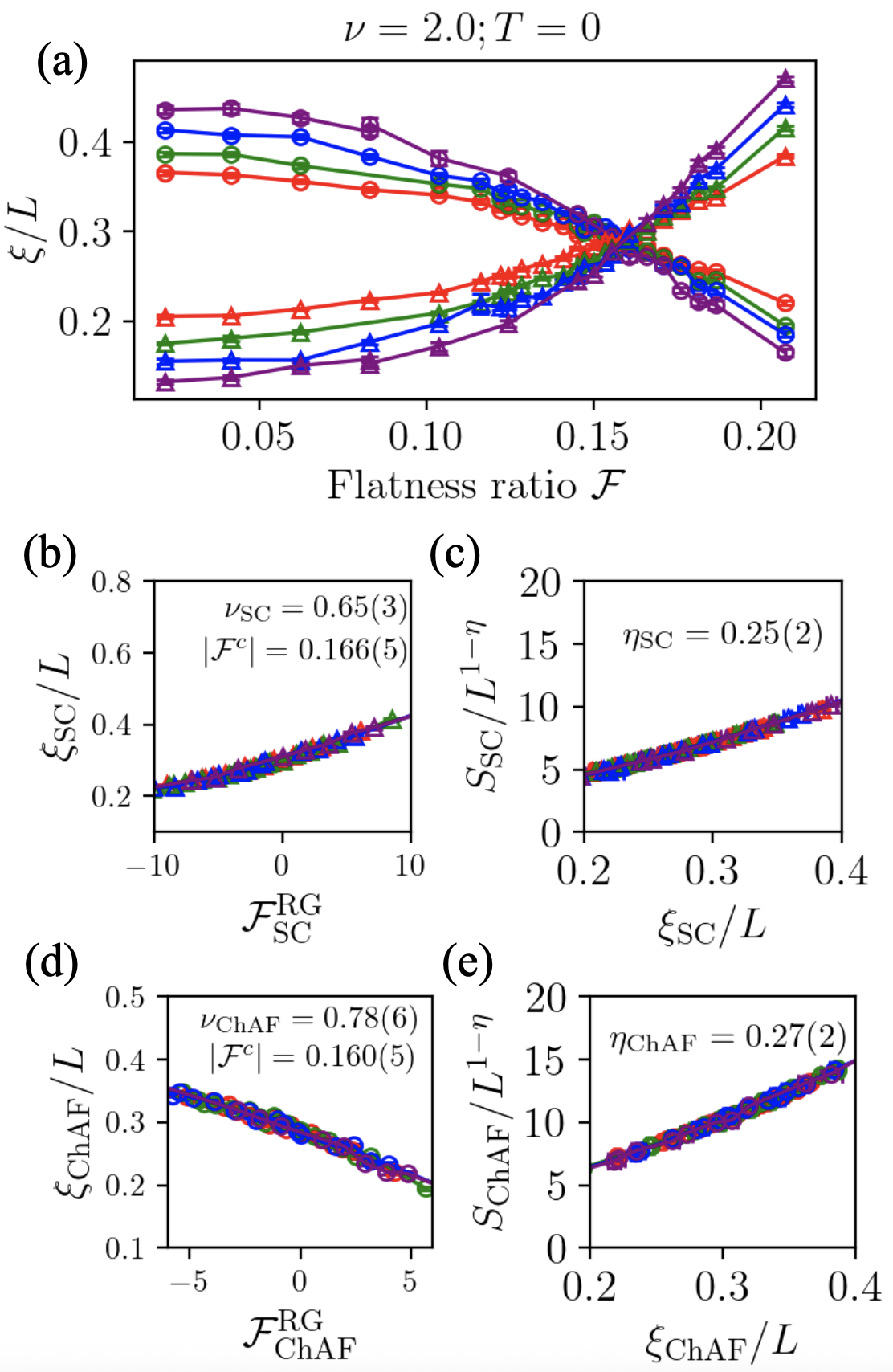}
\caption{\label{fig::flatness}
Bandwidth tuned transition at $\nu=2.0, T=0$ and $|\JH|/\JA=0.5$. (a) Correlation length $\xi_O$ for $O=\rm{ChAF, SC}$ for different system sizes. Scaling collapse analysis for (b) SC correlation length, where $\mathcal{F}^{\tn{RG}}_{\tn{O}}$ is defined in Eq.~\eqref{eq::collect_fit3}b. Scaling collapse analysis for (c) static SC susceptibility, (d) ChAF correlation length, and (e) static ChAF susceptibility. The critical exponents in (b)-(e) are obtained from collective fitting using Eq.~\eqref{eq::collect_fit3}a and Eq.~\eqref{eq::collect_fit4}.
}
\end{figure}
Let us now turn to the bandwidth-tuned transition, where we fix $|\JH|/\JA=0.5$ at $\nu=2$ and vary the bandwidth by tuning $t_5$ (fifth neighbor hopping), which does not modify the single-particle wavefunctions \cite{single_chern, svh}. We obtain results for values of $\mathcal{F}$ ranging from $0.01$ to $0.2$. A finite-size-scaling analysis of the RG-invariant correlation length, $\xi_{O}/L$, is shown in Fig.~\ref{fig::flatness}(a) for $O\equiv$ ChAF, SC. For $\mathcal{F}=0.01$, as noted previously, the ground state at $\nu=2$ is an insulating ChAF. With increasing $\mathcal{F}$, the ChAF order is lost with an onset of SC. To determine the location of the quantum critical point(s) and their associated universality, we revisit the scaling ansatz \cite{fakher_scaling},
\begin{subequations}\label{eq::collect_fit3}
\beq
\frac{\xi_O(\mathcal{F}, L)}{L} &=& \tilde{f}_\xi(\mathcal{F}_{\tn{O}}) + L^{-\omega} \tilde{g}_{\xi}(\mathcal{F}_{\tn{O}}),\\
\mathcal{F}^{\tn{RG}}_{\tn{O}} &=& [(\mathcal{F}-\mathcal{F}^{\tn{c}})/\mathcal{F}^{\tn{c}}]L^{1/\nu_{\tn{O}}},\label{eq::Jrg}
\eeq
\end{subequations}
for the correlation length and where the equal-time correlation function,
\beq\label{eq::collect_fit4}
S_{O}(\q=0;\mathcal{F},L) = L^{1-\eta_O}~\tilde{h}_{S,\xi}(\xi_O/L).
\eeq
As before, we assume a polynomial form for $\tilde{f}_\xi(J_{\tn{O}})$, $\tilde{g}_{\xi}(J_{\tn{O}})$ and $\tilde{h}_{S,\xi}(\xi_O/L)$.

The results show scaling-collapse for onset of SC in Fig.~\ref{fig::flatness}(b)-(c) and loss of ChAF in Fig.~\ref{fig::flatness}(d)-(e). 
For the bandwidth-tuned transition, the critical point for onset of SC, $\mathcal{F_{\tn{c1}}} = 0.166(5)$, and for loss of ChAF order, $\mathcal{F_{\tn{c2}}} = 0.160(5)$, are again coinciding, suggesting a direct transition between SC and ChAF. The correlation length critical exponent, $\nu_{\tn{SC}}=0.65(3)$, and anomalous dimension, $\eta_{\tn{SC}}=0.25(2)$, for onset of SC; 
similarly, $\nu_{\tn{ChAF}}=0.78(6)$ and  $\eta_{\tn{ChAF}}=0.27(2)$. Collectively, these exponents are quite clearly distinct from either a 3D-XY or 3D-O(3) transition, suggesting a distinct universality class.

\subsection{Interpretation of numerical results}\label{sub::interpretation}

Within our numerical resolution, we have observed a direct transition between the SC and ChAF phases, both when tuning the interaction anisotropy and the flatness ratio. \footnote{We acknowledge the numerical possibility of a tiny intermediate region between the ChAF and SC phases, either without any order or with coexistence of both; however, we have no direct numerical evidence or indication for such an intermediate regime and ignore it henceforth.} As is well known within the Ginzburg-Landau-Wilsonian paradigm, such a phase transition can either be first-order or has to be fine-tuned such that two independent continuous transitions coincide at a single point. Our data is consistent with a direct second order transition, and there are no indications of a \emph{strong} first order transition. Furthermore, the estimates of the critical exponents, specifically $\eta \simeq 0.25$, systematically deviate from the expectation for the $O(3)$ and $XY$ universality class, with $\eta_{O(3)}=0.037$, $\eta_{XY}=0.038$, respectively.
Instead, the exponents we find are in relative good agreement with previous numerical studies of deconfined quantum criticality with the same set of emergent symmetries, typically exhibiting $0.2 \lesssim  \eta \lesssim 0.3$ as well \cite{loop,assaad19,assaad21,assaad23,sandvik10, sandvik_science, Sandvik_2020}. 
Our estimates for the correlation length exponent, $\nu$, indicate $\nu \simeq 0.78$ (with one exception, $\nu_{SC}=0.65(3)$ for the bandwidth-tuned transition). Furthermore, our results are systematically larger than previous reports related to DQCP (see e.g., results for classical loop models \cite{loop}: $\nu=0.503,0.477$; QSH-SC transition \cite{assaad19,assaad21,assaad23}: $\nu=0.6,0.56$l; J-Q model \cite{sandvik10, sandvik_science, Sandvik_2020}: $\nu=0.455$.). Nevertheless, this is in line with the strong finite size drifts that were observed and likely due to the limited system size in our electronic model.

The low energy effective field theory derived above  displays an emergent $SO(5)$ symmetry that we substantiated numerically. Conformal bootstrap calculations \cite{bootstrap1, bootstrap2} put a lower bound on $\eta> 0.52$ for DQCPs with an emergent $SO(5)$ symmetry, assuming there is no relevant singlet operator which would render the DQCP multi-critical \cite{tri-crit, tri-crit2}. The topological level-1 WZW term encodes the intertwined nature of SC and ChAF, allows for a direct continuous phase transition between the two phases in principle. However, it is an irrelevant perturbation of the Gaussian fixed point which can only become relevant at strong coupling. Recently, a generalized non-linear sigma model with the equivalent level-1 WZW term has been implemented on the fuzzy sphere \cite{sphere1}. Interestingly, a stable strong coupling fixed point, describing the purported DQCP could not be identified. Instead, a pair of complex conformal field theories (CFT) from a fixed point collision may induce a {\it weakly} first-order transition. In the present setup, we conjecture that the RG flow slows down significantly in the vicinity of the complex CFT such that the SC-ChAF transition appears to be ``pseudo-critical'' and only much large lattices reveal the \emph{weakly} first-order transition \cite{sphere1, sandvik_science,Troyer08,Rychkov18,Gorbenko2018,WangC19}.

\section{Outlook}\label{sec::outlook}

We have focused here on a time-reversal symmetric model of partially filled spinful Chern bands with an additional ``valley" degree of freedom interacting via (anti-)ferromagnetic interactions. Using a combination of numerically exact DQMC and approximate Bethe-Salpeter equations based on the Hartree-Fock renormalized ground-state, we have analyzed the spectral features associated with the neutral collective modes and charged excitations. The latter include the renormalized electronic quasiparticles as well as charged spin$-3/2$ excitations. We have pointed out the regime in parameter space where DQMC is able to better capture the energetic competition between the distinct charged excitations compared to the approximate method. Specifically, while the Hartree-Fock analysis is able to capture accurately the interaction-induced insulating ground state at commensurate filling and the neutral magnon excitations, it fails to reproduce the energetics for the lower-energy spin$-3/2$ excitations. This highlights the important role played by quantum fluctuations going beyond the Bethe-Salpeter equations, which can contribute to the renormalization of the effective coupling between the renormalized electronic quasi-particles and the low-energy spin-flip excitations. Investigating these effects systematically is an interesting future direction.

Given the close competition between the insulating Chern antiferromagnet and a superconductor in the vicinity of $\nu=2$, we have also analyzed carefully the nature of the quantum phase transitions between the two phases as a function of two distinct tuning parameters. We have observed the transition to be controlled by a distinct universality class compared to either 3D-XY or 3D-O(3). Instead, there is a clear signature of an approximate emergent SO(5) symmetry in the vicinity of a direct continuous transition between the two ordered states, and is reminiscent of a deconfined quantum phase transition. These results not only demonstrate the rich physics emanating from the interplay between interaction and topology in a \textit{generic} Chern band setup, but also serve as a proof-of-principle building block that can help engineer emergent exotic excitations leading to superconductivity. Interestingly, even though the electrons are gapped on either side of the transition(s), the numerically extracted critical points differ quantitatively depending on the bandwidth vs. interaction-anisotropy-tuned transitions. Whether the charge fluctuations are indirectly responsible for this difference in the bandwidth-tuned transition remains an interesting question for the future.

In conjunction with our companion article \cite{skyr_short}, we have demonstrated explicitly the robustness of skyrmions, their role in stabilizing superconductivity, and as being fundamentally responsible for driving exotic quantum phase transitions in a broad regime of microscopic parameters. While our model shares a broad conceptual similarity with a ``solvable" regime of microscopic model of moir\'e graphene \cite{AV18, AV19, AV20_prx, SS20,TBG4,KangVafekPRL, Hofmann_tbg,chatterjee20, ashvin_skrymion, baby_skyrmion, chatterjee22}, we have explicitly suppressed the effect of charge Coulomb repulsion terms in our analysis. The latter are also responsible for reintroducing the sign-problem at a generic filling of the Chern bands. Developing an alternative framework that can incorporate the effects of these perturbations can shed new light on the robustness of our observed phenomenology. Finally, finding an experimental realization of the models studied here, either in the moir\'e or cold-atoms based setups \cite{Cooper_rmp19,Jotzu14,Aidelsburger13,Miyake13,Zhang2018} can enable a new frontier of studying exotic quantum critical points in a highly tunable platform.

\begin{acknowledgments}
We thank A. Vishwanath for useful discussions and J. F. Mendez-Valderrama for a related collaboration. D.C. is supported in part by a NSF CAREER grant (DMR-2237522), and a Sloan Research Fellowship. D.C. also acknowledges the hospitality of MPI-PKS, Dresden, and the support provided by the Aspen Center for Physics where this manuscript was finalized, which is supported by National Science Foundation grant PHY-2210452. This work used Expanse at the San Diego Supercomputer Center through allocation TG-PHY240209 from the Advanced Cyberinfrastructure Coordination Ecosystem: Services \& Support (ACCESS) program \cite{access}, which is supported by National Science Foundation grants \#2138259, \#2138286, \#2138307, \#2137603, and \#2138296.
The auxiliary field QMC simulations were carried out using the ALF package \cite{alf}.
This work was supported in part by the Deutsche Forschungsgemeinschaft DFG through the Würzburg-Dresden Cluster of Excellence on Complexity and Topology in Quantum Matter `ct.qmat' (EXC2147, project ID 390858490).
\end{acknowledgments}

\section*{DATA AVAILABILITY}
The data that supports the findings of this article cannot be made publicly available because the size of data set is prohibitively large to be deposited and hosted publicly. The data are available upon reasonable request from the authors.

\clearpage
\newpage
\newpage

\clearpage

\appendix
\newpage

\section{Additional details for Hartree-Fock computations}\label{App::eig}

\begin{figure}[htb]
\includegraphics[width=90mm,scale=1]{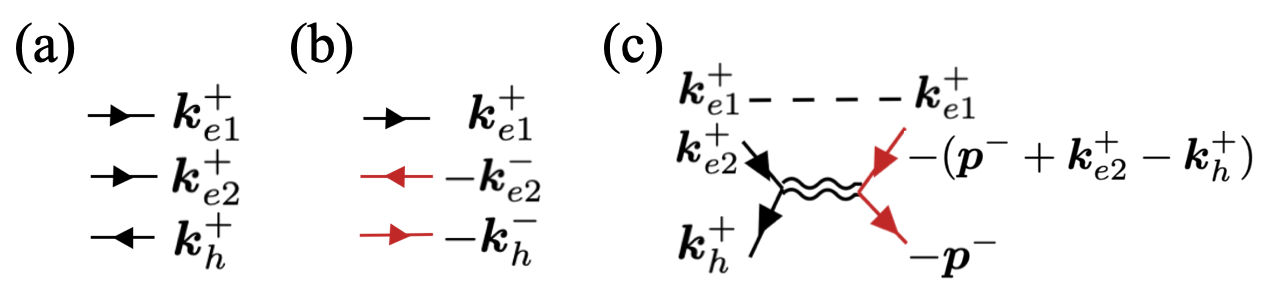}
\caption{\label{fig::momentum_assign} Momentum assignment for propagator of composite charged excitation $W_{\Sigma}'$ in (a) $(0,0)$-entry; (b) $(1,1)$-entry. (c) Example of diagram in the off-diagonal entry of $K_{BS}'$ (denoted as Eq.~\eqref{eq::bs_rep}).}
\end{figure}

In this appendix, we provide further technical details on the analysis of the Bethe-Salpeter kernel (Fig.~\ref{fig::feyn2}(b)) for the composite charged excitations. The computations for the collective modes can be formulated similarly in a straightforward fashion. Let us start by defining the basis states, assuming the Hartree-Fock ground state is $\ket{\tn{GS}}\equiv \ket{+\down}\otimes\ket{-\up}$. The $(0,0)$ entry of $\tilde{K}_{BS}^{(0,0)}$ and $\tilde{W}_{\Sigma}^{(0,0)}$ act on the Hilbert space spanned by $\ket{\vec{k}^{+}_{e1},\vec{k}^{+}_{e2},\vec{k}^{+}_{h}}\equiv c^{\dagger}_{\k_{e1};+\up} c^{\dagger}_{\k_{e2};+\up} c_{\k_{h};+\down}\ket{\tn{GS}}$ (denoted pictorially in Fig.~\ref{fig::momentum_assign}), where the momentum labels satisfy the relation, $\q = \k^{+}_{e1} + \k^{+}_{e2} -\k^{+}_{h}$. Here, we use $\k^{\tau}_{e(h)}$ to denote momentum of electron (hole) from valley $\tau$. Noting that the $c^{\dagger}_{\k_{e1};+\up}$ and $c^{\dagger}_{\k_{e2};+\up}$ represent indistinguishable states, the state $\ket{\vec{k}^{+}_{e1},\vec{k}^{+}_{e2},\vec{k}^{+}_{h}}$ needs to be anti-symmetrized between the first two entries, and the dimension of the Hilbert space is $N(N-1)/2$ for each external momentum $\q$ in the $(0,0)$-entry.

Similarly, $\tilde{K}_{BS}^{(1,1)}$ acts on the Hilbert space spanned by $\ket{\vec{k}^{+}_{e1},-\vec{k}^{-}_{e2},-\vec{k}^{-}_{h}}$. We intentionally assign an additional minus sign to electron and hole from $\tau=-$, such that the external momentum is still $\q = \k^{+}_{e1} + \k^{-}_{e2} -\k^{-}_{h}$ and the momentum conservation relation is automatically fulfilled by matrix multiplication. The dimension of Hilbert space is $N^2$ for each external momentum $\q$ in the $(1,1)$-entry.

Next, we show explicitly the matrix element for a representative diagram in Fig.~\ref{fig::momentum_assign}(c). Other diagrams can be encoded into matrices in a similar fashion. The matrix element represented by Fig.~\ref{fig::momentum_assign}(c) is obtained as,
\begin{widetext}
\begin{equation}\label{eq::bs_rep}
\tilde{K}_{BS~}^{\tn{rep}}=\JA\sum_{\vec{p}^{-}} u^{*}_{+}(\vec{k}^{+}_{h})u_{+}(\vec{k}^{+}_{e2})u_{+}^{*}(\vec{p}^{-}+\k_{e2}^{+}-\k_{h}^{+})u_{+}(\vec{p}^{-})~\ket{\vec{k}^{+}_{e1},\vec{k}^{+}_{e2},\vec{k}^{+}_{h}}\bra{\vec{k}^{+}_{e1}, -(\vec{p}^{-}+\k_{e2}^{+}-\k_{h}^{+}),-\vec{p}^{-}},
\end{equation}
\end{widetext}
where $u_{+}(\k)$ is the Bloch wavefunction for states from $\tau=+$ valley. We have used time-reversal symmetry, $u^{*}_{-}(-\k)=u_{+}(\k)$ in deriving Eq.~\eqref{eq::bs_rep}.

Having defined the basis and the Bethe-Salpeter kernel, we now turn to solving the generalized eigenvalue problem defined in Eq.~\eqref{eq::bse_caf} and Eq.~\eqref{eq::tw_eig}. For simplicity,
let us use the following notation,
\begin{equation}\label{eq::k_def}
K\equiv
\begin{bmatrix}
K^{(0,0)}_{n\times n} & K^{(0,1)}_{n\times m} \\
K^{(1,0)}_{m\times n} & K^{(1,1)}_{m\times m}
\end{bmatrix}
; ~~~\Lambda_z\equiv 
\begin{bmatrix}
\mathbb{I}_{n\times n} &  \\
 & -\mathbb{I}_{m\times m}
\end{bmatrix}.
\end{equation}
The generalized eigenvalue problem can be defined as solving for $\Omega$ in 
\begin{equation}\label{eq::gen_eigen}
\det[\Omega \Lambda_z - K]=0.
\end{equation}
Equivalently, the solution to Eq.~\eqref{eq::gen_eigen} can be obtained by diagonalizing the matrix $\Lambda_z K$. If denoting the eigenvalue and eigenvector of $\Lambda_z K$ as $\{ E_{\alpha}, \ket{\alpha}\}$, the solution to Eq.~\eqref{eq::gen_eigen} corresponds to those $E_{\alpha}$ with $\bra{\alpha}\Lambda_z\ket{\alpha}=1$.

\section{Effective low-energy theory}\label{app::so5}
In this appendix, we elaborate further on some of the technical details for the low-energy theory discussed in Section \ref{sec::so5}.

\subsection{Effective projected model in band basis}
Let us recall the Nambu-spinor in momentum space as, $\Psi_{\vec{k}}^\dagger\equiv \bigg[d^\dagger_{+\up}(\vec{k})~, d^\dagger_{+\down}(\vec{k})~, d_{-\up}(-\vec{k})~, d_{-\down}(-\vec{k})\bigg]$, where the operators $d^{\dagger}_{+\alpha}(\vec{k}),~d_{-\alpha}(-\vec{k})$ are related by a combined operation of complex-conjugation and time reversal $\mathcal{T}$, which leaves the Bloch wavefunction invariant, 
\beq\label{eq::ff_trs}
u_{+}(\vec{k}) = [u_{-}(-\vec{k})]^{*}
\eeq
Based on this observation, the form-factor $F_{\vec{k},\vec{p},\vec{q}}$ from different component of the Nambu-spinor should \textit{factorize}, i.e. only one independent form-factor will enter the projected Hamiltonian despite the presence of inter-valley interaction, where $F_{\vec{k},\vec{p},\vec{q}}$ is defined as
\beq
F_{\vec{k},\vec{p},\vec{q}} \equiv u
_{+}^{*}(\vec{k}+\vec{q}) u_{+}(\vec{k}) u_{+}^{*}(\vec{p}-\vec{q}) u_{+}(\vec{p}).
\eeq.
Namely, after projecting to the active Chern bands, the ferromagnetic exchange in $\tau=+$ and $\tau=-$ valley after projection are given by,
\beq\label{eq::h_jh_pm}
\begin{aligned}
H_{\JH;+} &= 
-\frac{3\JH}{2}  F_{\vec{k},\vec{p},\vec{q}}
\bigg[ d^{\dagger}_{\vec{k}+\vec{q};+\up}~d^{\dagger}_{\vec{p}-\vec{q};+\down}~d_{\vec{k};+\down}~d_{\vec{p};+\up} \bigg]\\
H_{\JH;-} &=-\frac{3\JH}{2}F_{\vec{k},\vec{p},\vec{q}}\bigg[ d^{\dagger}_{-\vec{k};-\up}~d^{\dagger}_{-\vec{p};-\down}~d_{-\vec{k}-\vec{q};+\down}~d_{-\vec{p}+\vec{q};+\up} \bigg].
\end{aligned}
\eeq
respectively, where repeated indices are summed over.

Such factorization immediately suggests that any model in atomic limit with emergent symmetries can inherit those emergent symmetries to the models with (topological) flat bands (i.e. emergent symmetries are fully determined by the commutation relations among fermion bilinears). For the sake of completeness, we will now explicitly derive the projected interaction and verify that the above result follows in the expected fashion.

After utilizing Fierz identities to decompose Eq.~\eqref{eq::h_jh_pm} into superconducting and particle-hole channel, we obtain
\begin{widetext}
\beq\label{eq::jh_k}
\begin{aligned}
H_{\JH} = \frac{3|\JH|}{8}
\sum_{\vec{k}\vec{p}\vec{q}}
F_{\vec{k},\vec{p},\vec{q}}
\bigg[&
\Psi^{\dagger}_{\vec{k}+\vec{q}} \Psi_{\vec{k}} 
\Psi^{\dagger}_{\vec{p}-\vec{q}} \Psi_{\vec{p}}
~~+~~
\Psi^{\dagger}_{\vec{k}+\vec{q}} L_{15} \Psi_{\vec{k}} 
\Psi^{\dagger}_{\vec{p}-\vec{q}} L_{15} \Psi_{\vec{p}}
~~-
\sum_{(\alpha\beta)=(23),(34),(24)}
\Psi^{\dagger}_{\vec{k}+\vec{q}} L_{\alpha\beta} \Psi_{\vec{k}} 
\Psi^{\dagger}_{\vec{p}-\vec{q}} L_{\alpha\beta} \Psi_{\vec{p}}\\&
+\sum_{\mu=2,3,4}
\Psi^{\dagger}_{\vec{k}+\vec{q}} N_{\mu} \Psi_{\vec{k}} 
\Psi^{\dagger}_{\vec{p}-\vec{q}} N_{\mu} \Psi_{\vec{p}}
\bigg]
~~+~~ \frac{3|\JH|}{16}\sum_{\alpha; \vec{k}\vec{p} }~\bigg[
|u_{+}(\vec{k})|^2|u_{+}(\vec{p})|^2 d^{\dagger}_{\vec{p};+\alpha} 
d_{\vec{p};+\alpha} 
\bigg].
\end{aligned}
\eeq
\end{widetext}
, where $N_\mu$ and $L_{\alpha\beta}$ are defined in the main text. The last term (referred as generalized uniform pairing condition in the main text) in Eq.~\ref{eq::jh_k} originates from normal ordering.

Similarly, the projected interaction originated from the anti-ferromagnetic exchange can be obtained as
\begin{widetext}
\beq\label{eq::ja_k}
\begin{aligned}
H_{\JA} &= \frac{\JA}{16} F_{\vec{k},\vec{p},\vec{q}}
\Bigg[
\sum_{(\alpha\beta)=(25),(35),(45),(12),(13),(14)}
\Psi^{\dagger}_{\vec{k}+\vec{q}} L_{\alpha\beta} \Psi_{\vec{k}} 
\Psi^{\dagger}_{\vec{p}-\vec{q}} L_{\alpha\beta} \Psi_{\vec{p}}
\ -\ 
3 \sum_{\mu=1,5}
\Psi^{\dagger}_{\vec{k}+\vec{q}} N_{\mu} \Psi_{\vec{k}} 
\Psi^{\dagger}_{\vec{p}-\vec{q}} N_{\mu} \Psi_{\vec{p}}
\Bigg]\\&
-\frac{\JA}{8}F_{\vec{k},\vec{p},\vec{q}}
\Bigg[
-\sum_{(\alpha\beta)=(23),(34),(24)}
\Psi^{\dagger}_{\vec{k}+\vec{q}} L_{\alpha\beta} \Psi_{\vec{k}} 
\Psi^{\dagger}_{\vec{p}-\vec{q}} L_{\alpha\beta} \Psi_{\vec{p}}
~~+
\sum_{\mu=2,3,4}
\Psi^{\dagger}_{\vec{k}+\vec{q}} N_{\mu} \Psi_{\vec{k}} 
\Psi^{\dagger}_{\vec{p}-\vec{q}} N_{\mu} \Psi_{\vec{p}}
\Bigg].
\end{aligned}
\eeq
\end{widetext}

Combining Eq.~\eqref{eq::jh_k} and Eq.~\ref{eq::ja_k}, we obtained the effective rotor model in Eq.~\eqref{eq::ham_rotor_k_main} of the main text.
 
We stress again that the emergent SO(5) symmetry of the projected effective theory manifest itself at the level of \textit{fermionic operators}. The derivations above by rearranging the fermion operators into SO(5) fermion bilinear are exact and are just for better demonstration of the emergent symmetry. One should obtain the same conclusion through calculating $[H_{\tn{int}}, L_{\alpha\beta}]$ without the rearrangement.

\subsection{Wess-Zumino-Witten term}
In addition to the effective rotor Hamiltonian in Eq.~\eqref{eq::ham_rotor_k_main}, an effective Wess-Zumino-Witten term can be generated in a fully-gapped topological system \cite{lee_sachdev1, lee_sachdev2,dqcp_prb_04,dqcp1,so5_3,TH05,SF06}, which depicts the intertwined nature of the couplings between the different symmetry-breaking orders. Let us start with the projected electronic Lagrangian with mean-field decomposition in the possible symmetry-breaking channel introduced previously as
\beq\label{eq::wzw_mf}
\mathcal{L} = \sum_{\tau,\alpha} d^{\dagger}_{\tau\alpha}\frac{\partial d_{\tau\alpha}}{\partial \tau} - \lambda \sum_{\vec{q}\vec{k};a} N^{a}_{\vec{q}}~\Lambda(\k,\q) \Psi^{\dagger}_{\vec{k}+\vec{q}} \Gamma_{a} \Psi_{\vec{k}},
\eeq
where the form factor $\Lambda(\k,\q)\equiv \braket{u_{\vec{k}+\vec{q}}}{u_{\vec{k}}}$, and $N^{a}_{\q}$ denotes the order parameters with $a=1...5$ with $\lambda$ denoting the effective Yukawa coupling strength. The intertwined orders are parameterized by $\Gamma$-matrices defined as 
\begin{equation}\label{eq::gamma_mat_supp}
\begin{aligned}
&
\Gamma_1 = \sigma_y\mu_y, ~\Gamma_2=\sigma_x, ~\Gamma_3 = \sigma_y\mu_z, \\&
~\Gamma_4 = \sigma_z, ~\Gamma_5 = \sigma_y\mu_x.
\end{aligned}
\end{equation}
Let us recall again the Nambu-spinor in momentum space $\Psi_{\vec{k}}^\dagger\equiv \bigg[d^\dagger_{+\up}(\vec{k})~, d^\dagger_{+\down}(\vec{k})~, d_{-\up}(-\vec{k})~, d_{-\down}(-\vec{k})\bigg]$, such that the flavor-dependence of the form-factor $\braket{u_{\vec{k}+\vec{q}}}{u_{\vec{k}}}$ \textit{factorize} in Eq.~\eqref{eq::wzw_mf} due to time-reversal symmetry.

We are interested in the effective action,
\beq
\label{eq::wzw}
S_{\tn{WZW}}[N^{a}] = K \int_{\tau, \vec{r}, u} ~\varepsilon_{abcde} N^{a}\partial_x N^b \partial_y N^c \partial_\tau N^d \partial_u N^e,\nn\\ 
\eeq 
where the pre-factor $K$ can be obtained by calculating the order-parameter fluctuations around the saddle point.
The leading order contribution to Eq.~\ref{eq::wzw} is given by,
\begin{equation}
\begin{aligned}
\tikzset{every picture/.style={line width=0.75pt}} 
S^{(4)}&=~~~~\begin{tikzpicture}[baseline={(0, -3.2cm)},x=0.75pt,y=0.75pt,yscale=-0.8,xscale=0.85]

\draw [color={rgb, 255:red, 0; green, 0; blue, 0 }  ,draw opacity=1 ]   (188.91,194.97) -- (281.08,194.9) ;

\draw  [color={rgb, 255:red, 0; green, 0; blue, 0 }  ,draw opacity=1 ][fill={rgb, 255:red, 0; green, 0; blue, 0 }  ,fill opacity=1 ] (230.78,195.12) -- (239.29,190.42) -- (239.12,199.09) -- cycle ;

\draw    (281.25,101.46) .. controls (281.38,99.11) and (282.62,97.99) .. (284.97,98.11) .. controls (287.32,98.24) and (288.56,97.12) .. (288.68,94.77) .. controls (288.81,92.42) and (290.05,91.3) .. (292.4,91.42) .. controls (294.75,91.54) and (295.99,90.42) .. (296.11,88.07) .. controls (296.23,85.72) and (297.47,84.6) .. (299.82,84.72) -- (300.3,84.29) -- (300.3,84.29) ;

\draw    (188.91,100.37) -- (188.91,194.97) ;

\draw  [color={rgb, 255:red, 0; green, 0; blue, 0 }  ,draw opacity=1 ][fill={rgb, 255:red, 0; green, 0; blue, 0 }  ,fill opacity=1 ] (189.3,145.23) -- (194.38,153.42) -- (185.57,153.58) -- cycle ;

\draw    (281.25,101.46) -- (281.08,194.9) ;

\draw  [color={rgb, 255:red, 0; green, 0; blue, 0 }  ,draw opacity=1 ][fill={rgb, 255:red, 0; green, 0; blue, 0 }  ,fill opacity=1 ] (281.64,153.51) -- (276.6,145.3) -- (285.41,145.17) -- cycle ;

\draw [color={rgb, 255:red, 0; green, 0; blue, 0 }  ,draw opacity=1 ]   (188.91,100.37) -- (281.25,101.46) ;

\draw  [color={rgb, 255:red, 0; green, 0; blue, 0 }  ,draw opacity=1 ][fill={rgb, 255:red, 0; green, 0; blue, 0 }  ,fill opacity=1 ] (239.28,100.55) -- (231,105.61) -- (230.77,96.94) -- cycle ;

\draw    (281.08,194.9) .. controls (283.43,194.8) and (284.66,195.93) .. (284.77,198.28) .. controls (284.88,200.63) and (286.11,201.76) .. (288.46,201.65) .. controls (290.81,201.54) and (292.04,202.67) .. (292.15,205.02) .. controls (292.26,207.37) and (293.49,208.5) .. (295.84,208.4) .. controls (298.19,208.29) and (299.42,209.42) .. (299.53,211.77) -- (301.78,213.82) -- (301.78,213.82) ;

\draw    (188.91,194.97) .. controls (188.83,197.33) and (187.61,198.47) .. (185.26,198.39) .. controls (182.91,198.31) and (181.69,199.45) .. (181.61,201.8) .. controls (181.53,204.15) and (180.31,205.29) .. (177.96,205.22) .. controls (175.61,205.15) and (174.39,206.29) .. (174.31,208.64) .. controls (174.23,210.99) and (173.01,212.13) .. (170.66,212.05) -- (167.21,215.28) -- (167.21,215.28) ;

\draw    (188.91,100.37) .. controls (186.57,100.64) and (185.27,99.6) .. (185,97.26) .. controls (184.73,94.92) and (183.42,93.88) .. (181.08,94.15) .. controls (178.74,94.42) and (177.44,93.38) .. (177.17,91.04) .. controls (176.9,88.69) and (175.6,87.65) .. (173.25,87.92) .. controls (170.91,88.19) and (169.61,87.15) .. (169.34,84.81) -- (168.69,84.29) -- (168.69,84.29) ;

\draw (148.37,60.29) node [anchor=north west][inner sep=0.75pt]    {$\pi _{i}  \ $};

\draw (308.19,64.66) node [anchor=north west][inner sep=0.75pt]    {$\pi _{j} \ \ $};

\draw (309.83,216.69) node [anchor=north west][inner sep=0.75pt]    {$\pi _{l} \ \ $};

\draw (143.38,220.14) node [anchor=north west][inner sep=0.75pt]    {$\pi _{m} \ \ $};

\draw (229,75.4) node [anchor=north west][inner sep=0.75pt]    {$k$};

\draw (293,141.4) node [anchor=north west][inner sep=0.75pt]    {$k+p_{1}$ };

\draw (200,203.4) node [anchor=north west][inner sep=0.75pt]    {$k+p_{1} +p_{2}$};

\draw (71,138.4) node [anchor=north west][inner sep=0.75pt]    {$k+p_{1} +p_{2} +p_{3}$};+ (10cm, 10cm)
\end{tikzpicture}
\\&  
=~ \frac{\lambda^4}{4} \pi_i\pi_j\pi_l\pi_m \sum_{\vec{k},i\omega} \Lambda^{\mathrm{box}}_{\k,\vec{p}_1,\vec{p}_2,\vec{p}_3}
\tn{Tr}\bigg[ G\Gamma^{i}G\Gamma^{j}G\Gamma^{l}G\Gamma^{m} \bigg],
\end{aligned}\label{eq::box_diagram}
\end{equation}
where the $\Lambda^{\mathrm{box}}$ is the form-factor for the box diagram in terms of overlaps of Bloch wavefunctions.

Let us assume $N^a$ is polarized along $(0,0,0,1,0)$ direction, and the fluctuation around this saddle point can be parametrized as $N^a = (\pi_1, \pi_2, \pi_3, 1, \pi_4)$. The fully gapped electronic Green's function is given by
\beq
G_{\omega}=(i\omega + \lambda \Gamma_4)^{-1},
\eeq
where we have assumed a perfectly flat band and thereby dropped the momentum-dependence arising from the dispersion. 
Since the trace part in Eq.~\eqref{eq::box_diagram} is independent of $\k$, the $\k$ summation can be performed only with respect to $\Lambda^{\mathrm{box}}_{\k,\vec{p}_1,\vec{p}_2,\vec{p}_3}$.
We note that the contribution from the form-factor $\Lambda^{\mathrm{box}}_{\k,\vec{p}_1,\vec{p}_2,\vec{p}_3}$ defined in Eq.~\eqref{eq::box_diagram} is gauge invariant because
\begin{widetext}
\beq
\sum_{\vec{k}} 
\Lambda^{\mathrm{box}}_{\k,\vec{p}_1,\vec{p}_2,\vec{p}_3}
= \sum_{\vec{k}} \tn{tr}\bigg[\hat{P}_{\vec{k}} \hat{P}_{\vec{k}+ \vec{p_1} + \vec{p_2} + \vec{p_3}} \hat{P}_{\vec{k} + \vec{p_1} + \vec{p_2}} \hat{P}_{\vec{k} + \vec{p_1}}\bigg],
\eeq
where $\hat{P}_{\vec{k}} \equiv \ket{u_{\vec{k}}}\bra{u_{\vec{k}}}$.
To proceed, we make use of the following small-momentum expansion of the form-factor:
\begin{subequations}
\beq
\Lambda(\k,\vec{p}) &\simeq& 1 + i a_{\mu}(\vec{k})p^{\mu} -\frac{1}{2}\gamma_{\mu\nu}(\vec{k})p^{\mu}p^{\nu}\\
\Lambda(\k+\vec{p},\q) &\simeq& 1 + i a_{\mu}(\vec{k})q^{\mu} -i\frac{F_{\mu\nu}(\k)}{2}p^{\mu}q^{\nu} -\frac{1}{2}\gamma_{\mu\nu}(\vec{k})q^{\mu}q^{\nu},
\eeq\label{eq::ff_lookup}
\end{subequations}
\end{widetext}
where $a_{\mu}(\vec{k})$ is the Berry connection, $F_{\mu\nu}(\vec{k})$ is the Berry curvature and $\gamma_{\mu\nu}(\vec{k})$ is the real-part of the quantum metric. We focus on the contribution $\sim \varepsilon_{\mu\nu}p_1^{\mu}p_2^{\nu}$, which is to be contracted with an anti-symmetric rank-2 tensor to have a finite contribution. Given that the only gauge independent anti-symmetric rank-2 tensor in Eq.~\ref{eq::ff_lookup} is the Berry curvature, after evaluating the momentum-integral in Eq.\ref{eq::box_diagram}, we obtain
\begin{widetext}
\beq
\begin{aligned}\label{eq::s4_contd}
S^{(4)}&=\frac{\lambda^4}{4} \pi_i\pi_j\pi_l\pi_m ~ 
\bigg(p_1^{\mu}p_2^{\nu} 
+ p_1^{\mu}p_3^{\nu} + p_2^{\mu}p_3^{\nu} \bigg)
~\int \frac{d^2\vec{k}}{(2\pi)^2}\frac{F_{\mu\nu}(\k)}{2}~\sum_{i\omega}\tn{Tr}\bigg[ G\Gamma^{i}G\Gamma^{j}G\Gamma^{l}G\Gamma^{m} \bigg]\\&
=\frac{\lambda^4 C}{16\pi} \varepsilon_{\mu\nu} 
\bigg(p_1^{\mu}p_2^{\nu} 
+ p_1^{\mu}p_3^{\nu} + p_2^{\mu}p_3^{\nu} \bigg)
~ \pi_i\pi_j\pi_l\pi_m ~\sum_{i\omega}\tn{Tr}\bigg[ G\Gamma^{i}G\Gamma^{j}G\Gamma^{l}G\Gamma^{m} \bigg],
\end{aligned}
\eeq
\end{widetext}
where $C$ is the Chern number of the underlying electronic band structure. Furthermore, to obtain the contribution proportional to Eq.~\eqref{eq::wzw}, we need the term proportional to $ \Omega_\alpha$, for $\alpha=1,2,3$ in Eq.\ref{eq::s4_contd} by expanding around $\Omega_\alpha=0$ as, 
\begin{widetext}
\beq
\begin{aligned}
\partial_{\Omega_3} S^{(4)} &=\pi_i\pi_j\pi_l\pi_m \frac{\lambda^4 C}{16\pi} \varepsilon_{\mu\nu} p_1^{\mu}p_2^{\nu}\Omega_3~  \sum_{\omega_n}\tn{Tr}\bigg[ G_{\omega_n}\Gamma_i
G_{\omega_n + \Omega_1}\Gamma_j
G_{\omega_n + \Omega_1+\Omega_2}
\Gamma_l
\partial_{\Omega_{3}} G_{\omega_n + \Omega_1+\Omega_2 + \Omega_3}\Gamma_m
\bigg]\bigg|_{\Omega_\alpha=0}\\&
=\frac{C}{16\pi} \varepsilon_{\mu\nu} p_1^{\mu}p_2^{\nu}\Omega_3 ~ \pi_i\pi_j\pi_l\pi_m ~
\Tr[\Gamma_i\Gamma_j\Gamma_l\Gamma_m\Gamma_4]
\sum_{\omega_n}\frac{\lambda^5}{(\omega_n^2+\lambda^2)^3} \\&
=\frac{3C}{64\pi} \varepsilon_{\mu\nu} p_1^{\mu}p_2^{\nu}\Omega_3 ~ \bigg(\pi_i\pi_j\pi_l\pi_m \varepsilon^{ijlm}\bigg)\label{eq::wzw_w1}
\end{aligned}
\eeq
\beq
\begin{aligned}
\partial_{\Omega_2} S^{(4)} &=\pi_i\pi_j\pi_l\pi_m \frac{\lambda^4 C}{16\pi} \varepsilon_{\mu\nu} p_1^{\mu}p_3^{\nu}\Omega_2~  \sum_{\omega_n}\tn{Tr}\bigg[ G_{\omega_n}\Gamma_i
G_{\omega_n + \Omega_1}\Gamma_j
\partial_{\Omega_{2}}G_{\omega_n + \Omega_1+\Omega_2}
\Gamma_l
G_{\omega_n + \Omega_1+\Omega_2 + \Omega_3}\Gamma_m
\bigg]\bigg|_{\Omega_\alpha=0} + \partial_{\Omega_3} S^{(4)}
\\& = 0\label{eq::wzw_w2}
\end{aligned}
\eeq
\beq
\begin{aligned}
\partial_{\Omega_1} S^{(4)} &=\pi_i\pi_j\pi_l\pi_m \frac{\lambda^4 C}{16\pi} \varepsilon_{\mu\nu} p_2^{\mu}p_3^{\nu}\Omega_1~  \sum_{\omega_n}\tn{Tr}\bigg[ G_{\omega_n}\Gamma_i
\partial_{\Omega_{1}}G_{\omega_n + \Omega_1}\Gamma_j
G_{\omega_n + \Omega_1+\Omega_2}
\Gamma_l
G_{\omega_n + \Omega_1+\Omega_2 + \Omega_3}\Gamma_m
\bigg]\bigg|_{\Omega_\alpha=0} + \partial_{\Omega_2} S^{(4)}
\\& = \frac{3C}{64\pi} \varepsilon_{\mu\nu} p_2^{\mu}p_3^{\nu}\Omega_1 ~ \bigg(\pi_i\pi_j\pi_l\pi_m \varepsilon^{ijlm}\bigg),\label{eq::wzw_w3}
\end{aligned}
\eeq
\end{widetext}
where the vanishing contribution in Eq.~\eqref{eq::wzw_w2} is due to the anti-commuting property of $\Gamma$-matrices. Combining Eq.~\eqref{eq::wzw_w1} - Eq.~\eqref{eq::wzw_w3} and focusing on terms proportional to $\varepsilon^{ijlm}\pi_i\partial_x\pi_j\partial_y\pi_l\partial_\tau\pi_m$, we obtain the topological term as
\beq
S_{\tn{WZW}}[\pi^{a}]=i\frac{3C}{16\pi}\varepsilon^{ijlm}\int_{x,y,\tau} \pi_i\partial_x\pi_j\partial_y\pi_l\partial_\tau\pi_m,
\eeq
or equivalently a level-$C$ WZW term,
\beq
\begin{aligned}
&
S_{\tn{WZW}} = 2\pi i C ~W[N_a]\\&
W[N_a] = \frac{3}{8\pi^2}\int_{\r,\tau, u} ~\varepsilon_{abcde} N^{a}\partial_x N^b \partial_y N^c \partial_\tau N^d \partial_u N^e.
\end{aligned}
\eeq

\begin{widetext}

\section{Additional Details on Emergent Symmetry}\label{App:emer_symm}
\subsection{Supplementary Data on $\chi_{\Delta_{\tn{T}}}(\tau)$}

\begin{figure*}[htb]
\includegraphics[width=180mm,scale=1]{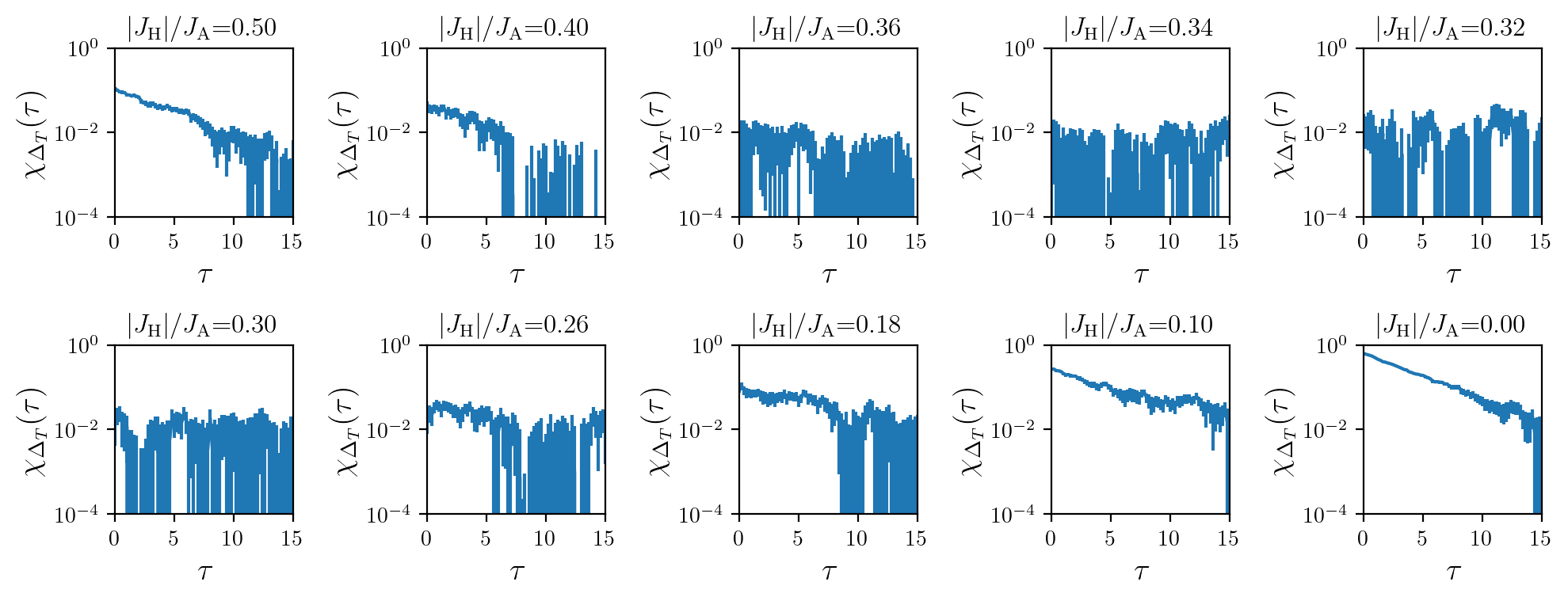}
\caption{\label{Fig::tauso5_supp}  Dynamical correlation function $\chi_{\Delta_{\tn{T}}}(\tau)$ (in a semi-log scale) defined in Eq.~\eqref{eq::dyn_so5}  evaluated at $L=8$ and different $|\JH|/\JA$ (labeled on top of each sub-plots).
}
\end{figure*}
\begin{figure*}[htb]
\includegraphics[width=180mm,scale=1]{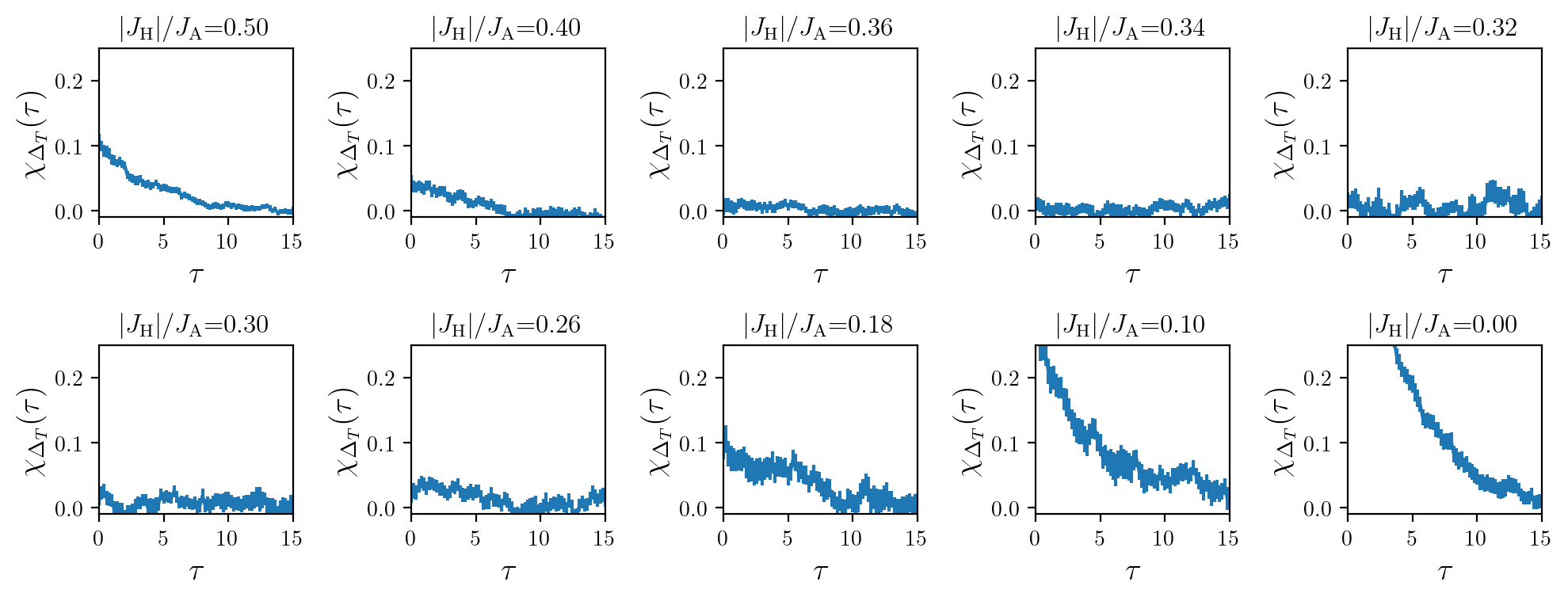}
\caption{\label{Fig::tauso5_nolog}   The same data of $\chi_{\Delta_{\tn{T}}}(\tau)$ as Fig.\ref{Fig::tauso5_nolog}, but in a normal scale.
}
\end{figure*}
\begin{figure*}[htb]
\includegraphics[width=180mm,scale=1]{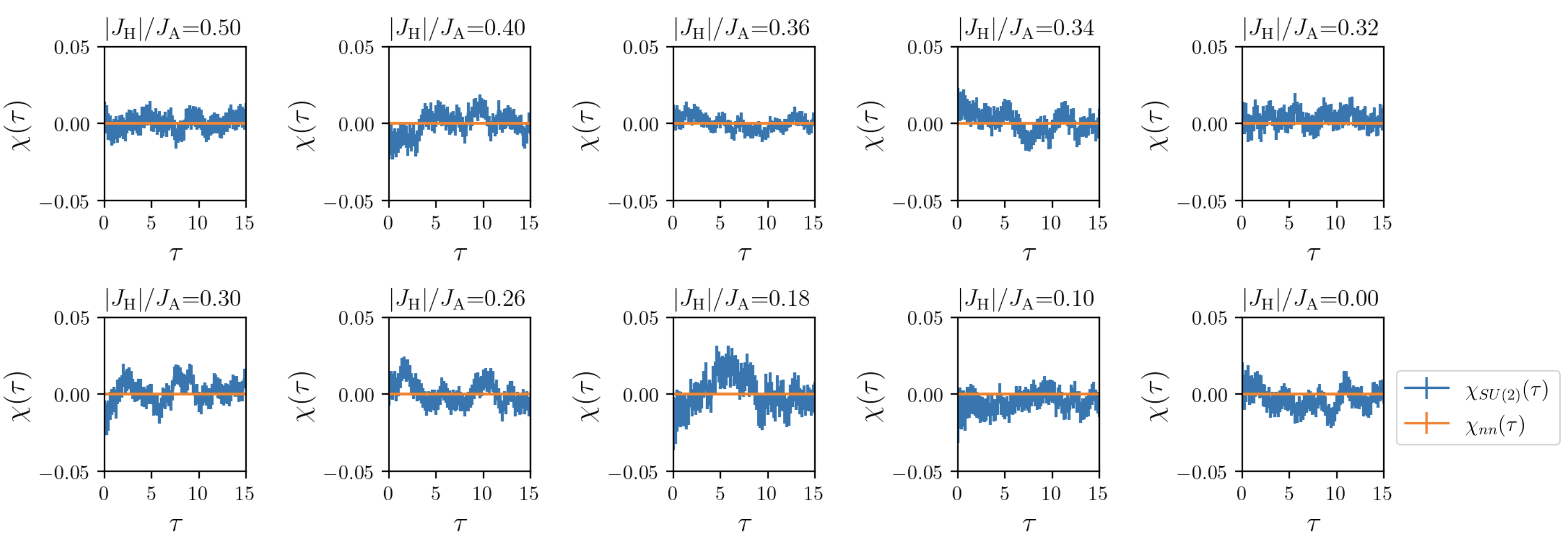}
\caption{\label{Fig::tauso5_nolog_su2} 
 Dynamical correlation function $\chi(\tau)$ for density (orange curves) and spin channels (blue curves), evaluated at $L=8$ and different $|\JH|/\JA$ (labeled on top of each sub-plots).
}
\end{figure*}

In this section, we provide additional data on $\chi_{\Delta_{\tn{T}}}(\tau)$ defined in Eq.~\eqref{eq::dyn_so5}. In Fig.\ref{Fig::tauso5_supp}, we plot $\chi_{\Delta_{\tn{T}}}(\tau)$ at different $|\JH|/\JA$ in a semi-log scale. As noted in the main text, the slope of $\chi_{\Delta_{\tn{T}}}(\tau)$ in the semi-log plot denotes the energy cost to rotate the order parameter between ChAF and SC, which serves as a probe for the emergent SO(5) symmetry. We found that the energy cost decrease either when decreasing $|\JH|/\JA$ from $|\JH|/\JA=0.5$ or when increasing $|\JH|/\JA$ from $|\JH|/\JA=0$; whereas near the quantum critical point(s), $\chi_{\Delta_{\tn{T}}}(\tau)$ behaves as noisy fluctuation near zero (see Fig.\ref{Fig::tauso5_nolog} for normal scale plots). We believe that such noisy fluctuation near zero are expected if the SO(5) symmetry does emerge near the critical point(s) due to the vanishing static susceptibility.

\subsection{Vanishing Static Correlation for the Emergent Conserved Charges }
Let us start from the low-energy effective rotor model after explicit projection to active Hilbert space $H_{\tn{rotor}}$ defined in Eq.\ref{eq::ham_rotor_k_main}. Let us define the conserved charge as
\beq
Q_{ab} \equiv \int_{\k,\q} \Psi^{\dagger}_{\k+\q} L_{ab} \Psi_{\k}.
\eeq
As we stressed in Appendix \ref{app::so5}, the SO(5) symmetry emerge near the critical point at fermionic level, namely $[Q_{ab},H_{\tn{rotor}}]=0$. Thus, it is natural to make the following assumption for the zero-temperature density matrix $\rho_0$ as
\beq
\rho_0 \equiv \lim_{\beta \rightarrow \infty} e^{-\beta H_{\tn{rotor}}}/\Tr[e^{-\beta H_{\tn{rotor}}}]; ~~~ [\rho_0, Q_{ab}] = 0
\eeq
, which also agrees with our PQMC simulation in the sense that neither of the SO(5) order parameters are ordered at the critical point. 

Since $Q_{ab}$'s fulfill the adjoint representation of SO(5), it is not hard to find specific symmetry transformation $U_{ab\rightarrow cd}$ in SO(5) group via standard group theoretical representation theory such that
\beq
U^{-1}_{ab\rightarrow cd} Q_{ab} U_{ab\rightarrow cd} = Q_{cd}.
\eeq
Specifically, for $(c,d)=(1,5)$, this immediately implies the following property of the static correlation function
\beq\label{eq::so5_static}
\Tr[\rho_0 Q_{ab}^2] = \Tr[\rho_0 Q_{15}^2], ~~~ \forall a<b.
\eeq
Since $Q_{15}\equiv \sum_{\r\tau\alpha}(n_{\r\tau\alpha}-2)$ and our PQMC simulation are performed in the canonical ensemble, the RHS of Eq.\ref{eq::so5_static} should vanish (orange curves in Fig.\ref{Fig::tauso5_nolog_su2}). This will imply that $\chi_{\Delta_{\tn{T}}}(\tau=0)=0$ at the critical point with emergent SO(5) symmetry, which agrees with the results in PQMC simulation where only noisy fluctuation near zero are observed. A small residue finite value of $\chi_{\Delta_{\tn{T}}}(\tau)$ at $|\JH|/\JA=0.3, 0.32, 0.34$ and $0.36$ may be due to the effect of remote bands (i.e. not in a strict projected limit) or finite size effect, and we will leave this open question for future study.

For completeness, we also show the dynamical correlation functions of other generators (defined in Eq.\eqref{eq::Angular_momentum}) of the $SO(5)$ group in Fig.\ref{Fig::tauso5_nolog_su2}, where orange curves are for $Q_{15}\equiv \tn{density}$ and blue curves are for a $SU(2)_{\tn{spin}}$-symmetrized correlator for $\{Q_{23}, Q_{34}, Q_{24} \}\equiv \tn{spin}$. We note the dynamical spin correlation functions are noisy fluctuations around zero for all $|\JH|/\JA$, since both ChAF and SC phases are effectively spin-singlet.

\section{Crossing point analysis for quantum phase transition(s)}\label{App::cross}

\begin{figure*}[htb]
\includegraphics[width=140mm,scale=1]{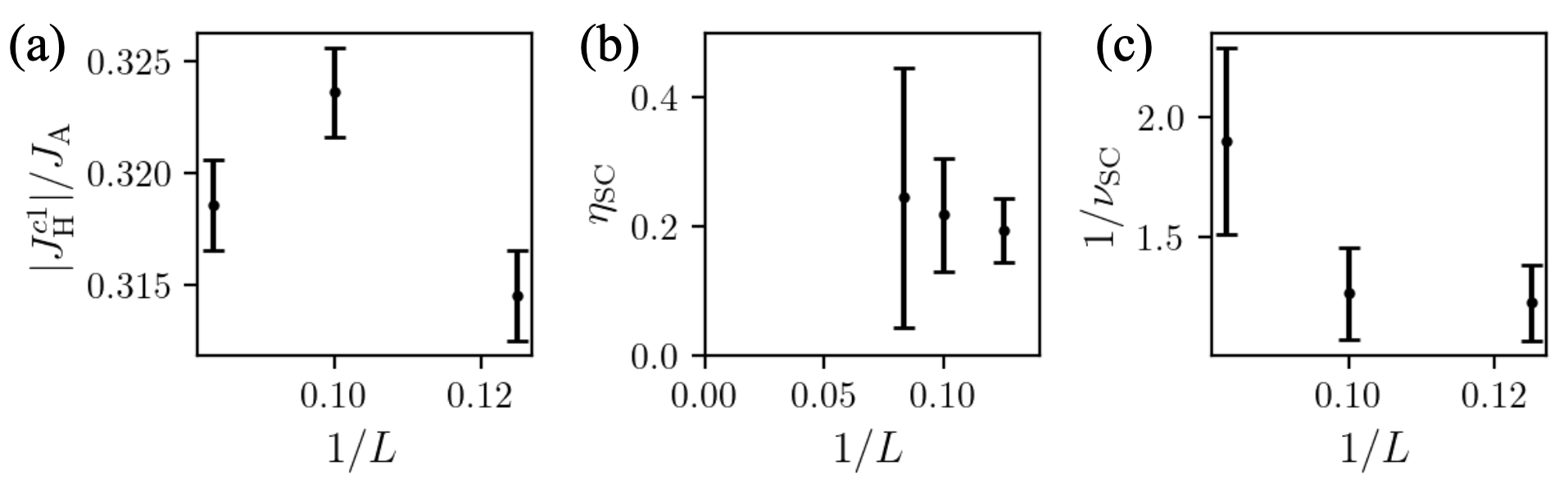}
\caption{\label{fig::cross_sc} Crossing point analysis of SC order parameter for interaction-anisotropy tuned transition. (a) Critical points for different linear system size, extracted from the crossing point for $\xi_{\tn{{SC}}}$ between $L$ and $L+2$. (b) Anomalous dimension $\eta_{\tn{{SC}}}$  extracted using Eq.~\eqref{eq::crossing_point_eta}. (c) $1/\nu_{\tn{{SC}}}$ extracted using Eq.~\eqref{eq::crossing_point_nu}.}
\end{figure*}
\begin{figure*}[htb]
\includegraphics[width=140mm,scale=1]{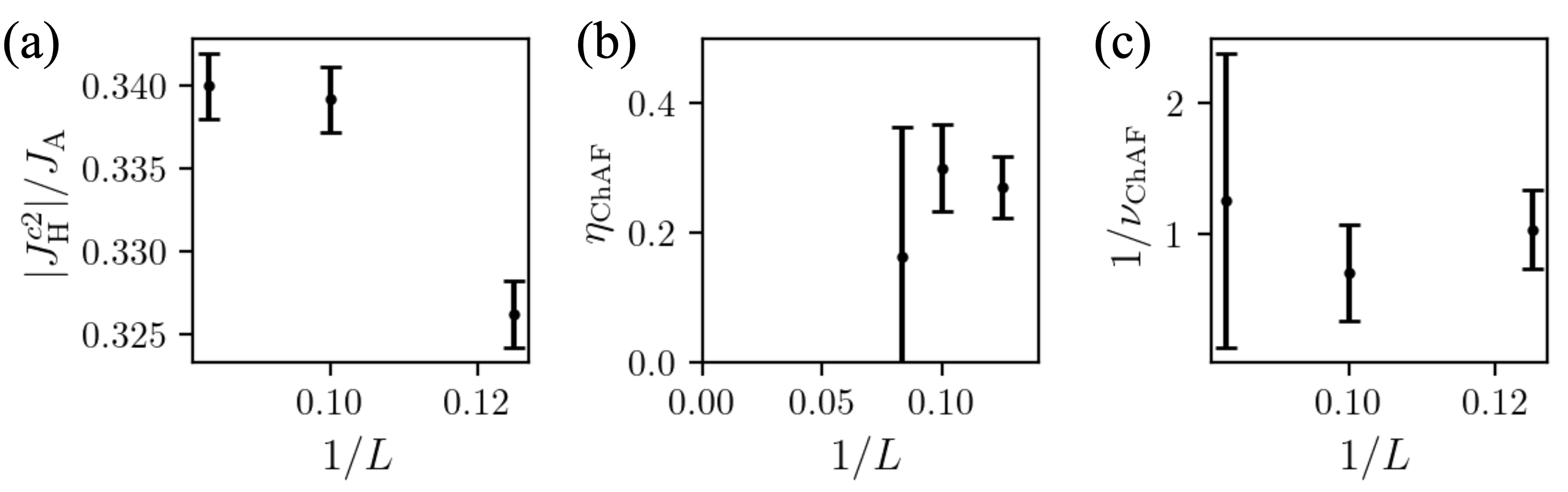}
\caption{\label{fig::cross_caf} Crossing point analysis of ChAF order parameter in interaction-anisotropy tuned transition. (a) critical points for different linear system size, extracted from the crossing point for $\xi_{\tn{{ChAF}}}$ between $L$ and $L+2$. (b) anomalous dimension $\eta_{\tn{{ChAF}}}$  extracted using Eq.~\eqref{eq::crossing_point_eta}. (c) $1/\nu_{\tn{{ChAF}}}$ extracted using Eq.~\eqref{eq::crossing_point_nu}}
\end{figure*}

In this appendix, we provide additional details for the interaction-anisotropy tuned transition(s) by carrying out an independent crossing point analysis. The (pseudo-)critical points, as discussed in the main text, for different linear system size $L$ are extracted from the crossing point of $\xi_{O}/L$ between $L$ and $L+2$, for $O=\tn{ChAF}, \tn{SC}$, respectively. The anomalous dimensions, $\eta_{O}$, for different $L$ can also be derived from the scaling ansatz Eq.~\eqref{eq::collect_fit2} 
\begin{equation}\label{eq::crossing_point_eta}
    \eta_O(L) \equiv \frac{1}{\log [L/(L+2)]} \log \frac{S_O(L+2)}{S_O(L)}\bigg|_{\JH=\JH^c} -1.
\end{equation}
Here, $S_O(L)$ in Eq.~\eqref{eq::crossing_point_eta} denotes the static susceptibility for $O$. The correlation length exponent, $\nu_O$, can also be extracted using the data for two different system sizes, $L$ and $L+2$, 
\begin{equation}\label{eq::crossing_point_nu}
    \frac{1}{\nu_O(L)} \equiv \frac{1}{\log [(L+2)/L]} \log \frac{\partial_{\JH}R_O(L+2)}{\partial_{\JH}R_O(L)}\bigg|_{\JH=\JH^c},
\end{equation}
where $R_O(L)\equiv \xi_O/L$ is the RG-invariant correlation length. The result of crossing point analysis for SC and ChAF are shown in Fig.~\ref{fig::cross_sc} and  Fig.~\ref{fig::cross_caf}, respectively.

\begin{figure*}[htb]
\includegraphics[width=180mm,scale=1]{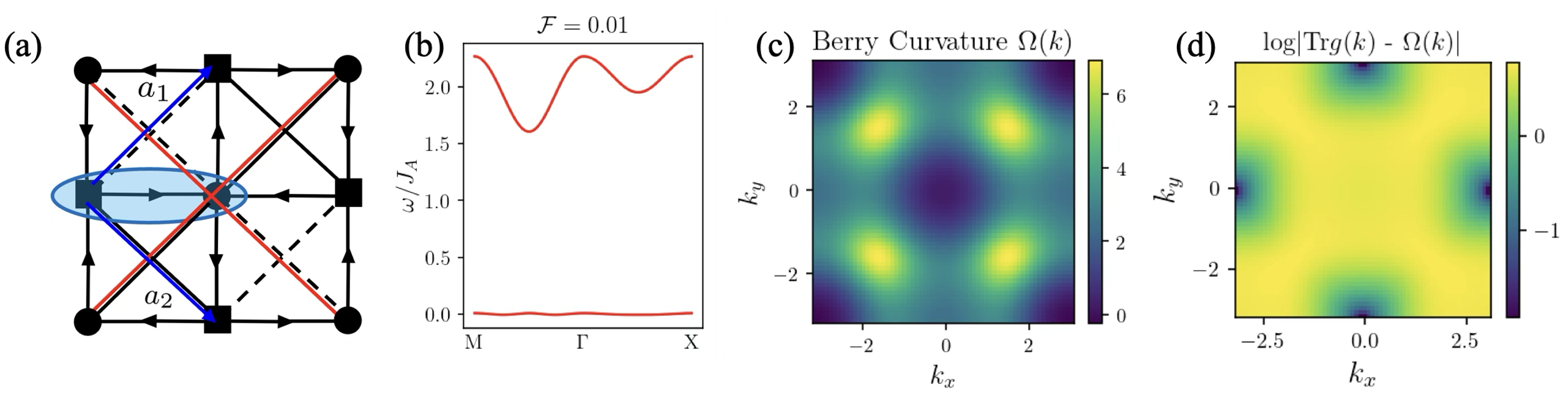}% Here is how to import EPS art
\caption{\label{Fig::band}
 (a) Schematics for lattice model, where sublattice $\eta=A$ are denoted by black circle and $\eta=B$ are denoted by black box. Black arrows denote the phase $\phi_{\r,\r'}$ in $t_1$ hopping in a $\pi$-flux lattice. Solid and dashed black lines denotes $s_{\langle \r, \r'\rangle_2}=1$ and $s_{\langle \r, \r'\rangle_2}=-1$ for hopping $t_2$, respectively. Red lines denotes hopping term $t_5$.
 (b) Non-interacting band structure along a high-symmetry cut in the Brillouin zone for $\mathcal{F} = 0.01$. (c) Berry curvature distribution $\Omega(\k)$, which is independent of $\mathcal{F}$ since $t_5$ couples to the identity matrix in sublattice space. (d) Deviation from ideal band geometry in a log-scale. Here $g(\k)$ denotes the quantum geometric tensor.
}
\end{figure*}

\end{widetext}

\section{Details on Lattice Hamiltonian and Absence of Sign-Problem}

For the sake of completeness, let us reiterate the detailed setup of our lattice Hamiltonian and the absence of fermion sign-problem. 

The non-interacting lattice model is given by,
\begin{equation}\label{eq::h_kin_real}
\begin{aligned}
H_{\tn{kin}}^{(\tau)}= & \bigg[
~-t_1 \sum_{\langle \r, \r'\rangle_1, \sigma} e^{i \phi_{\r \r'}^\tau} c_{\r, \sigma}^{\dagger} c_{\r', \sigma}
\\&
~~~-t_2 \sum_{\langle \r, \r'\rangle_2, \sigma} s_{\langle \r, \r'\rangle_2} c_{\r, \sigma}^{\dagger} c_{\r', \sigma}
\\&
~~~-t_5 \sum_{\langle \r, \r'\rangle_5, \sigma} c_{\r, \sigma}^{\dagger} c_{\r', \sigma}+\text { h.c. }\bigg]-\mu \sum_i n_i
\end{aligned}
\end{equation}
, where $t,~t_2,~t_5$ are the hopping parameters and $\langle \r, \r'\rangle_n$ denotes that $\r$ and $\r'$ belongs to the n-th nearest neighbor. Again, we stress that $\r=\vec{R}+\vec{\delta}_\eta$ is the combined coordinates of unit-cell and sublattice, where $\vec{\delta}_\eta$ is the coordinates of sublattice $\eta=A,B$ in a unit cell. 

A schematics of the lattice model is shown in Fig.\ref{Fig::band}a. Here, the sublattice $\eta=A$ are denoted by black circle and $\eta=B$ are denoted by black box. Black arrows denote the phase $\phi_{\r,\r'}$ in $t_1$ hopping in a $\pi$-flux lattice. Solid and dashed black lines denotes $s_{\langle \r, \r'\rangle_2}=1$ and $s_{\langle \r, \r'\rangle_2}=-1$ for hopping $t_2$, respectively.
The unit cell are spanned by blue arrow in Fig.\ref{Fig::band}a, where $\vec{a}_1 = (1,1)$ and $\vec{a}_2=(1,-1)$. The coordinates of sublattice $\eta$ within unit-cell are given by $\vec{\delta}_A=(1,0)$ and $\vec{\delta}_B=(0,0)$. Here, we choose a $C_4$-symmetric assignment of sublattice location. For the simulations with $\mathcal{F}=0.01$, the hopping parameters are given by $t_2 = t/\sqrt{2}$ and $t_5 = (1-\sqrt{2})t/4 $ \cite{single_chern}; whereas for the simulations with $\mathcal{F}=0.2$, the hopping parameters are given by  $t_2 = t/\sqrt{2}$ and $t_5 = 0 $, respectively. 
The non-interacting band structure, associated Berry curvature distribution and deviation from ideal band geometry in the Brillouin zone are shown in Fig.~\ref{Fig::band}.

In the DQMC simulation, the presence of an anti-unitary symmetry $\mathcal{T}' = i\tau_x \sigma_y\mathcal{K}$ guarantees the absence of fermion sign-problem \cite{congjun}. Let us re-write the interaction  Hamiltonian, $H_{\tn{interaction}}$, in Eqn.~\ref{eq::ham_int} as,

\beq
H_{\tn{int}} &= \sum_{\vec{r}} \bigg[J_1 (\vec{S}^{+}_{\vec{r}} + \vec{S}^{-}_{\vec{r}})^2 - J_2 (\vec{S}^{+}_{\vec{r}} - \vec{S}^{-}_{\vec{r}})^2 \bigg],
\eeq
where $J_1 = J_{\tn{A}}/4 + J_{\tn{H}}/2$ and $J_2 = J_{\tn{A}}/4 - J_{\tn{H}}/2$, respectively. Introducing a Hubbard-Stratonovich decomposition using auxiliary fields, $\vec{\phi}_{1\r}$, $\vec{\phi}_{2\r}$, in the spin channel, at leading order in $\Delta \tau$, the imaginary-time evolution operator is given by,
\begin{widetext}
\begin{equation}\label{eq:hint_hs}
\begin{aligned}
e^{-\Delta\tau H_{\tn{int}}} &= \sum_{\{\vec{\phi}_{1\r},\vec{\phi}_{2\r}\}} \gamma_{\vec{\phi}_{1\r}} \gamma_{\vec{\phi}_{2\r}} e^{\sum_{\vec{r}} \bigg[\sqrt{-\Delta\tau J_1} ~ \vec{\phi}_{1\r} \cdot (\vec{S}^{+}_{\vec{r}} + \vec{S}^{-}_{\vec{r}}) - i \sqrt{-\Delta\tau J_2} \vec{\phi}_{2\r} \cdot(\vec{S}^{+}_{\vec{r}} - \vec{S}^{-}_{\vec{r}})\bigg]}
\end{aligned}
\end{equation}
\end{widetext}
, where $\gamma_{\vec{\phi}_{i\r}}$ is the associated weight for Gaussian quadrature decomposition. The spin operators in each valley transform under $\mathcal{T}'$ as $\mathcal{T}' \vec{S}^{\pm}_{\vec{r}} \mathcal{T}'^{-1} = -\vec{S}^{\mp}_{\vec{r}}$. It is straightforward to show that if $J_1>0$ and $J_2>0$, $e^{-\Delta\tau H_{\tn{int}}}$ is invariant under $\mathcal{T}'$, which ensures the eigenstates of $H_{\tn{interaction}}$ can be grouped into Kramer doublets, guaranteeing the positive-definiteness of the partition function \cite{congjun}. Similar arguments also hold for $H_{\tn{kin}}$ in the sense that $\mathcal{T}' H_{\tn{kin}} \mathcal{T}'^{-1} = H_{\tn{kin}}$, and thus the model is sign-problem-free.

\section{Projective quantum Monte Carlo}
\label{app:pqmc}

We use the zero-temperature projective quantum Monte Carlo
(PQMC) method to evaluate ground-state properties of the interacting
fermion model. For a detailed overview of the finite-temperature and
projective auxiliary-field formulation, see Ref.~\cite{alf, AssaadEvertz2008}.

% \ket{\Psi^{\Theta}_{\rm{G}}}\equiv e^{-\Theta H}\ket{\Psi_{\rm{trial}}}
Let $\ket{\Psi_{\rm{trial}}}$ denote an $N_p$-particle Slater determinant,
\begin{equation}
\ket{\Psi_{\rm{trial}}} =
\prod_{a=1}^{N_p}
\left(
\sum_i [\Phi_T]_{ia} c_i^\dagger
\right)|0\rangle ,
\end{equation}
where $i$ labels the single-particle quantum numbers, including unit cell,
sublattice, valley, and spin, and $\Phi_T$ is an
$N_{\rm ham}\times N_p$ matrix ($N_p\equiv \tn{particle number}$) whose columns specify the occupied
single-particle orbitals of the trial wavefunction. In the simulations at
$\nu=2$, we work in the canonical ensemble with fixed particle number
$N_p=2L^2$. The trial state is chosen to have nonzero overlap with the
ground state and to preserve the microscopic
$U(1)_{\rm charge}\times U(1)_{\rm valley}\times SU(2)_{\rm spin}$
symmetry.

Ground-state expectation values are obtained from
\begin{equation}
\langle O\rangle_0 =
\lim_{\Theta\rightarrow\infty}
\frac{
\langle \Psi_{\rm{trial}}|
e^{-\Theta H}
O
e^{-\Theta H}
|\Psi_{\rm{trial}}\rangle
}{
\langle \Psi_{\rm{trial}}|
e^{-2\Theta H}
|\Psi_{\rm{trial}}\rangle
}.
\label{eq:pqmc_expectation}
\end{equation}
In practice, the projection length is discretized as
$2\Theta=L_\tau\Delta\tau$. We use a Trotter decomposition,
\begin{equation}
e^{-\Delta\tau H}
=
e^{-\Delta\tau H_{\rm kin}}
e^{-\Delta\tau H_{\rm int}}
+\mathcal{O}(\Delta\tau^2),
\label{eq:pqmc_trotter}
\end{equation}
and decouple the interaction term by a Hubbard-Stratonovich transformation
in the spin channel, as defined in Eq.~\eqref{eq:hint_hs}. For each time slice $\ell$ one obtains a quadratic
fermion evolution operator
\begin{equation}
\hat B_\ell[s_\ell]
=
e^{-\Delta\tau \hat H_{\rm kin}}
e^{\hat V[s_\ell]},
\label{eq:pqmc_Bhat}
\end{equation}
where $s_\ell\equiv\{ \phi_{i,\r}^{(\ell)} \}$ denotes the collection of auxiliary fields on time slice
$\ell$. 

For a fixed HS configuration
$s\equiv\{s_1,\ldots,s_{L_\tau}\}$, the fermions are quadratic and can be
integrated out exactly. Let $B_\ell[s_\ell]$ denote the single-particle
matrix representation of $\hat B_\ell[s_\ell]$, and define
\begin{equation}
\mathcal{B}[s]
=
B_{L_\tau}[s_{L_\tau}]
\cdots
B_2[s_2]
B_1[s_1].
\end{equation}
The projected normalization becomes
\begin{equation}
Z_{\rm PQMC}(\Theta)
\equiv
\langle \Psi_{\rm trial}|e^{-2\Theta H}|\Psi_{\rm trial}\rangle
=
\sum_{\{s\}}
W[s],
\label{eq:pqmc_Z}
\end{equation}
with
\begin{equation}
W[s]
=
P_{\rm HS}[s]\
\det
\left[
\Phi_T^\dagger
\mathcal{B}[s]
\Phi_T
\right].
\label{eq:pqmc_weight}
\end{equation}
Here $P_{\rm HS}[s]$ contains the product of the bare HS-field weights and
normalization factors. 

Expectation values of observables are obtained by averaging over HS configurations,
\begin{equation}
\langle O\rangle_0
=
\frac{
\sum_{\{s\}}
W[s]\,
O[s]
}{
\sum_{\{s\}}
W[s]
},
\label{eq:pqmc_average}
\end{equation}
where $O[s]$ denotes the estimator for the observable $O$ computed from the
equal-time Green's function in the HS-field configuration $s$. For correlation functions observables, $O[s]$ can be evaluated using Wick's theorem since the fermions are effectively non-interacting for fixed $s$.

\bibliographystyle{apsrev4-1_custom}
\bibliography{draftv1.bib}

@article{svh,
  title = {Intertwined magnetism and superconductivity in isolated correlated flat bands},
  author = {Wang, Xuepeng and Mendez-Valderrama, J. F. and Hofmann, Johannes S. and Chowdhury, Debanjan},
  journal = {Phys. Rev. B},
  volume = {110},
  issue = {4},
  pages = {L041105},
  numpages = {6},
  year = {2024},
  month = {Jul},
  publisher = {American Physical Society},
  doi = {10.1103/PhysRevB.110.L041105},
  url = {https://link.aps.org/doi/10.1103/PhysRevB.110.L041105}
}

@article{Gorbenko2018,
   author = {Victor Gorbenko and Slava Rychkov and Bernardo Zan},
   doi = {10.21468/SCIPOSTPHYS.5.5.050/PDF},
   issn = {25424653},
   issue = {5},
   journal = {SciPost Physics},
   month = {11},
   pages = {050},
   publisher = {SciPost Foundation},
   title = {Walking, weak first-order transitions, and complex CFTs II. Two-dimensional Potts model at Q > 4},
   volume = {5},
   year = {2018}
}

@article{Rychkov18,
   author = {Victor Gorbenko and Slava Rychkov and Bernardo Zan},
   doi = {10.1007/JHEP10(2018)108},
   issn = {1029-8479},
   issue = {10},
   journal = {Journal of High Energy Physics},
   month = {10},
   pages = {108},
   title = {Walking, weak first-order transitions, and complex CFTs},
   volume = {2018},
   url = {http://link.springer.com/10.1007/JHEP10(2018)108},
   year = {2018}
}

@article{WangC19,
   author = {Ruochen Ma and Chong Wang},
   doi = {10.1103/PhysRevB.102.020407},
   issn = {2469-9950},
   issue = {2},
   journal = {Physical Review B},
   month = {7},
   pages = {020407},
   title = {Theory of deconfined pseudocriticality},
   volume = {102},
   url = {https://link.aps.org/doi/10.1103/PhysRevB.102.020407},
   year = {2020}
}

@article{chatterjee20,
  title = {Symmetry breaking and skyrmionic transport in twisted bilayer graphene},
  author = {Chatterjee, Shubhayu and Bultinck, Nick and Zaletel, Michael P.},
  journal = {Phys. Rev. B},
  volume = {101},
  issue = {16},
  pages = {165141},
  numpages = {25},
  year = {2020},
  month = {Apr},
  publisher = {American Physical Society},
  doi = {10.1103/PhysRevB.101.165141},
  url = {https://link.aps.org/doi/10.1103/PhysRevB.101.165141}
}

@article{intertwined,
  title = {Colloquium: Theory of intertwined orders in high temperature superconductors},
  author = {Fradkin, Eduardo and Kivelson, Steven A. and Tranquada, John M.},
  journal = {Rev. Mod. Phys.},
  volume = {87},
  issue = {2},
  pages = {457--482},
  numpages = {26},
  year = {2015},
  month = {May},
  publisher = {American Physical Society},
  doi = {10.1103/RevModPhys.87.457},
  url = {https://link.aps.org/doi/10.1103/RevModPhys.87.457}
}

@article{
YC_nematic,
author = {Yuan Cao  and Daniel Rodan-Legrain  and Jeong Min Park  and Noah F. Q. Yuan  and Kenji Watanabe  and Takashi Taniguchi  and Rafael M. Fernandes  and Liang Fu  and Pablo Jarillo-Herrero },
title = {Nematicity and competing orders in superconducting magic-angle graphene},
journal = {Science},
volume = {372},
number = {6539},
pages = {264-271},
year = {2021},
doi = {10.1126/science.abc2836},
URL = {https://www.science.org/doi/abs/10.1126/science.abc2836},
eprint = {https://www.science.org/doi/pdf/10.1126/science.abc2836}}

@article{Efetov21,
	author = {Das, Ipsita and Lu, Xiaobo and Herzog-Arbeitman, Jonah and Song, Zhi-Da and Watanabe, Kenji and Taniguchi, Takashi and Bernevig, B. Andrei and Efetov, Dmitri K.},
	journal = {Nature Physics},
	number = {6},
	pages = {710--714},
	title = {Symmetry-broken Chern insulators and Rashba-like Landau-level crossings in magic-angle bilayer graphene},
	volume = {17},
	year = {2021}}

@article{
dean19_tbg,
author = {Matthew Yankowitz  and Shaowen Chen  and Hryhoriy Polshyn  and Yuxuan Zhang  and K. Watanabe  and T. Taniguchi  and David Graf  and Andrea F. Young  and Cory R. Dean },
title = {Tuning superconductivity in twisted bilayer graphene},
journal = {Science},
volume = {363},
number = {6431},
pages = {1059-1064},
year = {2019},
doi = {10.1126/science.aav1910},
URL = {https://www.science.org/doi/abs/10.1126/science.aav1910},
eprint = {https://www.science.org/doi/pdf/10.1126/science.aav1910}}

@article{Lu19,
	author = {Lu, Xiaobo and Stepanov, Petr and Yang, Wei and Xie, Ming and Aamir, Mohammed Ali and Das, Ipsita and Urgell, Carles and Watanabe, Kenji and Taniguchi, Takashi and Zhang, Guangyu and Bachtold, Adrian and MacDonald, Allan H. and Efetov, Dmitri K.},
	journal = {Nature},
	number = {7780},
	pages = {653--657},
	title = {Superconductors, orbital magnets and correlated states in magic-angle bilayer graphene},
	volume = {574},
	year = {2019}}

@article{Park21_tbg,
	author = {Park, Jeong Min and Cao, Yuan and Watanabe, Kenji and Taniguchi, Takashi and Jarillo-Herrero, Pablo},
	journal = {Nature},
	number = {7852},
	pages = {43--48},
	title = {Flavour Hund's coupling, Chern gaps and charge diffusivity in moir{\'e}graphene},
	volume = {592},
	year = {2021}}

@article{Zeldov22,
	author = {Grover, Sameer and Bocarsly, Matan and Uri, Aviram and Stepanov, Petr and Di Battista, Giorgio and Roy, Indranil and Xiao, Jiewen and Meltzer, Alexander Y. and Myasoedov, Yuri and Pareek, Keshav and Watanabe, Kenji and Taniguchi, Takashi and Yan, Binghai and Stern, Ady and Berg, Erez and Efetov, Dmitri K. and Zeldov, Eli},
	journal = {Nature Physics},
	number = {8},
	pages = {885--892},
	title = {Chern mosaic and Berry-curvature magnetism in magic-angle graphene},
	volume = {18},
	year = {2022}}

@article{Young20,
	author = {Saito, Yu and Ge, Jingyuan and Watanabe, Kenji and Taniguchi, Takashi and Young, Andrea F.},
	journal = {Nature Physics},
	number = {9},
	pages = {926--930},
	title = {Independent superconductors and correlated insulators in twisted bilayer graphene},
	volume = {16},
	year = {2020}}

@article{Ghiotto_2021, title={Quantum criticality in twisted transition metal dichalcogenides}, volume={597}, ISSN={1476-4687}, url={http://dx.doi.org/10.1038/s41586-021-03815-6}, DOI={10.1038/s41586-021-03815-6}, number={7876}, journal={Nature}, publisher={Springer Science and Business Media LLC}, author={Ghiotto, Augusto and Shih, En-Min and Pereira, Giancarlo S. S. G. and Rhodes, Daniel A. and Kim, Bumho and Zang, Jiawei and Millis, Andrew J. and Watanabe, Kenji and Taniguchi, Takashi and Hone, James C. and Wang, Lei and Dean, Cory R. and Pasupathy, Abhay N.}, year={2021}, month=sep, pages={345–349} }

@article{young21,
	author = {Kennes, Dante M. and Claassen, Martin and Xian, Lede and Georges, Antoine and Millis, Andrew J. and Hone, James and Dean, Cory R. and Basov, D. N. and Pasupathy, Abhay N. and Rubio, Angel},
	journal = {Nature Physics},
	number = {2},
	pages = {155--163},
	title = {Moir{\'e}heterostructures as a condensed-matter quantum simulator},
	volume = {17},
	year = {2021}}

@article{Li2021,
	author = {Li, Tingxin and Jiang, Shengwei and Li, Lizhong and Zhang, Yang and Kang, Kaifei and Zhu, Jiacheng and Watanabe, Kenji and Taniguchi, Takashi and Chowdhury, Debanjan and Fu, Liang and Shan, Jie and Mak, Kin Fai},
	journal = {Nature},
	number = {7876},
	pages = {350--354},
	title = {Continuous Mott transition in semiconductor moir{\'e}superlattices},
	volume = {597},
	year = {2021}}

@article{tmdsc25,
	author = {Kim, Sunghoon and Mendez-Valderrama, Juan Felipe and Wang, Xuepeng and Chowdhury, Debanjan},
	date = {2025/02/17},
	doi = {10.1038/s41467-025-56816-8},
	id = {Kim2025},
	isbn = {2041-1723},
	journal = {Nature Communications},
	number = {1},
	pages = {1701},
	title = {Theory of correlated insulators and superconductor at ν= 1 in twisted WSe2},
	url = {https://doi.org/10.1038/s41467-025-56816-8},
	volume = {16},
	year = {2025}}

@misc{zhang2025,
      title={Identifying Instabilities with Quantum Geometry in Flat Band Systems}, 
      author={Jia-Xin Zhang and Wen O. Wang and Leon Balents and Lucile Savary},
      year={2025},
      eprint={2504.03882},
      archivePrefix={arXiv},
      primaryClass={cond-mat.str-el},
      url={https://arxiv.org/abs/2504.03882}, 
}

@article{sphere1,
  title = {SO(5) Deconfined Phase Transition under the Fuzzy-Sphere Microscope: Approximate Conformal Symmetry, Pseudo-Criticality, and Operator Spectrum},
  author = {Zhou, Zheng and Hu, Liangdong and Zhu, W. and He, Yin-Chen},
  journal = {Phys. Rev. X},
  volume = {14},
  issue = {2},
  pages = {021044},
  numpages = {22},
  year = {2024},
  month = {Jun},
  publisher = {American Physical Society},
  doi = {10.1103/PhysRevX.14.021044},
  url = {https://link.aps.org/doi/10.1103/PhysRevX.14.021044}
}

@article{tri-crit,
  title = {Bootstrapping Deconfined Quantum Tricriticality},
  author = {Chester, Shai M. and Su, Ning},
  journal = {Phys. Rev. Lett.},
  volume = {132},
  issue = {11},
  pages = {111601},
  numpages = {7},
  year = {2024},
  month = {Mar},
  publisher = {American Physical Society},
  doi = {10.1103/PhysRevLett.132.111601},
  url = {https://link.aps.org/doi/10.1103/PhysRevLett.132.111601}
}

@article{sphere2,
  title = {Phases of $(2+1)\mathrm{D}$ SO(5) Nonlinear Sigma Model with a Topological Term on a Sphere: Multicritical Point and Disorder Phase},
  author = {Chen, Bin-Bin and Zhang, Xu and Wang, Yuxuan and Sun, Kai and Meng, Zi Yang},
  journal = {Phys. Rev. Lett.},
  volume = {132},
  issue = {24},
  pages = {246503},
  numpages = {7},
  year = {2024},
  month = {Jun},
  publisher = {American Physical Society},
  doi = {10.1103/PhysRevLett.132.246503},
  url = {https://link.aps.org/doi/10.1103/PhysRevLett.132.246503}
}

@article{sphere3,
  title = {Emergent conformal symmetry at the multicritical point of $(2+1)\mathrm{D}$ SO(5) model with Wess-Zumino-Witten term on a sphere},
  author = {Chen, Bin-Bin and Zhang, Xu and Meng, Zi Yang},
  journal = {Phys. Rev. B},
  volume = {110},
  issue = {12},
  pages = {125153},
  numpages = {7},
  year = {2024},
  month = {Sep},
  publisher = {American Physical Society},
  doi = {10.1103/PhysRevB.110.125153},
  url = {https://link.aps.org/doi/10.1103/PhysRevB.110.125153}
}

@article{Oh21,
	author = {Oh, Myungchul and Nuckolls, Kevin P. and Wong, Dillon and Lee, Ryan L. and Liu, Xiaomeng and Watanabe, Kenji and Taniguchi, Takashi and Yazdani, Ali},
	journal = {Nature},
	number = {7888},
	pages = {240--245},
	title = {Evidence for unconventional superconductivity in twisted bilayer graphene},
	volume = {600},
	year = {2021}}

@article{Wu21,
	author = {Wu, Shuang and Zhang, Zhenyuan and Watanabe, K. and Taniguchi, T. and Andrei, Eva Y.},
	journal = {Nature Materials},
	number = {4},
	pages = {488--494},
	title = {Chern insulators, van Hove singularities and topological flat bands in magic-angle twisted bilayer graphene},
	volume = {20},
	year = {2021}}

@article{Stepanov20,
	author = {Stepanov, Petr and Das, Ipsita and Lu, Xiaobo and Fahimniya, Ali and Watanabe, Kenji and Taniguchi, Takashi and Koppens, Frank H. L. and Lischner, Johannes and Levitov, Leonid and Efetov, Dmitri K.},
	journal = {Nature},
	number = {7816},
	pages = {375--378},
	title = {Untying the insulating and superconducting orders in magic-angle graphene},
	volume = {583},
	year = {2020}}

@article{
Sharpe19,
author = {Aaron L. Sharpe  and Eli J. Fox  and Arthur W. Barnard  and Joe Finney  and Kenji Watanabe  and Takashi Taniguchi  and M. A. Kastner  and David Goldhaber-Gordon },
title = {Emergent ferromagnetism near three-quarters filling in twisted bilayer graphene},
journal = {Science},
volume = {365},
number = {6453},
pages = {605-608},
year = {2019},
doi = {10.1126/science.aaw3780},
URL = {https://www.science.org/doi/abs/10.1126/science.aaw3780},
eprint = {https://www.science.org/doi/pdf/10.1126/science.aaw3780}}

@article{Schindler22,
  title = {Trions in twisted bilayer graphene},
  author = {Schindler, Frank and Vafek, Oskar and Bernevig, B. Andrei},
  journal = {Phys. Rev. B},
  volume = {105},
  issue = {15},
  pages = {155135},
  numpages = {23},
  year = {2022},
  month = {Apr},
  publisher = {American Physical Society},
  doi = {10.1103/PhysRevB.105.155135},
  url = {https://link.aps.org/doi/10.1103/PhysRevB.105.155135}
}

@ARTICLE{skyr_short,
       author = {{Wang}, Xuepeng and {Mendez-Valderrama}, J.~F. and {Hofmann}, Johannes S. and {Chowdhury}, Debanjan},
        title = "{Spin-Polaron Mediated Superconductivity in Doped Chern Antiferromagnets}",
      journal = {arXiv e-prints},
         year = 2025,
        month = jul,
          eid = {arXiv:2507.22971},
        pages = {arXiv:2507.22971},
          doi = {10.48550/arXiv.2507.22971},
archivePrefix = {arXiv},
       eprint = {2507.22971},
 primaryClass = {cond-mat.str-el},
       adsurl = {https://ui.adsabs.harvard.edu/abs/2025arXiv250722971W}
}

@article{baby_skyrmion,
	author = {Khalaf, Eslam and Vishwanath, Ashvin},
	date = {2022/10/21},
	doi = {10.1038/s41467-022-33673-3},
	id = {Khalaf2022},
	isbn = {2041-1723},
	journal = {Nature Communications},
	number = {1},
	pages = {6245},
	title = {Baby skyrmions in Chern ferromagnets and topological mechanism for spin-polaron formation in twisted bilayer graphene},
	url = {https://doi.org/10.1038/s41467-022-33673-3},
	volume = {13},
	year = {2022}}

@article{
dqcp1,
author = {T. Senthil  and Ashvin Vishwanath  and Leon Balents  and Subir Sachdev  and Matthew P. A. Fisher },
title = {Deconfined Quantum Critical Points},
journal = {Science},
volume = {303},
number = {5663},
pages = {1490-1494},
year = {2004},
doi = {10.1126/science.1091806},
URL = {https://www.science.org/doi/abs/10.1126/science.1091806},
eprint = {https://www.science.org/doi/pdf/10.1126/science.1091806}}

@article{dqcp2,
  title = {Deconfined Quantum Critical Points: Symmetries and Dualities},
  author = {Wang, Chong and Nahum, Adam and Metlitski, Max A. and Xu, Cenke and Senthil, T.},
  journal = {Phys. Rev. X},
  volume = {7},
  issue = {3},
  pages = {031051},
  numpages = {45},
  year = {2017},
  month = {Sep},
  publisher = {American Physical Society},
  doi = {10.1103/PhysRevX.7.031051},
  url = {https://link.aps.org/doi/10.1103/PhysRevX.7.031051}
}

@article{loop,
  title = {Deconfined Quantum Criticality, Scaling Violations, and Classical Loop Models},
  author = {Nahum, Adam and Chalker, J. T. and Serna, P. and Ortu\~no, M. and Somoza, A. M.},
  journal = {Phys. Rev. X},
  volume = {5},
  issue = {4},
  pages = {041048},
  numpages = {21},
  year = {2015},
  month = {Dec},
  publisher = {American Physical Society},
  doi = {10.1103/PhysRevX.5.041048},
  url = {https://link.aps.org/doi/10.1103/PhysRevX.5.041048}
}

@article{bootstrap1,
  title = {Necessary Condition for Emergent Symmetry from the Conformal Bootstrap},
  author = {Nakayama, Yu and Ohtsuki, Tomoki},
  journal = {Phys. Rev. Lett.},
  volume = {117},
  issue = {13},
  pages = {131601},
  numpages = {6},
  year = {2016},
  month = {Sep},
  publisher = {American Physical Society},
  doi = {10.1103/PhysRevLett.117.131601},
  url = {https://link.aps.org/doi/10.1103/PhysRevLett.117.131601}
}

@article{bootstrap2,
  title = {The conformal bootstrap: Theory, numerical techniques, and applications},
  author = {Poland, David and Rychkov, Slava and Vichi, Alessandro},
  journal = {Rev. Mod. Phys.},
  volume = {91},
  issue = {1},
  pages = {015002},
  numpages = {74},
  year = {2019},
  month = {Jan},
  publisher = {American Physical Society},
  doi = {10.1103/RevModPhys.91.015002},
  url = {https://link.aps.org/doi/10.1103/RevModPhys.91.015002}
}

@misc{tri-crit2,
      title={SO(5) multicriticality in two-dimensional quantum magnets}, 
      author={Jun Takahashi and Hui Shao and Bowen Zhao and Wenan Guo and Anders W. Sandvik},
      year={2024},
      eprint={2405.06607},
      archivePrefix={arXiv},
      primaryClass={cond-mat.str-el},
      url={https://arxiv.org/abs/2405.06607}, 
}

@misc{prembabu2025spinpolaronsflatband,
      title={Spin Polarons in Flat Band Ferromagnets}, 
      author={Saranesh Prembabu and Rahul Sahay and Stefan Divic and Ashvin Vishwanath},
      year={2025},
      eprint={2510.26798},
      archivePrefix={arXiv},
      primaryClass={cond-mat.str-el},
      url={https://arxiv.org/abs/2510.26798}, 
}

@incollection{AssaadEvertz2008,
  author    = {Assaad, F. F. and Evertz, H. G.},
  title     = {{World-line and Determinantal Quantum Monte Carlo Methods for Spins, Phonons and Electrons}},
  booktitle = {Computational Many-Particle Physics},
  series    = {Lecture Notes in Physics},
  volume    = {739},
  publisher = {Springer},
  address   = {Berlin, Heidelberg},
  pages     = {277--356},
  year      = {2008},
  doi       = {10.1007/978-3-540-74686-7_10}
}

@article{noether1,
  title = {Role of Noether's Theorem at the Deconfined Quantum Critical Point},
  author = {Ma, Nvsen and You, Yi-Zhuang and Meng, Zi Yang},
  journal = {Phys. Rev. Lett.},
  volume = {122},
  issue = {17},
  pages = {175701},
  numpages = {6},
  year = {2019},
  month = {May},
  publisher = {American Physical Society},
  doi = {10.1103/PhysRevLett.122.175701},
  url = {https://link.aps.org/doi/10.1103/PhysRevLett.122.175701}
}

@article{noether2,
  title = {Simulation of fermionic and bosonic critical points with emergent SO(5) symmetry},
  author = {Sato, Toshihiro and Wang, Zhenjiu and Liu, Yuhai and Hou, Disha and Hohenadler, Martin and Guo, Wenan and Assaad, Fakher F.},
  journal = {Phys. Rev. B},
  volume = {108},
  issue = {12},
  pages = {L121111},
  numpages = {5},
  year = {2023},
  month = {Sep},
  publisher = {American Physical Society},
  doi = {10.1103/PhysRevB.108.L121111},
  url = {https://link.aps.org/doi/10.1103/PhysRevB.108.L121111}
}

@article{uniform-pairing,
  title = {Effective theory and emergent $\text{SU}(2)$ symmetry in the flat bands of attractive Hubbard models},
  author = {Tovmasyan, Murad and Peotta, Sebastiano and T\"orm\"a, P\"aivi and Huber, Sebastian D.},
  journal = {Phys. Rev. B},
  volume = {94},
  issue = {24},
  pages = {245149},
  numpages = {17},
  year = {2016},
  month = {Dec},
  publisher = {American Physical Society},
  doi = {10.1103/PhysRevB.94.245149},
  url = {https://link.aps.org/doi/10.1103/PhysRevB.94.245149}
}

@article{lee_sachdev1,
  title = {Deconfined criticality in bilayer graphene},
  author = {Lee, Junhyun and Sachdev, Subir},
  journal = {Phys. Rev. B},
  volume = {90},
  issue = {19},
  pages = {195427},
  numpages = {12},
  year = {2014},
  month = {Nov},
  publisher = {American Physical Society},
  doi = {10.1103/PhysRevB.90.195427},
  url = {https://link.aps.org/doi/10.1103/PhysRevB.90.195427}
}

@article{lee_sachdev2,
  title = {Wess-Zumino-Witten Terms in Graphene Landau Levels},
  author = {Lee, Junhyun and Sachdev, Subir},
  journal = {Phys. Rev. Lett.},
  volume = {114},
  issue = {22},
  pages = {226801},
  numpages = {5},
  year = {2015},
  month = {Jun},
  publisher = {American Physical Society},
  doi = {10.1103/PhysRevLett.114.226801},
  url = {https://link.aps.org/doi/10.1103/PhysRevLett.114.226801}
}

@article{so5_param1,
  title = {SO(5) symmetric ladder},
  author = {Scalapino, D. and Zhang, Shou-Cheng and Hanke, W.},
  journal = {Phys. Rev. B},
  volume = {58},
  issue = {1},
  pages = {443--452},
  numpages = {0},
  year = {1998},
  month = {Jul},
  publisher = {American Physical Society},
  doi = {10.1103/PhysRevB.58.443},
  url = {https://link.aps.org/doi/10.1103/PhysRevB.58.443}
}

@article{so5_demler,
  title = {$\mathit{SO}(5)$ theory of antiferromagnetism and superconductivity},
  author = {Demler, Eugene and Hanke, Werner and Zhang, Shou-Cheng},
  journal = {Rev. Mod. Phys.},
  volume = {76},
  issue = {3},
  pages = {909--974},
  numpages = {0},
  year = {2004},
  month = {Nov},
  publisher = {American Physical Society},
  doi = {10.1103/RevModPhys.76.909},
  url = {https://link.aps.org/doi/10.1103/RevModPhys.76.909}
}

@article{single_chern,
  title = {Superconductivity, pseudogap, and phase separation in topological flat bands},
  author = {Hofmann, Johannes S. and Berg, Erez and Chowdhury, Debanjan},
  journal = {Phys. Rev. B},
  volume = {102},
  issue = {20},
  pages = {201112},
  numpages = {6},
  year = {2020},
  month = {Nov},
  publisher = {American Physical Society},
  doi = {10.1103/PhysRevB.102.201112},
  url = {https://link.aps.org/doi/10.1103/PhysRevB.102.201112}
}

@misc{si, title={See supplementary material for additional details on the numerical procedures and data.}}

@inproceedings{access,
author = {Boerner, Timothy J. and Deems, Stephen and Furlani, Thomas R. and Knuth, Shelley L. and Towns, John},
title = {ACCESS: Advancing Innovation: NSF’s Advanced Cyberinfrastructure Coordination Ecosystem: Services \& Support},
year = {2023},
isbn = {9781450399852},
publisher = {Association for Computing Machinery},
address = {New York, NY, USA},
url = {https://doi.org/10.1145/3569951.3597559},
doi = {10.1145/3569951.3597559},
booktitle = {Practice and Experience in Advanced Research Computing},
pages = {173–176},
numpages = {4},
location = {Portland, OR, USA},
series = {PEARC '23}
}

@article{LNW,
  title = {Doping a Mott insulator: Physics of high-temperature superconductivity},
  author = {Lee, Patrick A. and Nagaosa, Naoto and Wen, Xiao-Gang},
  journal = {Rev. Mod. Phys.},
  volume = {78},
  issue = {1},
  pages = {17--85},
  numpages = {0},
  year = {2006},
  month = {Jan},
  publisher = {American Physical Society},
  doi = {10.1103/RevModPhys.78.17},
  url = {https://link.aps.org/doi/10.1103/RevModPhys.78.17}
}

@Article{Mak2022,
author={Mak, Kin Fai
and Shan, Jie},
title={Semiconductor moir{\'e} materials},
journal={Nature Nanotechnology},
year={2022},
month={Jul},
day={01},
volume={17},
number={7},
pages={686-695},
issn={1748-3395},
doi={10.1038/s41565-022-01165-6},
url={https://doi.org/10.1038/s41565-022-01165-6}
}

@article{barrett,
  title = {Optically Pumped NMR Evidence for Finite-Size Skyrmions in GaAs Quantum Wells near Landau Level Filling $\mathit{\ensuremath{\nu}}\phantom{\rule{0ex}{0ex}}=\phantom{\rule{0ex}{0ex}}1$},
  author = {Barrett, S. E. and Dabbagh, G. and Pfeiffer, L. N. and West, K. W. and Tycko, R.},
  journal = {Phys. Rev. Lett.},
  volume = {74},
  issue = {25},
  pages = {5112--5115},
  numpages = {0},
  year = {1995},
  month = {Jun},
  publisher = {American Physical Society},
  doi = {10.1103/PhysRevLett.74.5112},
  url = {https://link.aps.org/doi/10.1103/PhysRevLett.74.5112}
}

@article{sondhi,
  title = {Skyrmions and the crossover from the integer to fractional quantum Hall effect at small Zeeman energies},
  author = {Sondhi, S. L. and Karlhede, A. and Kivelson, S. A. and Rezayi, E. H.},
  journal = {Phys. Rev. B},
  volume = {47},
  issue = {24},
  pages = {16419--16426},
  numpages = {0},
  year = {1993},
  month = {Jun},
  publisher = {American Physical Society},
  doi = {10.1103/PhysRevB.47.16419},
  url = {https://link.aps.org/doi/10.1103/PhysRevB.47.16419}
}

@Article{Andrei2021,
author={Andrei, Eva Y.
and Efetov, Dmitri K.
and Jarillo-Herrero, Pablo
and MacDonald, Allan H.
and Mak, Kin Fai
and Senthil, T.
and Tutuc, Emanuel
and Yazdani, Ali
and Young, Andrea F.},
title={The marvels of moir{\'e} materials},
journal={Nature Reviews Materials},
year={2021},
month={Mar},
day={01},
volume={6},
number={3},
pages={201-206},
issn={2058-8437},
doi={10.1038/s41578-021-00284-1},
url={https://doi.org/10.1038/s41578-021-00284-1}
}

@article{so5_1,
  title = {$\mathit{SO}(5)$ theory of antiferromagnetism and superconductivity},
  author = {Demler, Eugene and Hanke, Werner and Zhang, Shou-Cheng},
  journal = {Rev. Mod. Phys.},
  volume = {76},
  issue = {3},
  pages = {909--974},
  numpages = {0},
  year = {2004},
  month = {Nov},
  publisher = {American Physical Society},
  doi = {10.1103/RevModPhys.76.909},
  url = {https://link.aps.org/doi/10.1103/RevModPhys.76.909}
}

@article{so5_2,
author = {Shou-Cheng Zhang },
title = {A Unified Theory Based on SO(5) Symmetry of Superconductivity and Antiferromagnetism},
journal = {Science},
volume = {275},
number = {5303},
pages = {1089-1096},
year = {1997},
doi = {10.1126/science.275.5303.1089}}

@article{so5_3,
  title = {Topological Spin Hall States, Charged Skyrmions, and Superconductivity in Two Dimensions},
  author = {Grover, Tarun and Senthil, T.},
  journal = {Phys. Rev. Lett.},
  volume = {100},
  issue = {15},
  pages = {156804},
  numpages = {4},
  year = {2008},
  month = {Apr},
  publisher = {American Physical Society},
  doi = {10.1103/PhysRevLett.100.156804},
  url = {https://link.aps.org/doi/10.1103/PhysRevLett.100.156804}
}

@article{Yves22,
  title = {Skyrmions in Twisted Bilayer Graphene: Stability, Pairing, and Crystallization},
  author = {Kwan, Yves H. and Wagner, Glenn and Bultinck, Nick and Simon, Steven H. and Parameswaran, S. A.},
  journal = {Phys. Rev. X},
  volume = {12},
  issue = {3},
  pages = {031020},
  numpages = {22},
  year = {2022},
  month = {Jul},
  publisher = {American Physical Society},
  doi = {10.1103/PhysRevX.12.031020},
  url = {https://link.aps.org/doi/10.1103/PhysRevX.12.031020}
}

@article{PhysRevB.102.035136,
  title = {Correlated insulating phases of twisted bilayer graphene at commensurate filling fractions: A Hartree-Fock study},
  author = {Zhang, Yi and Jiang, Kun and Wang, Ziqiang and Zhang, Fuchun},
  journal = {Phys. Rev. B},
  volume = {102},
  issue = {3},
  pages = {035136},
  numpages = {9},
  year = {2020},
  month = {Jul},
  publisher = {American Physical Society},
  doi = {10.1103/PhysRevB.102.035136},
  url = {https://link.aps.org/doi/10.1103/PhysRevB.102.035136}
}

@article{MacDonald21,
  title = {Weak-Field Hall Resistivity and Spin-Valley Flavor Symmetry Breaking in Magic-Angle Twisted Bilayer Graphene},
  author = {Xie, Ming and MacDonald, A. H.},
  journal = {Phys. Rev. Lett.},
  volume = {127},
  issue = {19},
  pages = {196401},
  numpages = {6},
  year = {2021},
  month = {Nov},
  publisher = {American Physical Society},
  doi = {10.1103/PhysRevLett.127.196401},
  url = {https://link.aps.org/doi/10.1103/PhysRevLett.127.196401}
}

@article{MacDonald20,
  title = {Nature of the Correlated Insulator States in Twisted Bilayer Graphene},
  author = {Xie, Ming and MacDonald, A. H.},
  journal = {Phys. Rev. Lett.},
  volume = {124},
  issue = {9},
  pages = {097601},
  numpages = {6},
  year = {2020},
  month = {Mar},
  publisher = {American Physical Society},
  doi = {10.1103/PhysRevLett.124.097601},
  url = {https://link.aps.org/doi/10.1103/PhysRevLett.124.097601}
}

@article{dai_prb21,
  title = {Theories for the correlated insulating states and quantum anomalous Hall effect phenomena in twisted bilayer graphene},
  author = {Liu, Jianpeng and Dai, Xi},
  journal = {Phys. Rev. B},
  volume = {103},
  issue = {3},
  pages = {035427},
  numpages = {14},
  year = {2021},
  month = {Jan},
  publisher = {American Physical Society},
  doi = {10.1103/PhysRevB.103.035427},
  url = {https://link.aps.org/doi/10.1103/PhysRevB.103.035427}
}

@article{PhysRevB.102.045107,
  title = {Band structure and insulating states driven by Coulomb interaction in twisted bilayer graphene},
  author = {Cea, Tommaso and Guinea, Francisco},
  journal = {Phys. Rev. B},
  volume = {102},
  issue = {4},
  pages = {045107},
  numpages = {6},
  year = {2020},
  month = {Jul},
  publisher = {American Physical Society},
  doi = {10.1103/PhysRevB.102.045107},
  url = {https://link.aps.org/doi/10.1103/PhysRevB.102.045107}
}

@article{PhysRevB.98.081102,
  title = {Possible correlated insulating states in magic-angle twisted bilayer graphene under strongly competing interactions},
  author = {Ochi, Masayuki and Koshino, Mikito and Kuroki, Kazuhiko},
  journal = {Phys. Rev. B},
  volume = {98},
  issue = {8},
  pages = {081102},
  numpages = {5},
  year = {2018},
  month = {Aug},
  publisher = {American Physical Society},
  doi = {10.1103/PhysRevB.98.081102},
  url = {https://link.aps.org/doi/10.1103/PhysRevB.98.081102}
}

@article{PhysRevB.103.195127,
  title = {Importance of long-ranged electron-electron interactions for the magnetic phase diagram of twisted bilayer graphene},
  author = {Klebl, Lennart and Goodwin, Zachary A. H. and Mostofi, Arash A. and Kennes, Dante M. and Lischner, Johannes},
  journal = {Phys. Rev. B},
  volume = {103},
  issue = {19},
  pages = {195127},
  numpages = {7},
  year = {2021},
  month = {May},
  publisher = {American Physical Society},
  doi = {10.1103/PhysRevB.103.195127},
  url = {https://link.aps.org/doi/10.1103/PhysRevB.103.195127}
}

@article{wagner22,
  title = {Global Phase Diagram of the Normal State of Twisted Bilayer Graphene},
  author = {Wagner, Glenn and Kwan, Yves H. and Bultinck, Nick and Simon, Steven H. and Parameswaran, S. A.},
  journal = {Phys. Rev. Lett.},
  volume = {128},
  issue = {15},
  pages = {156401},
  numpages = {7},
  year = {2022},
  month = {Apr},
  publisher = {American Physical Society},
  doi = {10.1103/PhysRevLett.128.156401},
  url = {https://link.aps.org/doi/10.1103/PhysRevLett.128.156401}
}

@article{tmdsc_fai,
	author = {Xia, Yiyu and Han, Zhongdong and Watanabe, Kenji and Taniguchi, Takashi and Shan, Jie and Mak, Kin Fai},
	journal = {Nature},
	number = {8047},
	pages = {833--838},
	title = {Superconductivity in twisted bilayer WSe2},
	volume = {637},
	year = {2025}}

@article{tmdsc_dean,
	author = {Guo, Yinjie and Pack, Jordan and Swann, Joshua and Holtzman, Luke and Cothrine, Matthew and Watanabe, Kenji and Taniguchi, Takashi and Mandrus, David G. and Barmak, Katayun and Hone, James and Millis, Andrew J. and Pasupathy, Abhay and Dean, Cory R.},
	journal = {Nature},
	number = {8047},
	pages = {839--845},
	title = {Superconductivity in 5.0$\,^{\circ}$twisted bilayer WSe2},
	volume = {637},
	year = {2025}}

@article{
Hao21,
author = {Zeyu Hao  and A. M. Zimmerman  and Patrick Ledwith  and Eslam Khalaf  and Danial Haie Najafabadi  and Kenji Watanabe  and Takashi Taniguchi  and Ashvin Vishwanath  and Philip Kim },
title = {Electric field–tunable superconductivity in alternating-twist magic-angle trilayer graphene},
journal = {Science},
volume = {371},
number = {6534},
pages = {1133-1138},
year = {2021},
doi = {10.1126/science.abg0399},
URL = {https://www.science.org/doi/abs/10.1126/science.abg0399},
eprint = {https://www.science.org/doi/pdf/10.1126/science.abg0399}}

@article{jfmv_prl,
  title = {Low-Energy Optical Sum Rule in Moir\'e Graphene},
  author = {Mendez-Valderrama, J. F. and Mao, Dan and Chowdhury, Debanjan},
  journal = {Phys. Rev. Lett.},
  volume = {133},
  issue = {19},
  pages = {196501},
  numpages = {10},
  year = {2024},
  month = {Nov},
  publisher = {American Physical Society},
  doi = {10.1103/PhysRevLett.133.196501},
  url = {https://link.aps.org/doi/10.1103/PhysRevLett.133.196501}
}

@article{park21,
	author = {Park, Jeong Min and Cao, Yuan and Watanabe, Kenji and Taniguchi, Takashi and Jarillo-Herrero, Pablo},
	journal = {Nature},
	number = {7845},
	pages = {249--255},
	title = {Tunable strongly coupled superconductivity in magic-angle twisted trilayer graphene},
	volume = {590},
	year = {2021}}

@article{dc21,
  title = {Pairing in magic-angle twisted bilayer graphene: Role of phonon and plasmon umklapp},
  author = {Lewandowski, Cyprian and Chowdhury, Debanjan and Ruhman, Jonathan},
  journal = {Phys. Rev. B},
  volume = {103},
  issue = {23},
  pages = {235401},
  numpages = {18},
  year = {2021},
  month = {Jun},
  publisher = {American Physical Society},
  doi = {10.1103/PhysRevB.103.235401},
  url = {https://link.aps.org/doi/10.1103/PhysRevB.103.235401}
}

@article{PhysRevLett.127.247703,
  title = {Theory of Correlated Insulators and Superconductivity in Twisted Bilayer Graphene},
  author = {Shavit, Gal and Berg, Erez and Stern, Ady and Oreg, Yuval},
  journal = {Phys. Rev. Lett.},
  volume = {127},
  issue = {24},
  pages = {247703},
  numpages = {6},
  year = {2021},
  month = {Dec},
  publisher = {American Physical Society},
  doi = {10.1103/PhysRevLett.127.247703},
  url = {https://link.aps.org/doi/10.1103/PhysRevLett.127.247703}
}

@article{Yves_21prx,
  title = {Kekul\'e Spiral Order at All Nonzero Integer Fillings in Twisted Bilayer Graphene},
  author = {Kwan, Y. H. and Wagner, G. and Soejima, T. and Zaletel, M. P. and Simon, S. H. and Parameswaran, S. A. and Bultinck, N.},
  journal = {Phys. Rev. X},
  volume = {11},
  issue = {4},
  pages = {041063},
  numpages = {23},
  year = {2021},
  month = {Dec},
  publisher = {American Physical Society},
  doi = {10.1103/PhysRevX.11.041063},
  url = {https://link.aps.org/doi/10.1103/PhysRevX.11.041063}
}

@article{
ashvin_skrymion,
author = {Eslam Khalaf  and Shubhayu Chatterjee  and Nick Bultinck  and Michael P. Zaletel  and Ashvin Vishwanath },
title = {Charged skyrmions and topological origin of superconductivity in magic-angle graphene},
journal = {Science Advances},
volume = {7},
number = {19},
pages = {eabf5299},
year = {2021},
doi = {10.1126/sciadv.abf5299}}

@article{chatterjee22,
  title = {Skyrmion superconductivity: DMRG evidence for a topological route to superconductivity},
  author = {Chatterjee, Shubhayu and Ippoliti, Matteo and Zaletel, Michael P.},
  journal = {Phys. Rev. B},
  volume = {106},
  issue = {3},
  pages = {035421},
  numpages = {19},
  year = {2022},
  month = {Jul},
  publisher = {American Physical Society},
  doi = {10.1103/PhysRevB.106.035421},
  url = {https://link.aps.org/doi/10.1103/PhysRevB.106.035421}
}

@article{Bultinck20_prl,
  title = {Mechanism for Anomalous Hall Ferromagnetism in Twisted Bilayer Graphene},
  author = {Bultinck, Nick and Chatterjee, Shubhayu and Zaletel, Michael P.},
  journal = {Phys. Rev. Lett.},
  volume = {124},
  issue = {16},
  pages = {166601},
  numpages = {7},
  year = {2020},
  month = {Apr},
  publisher = {American Physical Society},
  doi = {10.1103/PhysRevLett.124.166601},
  url = {https://link.aps.org/doi/10.1103/PhysRevLett.124.166601}
}

@article{
SS20,
author = {Maine Christos  and Subir Sachdev  and Mathias S. Scheurer },
title = {Superconductivity, correlated insulators, and Wess–Zumino–Witten terms in twisted bilayer graphene},
journal = {Proceedings of the National Academy of Sciences},
doi = {10.1073/pnas.2014691117},
volume = {117},
number = {47},
pages = {29543-29554},
year = {2020}}

@article{congjun,
  title = {Sufficient condition for absence of the sign problem in the fermionic quantum Monte Carlo algorithm},
  author = {Wu, Congjun and Zhang, Shou-Cheng},
  journal = {Phys. Rev. B},
  volume = {71},
  issue = {15},
  pages = {155115},
  numpages = {14},
  year = {2005},
  month = {Apr},
  publisher = {American Physical Society},
  doi = {10.1103/PhysRevB.71.155115},
  url = {https://link.aps.org/doi/10.1103/PhysRevB.71.155115}
}

@book{qptbook, place={Cambridge}, edition={2}, title={Quantum Phase Transitions}, DOI={10.1017/CBO9780511973765}, publisher={Cambridge University Press}, author={Sachdev, Subir}, year={2011}}

@misc{auerbach1998quantum,
      title={Quantum Magnetism Approaches to Strongly Correlated Electrons}, 
      author={A. Auerbach},
      year={1998},
      eprint={cond-mat/9801294},
      archivePrefix={arXiv},
      primaryClass={cond-mat.str-el}
}

@article{fierz,
  title = {Half-filled Landau levels: A continuum and sign-free regularization for three-dimensional quantum critical points},
  author = {Ippoliti, Matteo and Mong, Roger S. K. and Assaad, Fakher F. and Zaletel, Michael P.},
  journal = {Phys. Rev. B},
  volume = {98},
  issue = {23},
  pages = {235108},
  numpages = {9},
  year = {2018},
  month = {Dec},
  publisher = {American Physical Society},
  doi = {10.1103/PhysRevB.98.235108},
  url = {https://link.aps.org/doi/10.1103/PhysRevB.98.235108}
}

@article{Hofmann_tbg,
  title = {Fermionic Monte Carlo Study of a Realistic Model of Twisted Bilayer Graphene},
  author = {Hofmann, Johannes S. and Khalaf, Eslam and Vishwanath, Ashvin and Berg, Erez and Lee, Jong Yeon},
  journal = {Phys. Rev. X},
  volume = {12},
  issue = {1},
  pages = {011061},
  numpages = {32},
  year = {2022},
  month = {Mar},
  publisher = {American Physical Society},
  doi = {10.1103/PhysRevX.12.011061},
  url = {https://link.aps.org/doi/10.1103/PhysRevX.12.011061}
}

@article{AV18,
  title = {Origin of Mott Insulating Behavior and Superconductivity in Twisted Bilayer Graphene},
  author = {Po, Hoi Chun and Zou, Liujun and Vishwanath, Ashvin and Senthil, T.},
  journal = {Phys. Rev. X},
  volume = {8},
  issue = {3},
  pages = {031089},
  numpages = {31},
  year = {2018},
  month = {Sep},
  publisher = {American Physical Society},
  doi = {10.1103/PhysRevX.8.031089},
  url = {https://link.aps.org/doi/10.1103/PhysRevX.8.031089}
}

@article{AV19,
  title = {Origin of Magic Angles in Twisted Bilayer Graphene},
  author = {Tarnopolsky, Grigory and Kruchkov, Alex Jura and Vishwanath, Ashvin},
  journal = {Phys. Rev. Lett.},
  volume = {122},
  issue = {10},
  pages = {106405},
  numpages = {6},
  year = {2019},
  month = {Mar},
  publisher = {American Physical Society},
  doi = {10.1103/PhysRevLett.122.106405},
  url = {https://link.aps.org/doi/10.1103/PhysRevLett.122.106405}
}

@article{AV20_prx,
  title = {Ground State and Hidden Symmetry of Magic-Angle Graphene at Even Integer Filling},
  author = {Bultinck, Nick and Khalaf, Eslam and Liu, Shang and Chatterjee, Shubhayu and Vishwanath, Ashvin and Zaletel, Michael P.},
  journal = {Phys. Rev. X},
  volume = {10},
  issue = {3},
  pages = {031034},
  numpages = {13},
  year = {2020},
  month = {Aug},
  publisher = {American Physical Society},
  doi = {10.1103/PhysRevX.10.031034},
  url = {https://link.aps.org/doi/10.1103/PhysRevX.10.031034}
}

@article{peri2021fragile,
  title = {Fragile Topology and Flat-Band Superconductivity in the Strong-Coupling Regime},
  author = {Peri, Valerio and Song, Zhi-Da and Bernevig, B. Andrei and Huber, Sebastian D.},
  journal = {Phys. Rev. Lett.},
  volume = {126},
  issue = {2},
  pages = {027002},
  numpages = {6},
  year = {2021},
  month = {Jan},
  publisher = {American Physical Society},
  doi = {10.1103/PhysRevLett.126.027002},
  url = {https://link.aps.org/doi/10.1103/PhysRevLett.126.027002}
}

@article{Bernevig21,
author = {Herzog-Arbeitman, Jonah and Peri, Valerio and Schindler, Frank and Huber, Sebastian D and Bernevig, B Andrei},
doi = {10.1103/PhysRevLett.128.087002},
issn = {0031-9007},
journal = {Physical Review Letters},
month = {feb},
number = {8},
pages = {087002},
pmid = {35275691},
title = {{Superfluid Weight Bounds from Symmetry and Quantum Geometry in Flat Bands}},
url = { https://link.aps.org/doi/10.1103/PhysRevLett.128.087002},
volume = {128},
year = {2022}
}

@article{sandvik07,
  title = {Evidence for Deconfined Quantum Criticality in a Two-Dimensional Heisenberg Model with Four-Spin Interactions},
  author = {Sandvik, Anders W.},
  journal = {Phys. Rev. Lett.},
  volume = {98},
  issue = {22},
  pages = {227202},
  numpages = {4},
  year = {2007},
  month = {Jun},
  publisher = {American Physical Society},
  doi = {10.1103/PhysRevLett.98.227202},
  url = {https://link.aps.org/doi/10.1103/PhysRevLett.98.227202}
}

@article{sandvik09,
  title = {Antiferromagnetic to valence-bond-solid transitions in two-dimensional $\text{SU}(N)$ Heisenberg models with multispin interactions},
  author = {Lou, Jie and Sandvik, Anders W. and Kawashima, Naoki},
  journal = {Phys. Rev. B},
  volume = {80},
  issue = {18},
  pages = {180414},
  numpages = {4},
  year = {2009},
  month = {Nov},
  publisher = {American Physical Society},
  doi = {10.1103/PhysRevB.80.180414},
  url = {https://link.aps.org/doi/10.1103/PhysRevB.80.180414}
}

@article{Sandvik_2020,
   title={Consistent Scaling Exponents at the Deconfined Quantum-Critical Point*},
   volume={37},
   ISSN={1741-3540},
   url={http://dx.doi.org/10.1088/0256-307X/37/5/057502},
   DOI={10.1088/0256-307x/37/5/057502},
   number={5},
   journal={Chinese Physics Letters},
   publisher={IOP Publishing},
   author={Sandvik, Anders W. and Zhao, Bowen},
   year={2020},
   month=may, pages={057502} }

@article{sandvik10,
  title = {Continuous Quantum Phase Transition between an Antiferromagnet and a Valence-Bond Solid in Two Dimensions: Evidence for Logarithmic Corrections to Scaling},
  author = {Sandvik, Anders W.},
  journal = {Phys. Rev. Lett.},
  volume = {104},
  issue = {17},
  pages = {177201},
  numpages = {4},
  year = {2010},
  month = {Apr},
  publisher = {American Physical Society},
  doi = {10.1103/PhysRevLett.104.177201},
  url = {https://link.aps.org/doi/10.1103/PhysRevLett.104.177201}
}

@article{hkin,
  title = {Fractional Quantum Hall States at Zero Magnetic Field},
  author = {Neupert, Titus and Santos, Luiz and Chamon, Claudio and Mudry, Christopher},
  journal = {Phys. Rev. Lett.},
  volume = {106},
  issue = {23},
  pages = {236804},
  numpages = {4},
  year = {2011},
  month = {Jun},
  publisher = {American Physical Society},
  doi = {10.1103/PhysRevLett.106.236804},
  url = {https://link.aps.org/doi/10.1103/PhysRevLett.106.236804}
}

@article{YahuiTS,
  title = {Bridging Hubbard model physics and quantum Hall physics in trilayer $\text{graphene}/h\ensuremath{-}\mathrm{BN}$ moir\'e superlattice},
  author = {Zhang, Ya-Hui and Senthil, T.},
  journal = {Phys. Rev. B},
  volume = {99},
  issue = {20},
  pages = {205150},
  numpages = {17},
  year = {2019},
  month = {May},
  publisher = {American Physical Society},
  doi = {10.1103/PhysRevB.99.205150},
  url = {https://link.aps.org/doi/10.1103/PhysRevB.99.205150}
}

@article{BABHF,
  title = {Magic-Angle Twisted Bilayer Graphene as a Topological Heavy Fermion Problem},
  author = {Song, Zhi-Da and Bernevig, B. Andrei},
  journal = {Phys. Rev. Lett.},
  volume = {129},
  issue = {4},
  pages = {047601},
  numpages = {10},
  year = {2022},
  month = {Jul},
  publisher = {American Physical Society},
  doi = {10.1103/PhysRevLett.129.047601},
  url = {https://link.aps.org/doi/10.1103/PhysRevLett.129.047601}
}

@article{LedwithPRX,
  title = {Nonlocal Moments and Mott Semimetal in the Chern Bands of Twisted Bilayer Graphene},
  author = {Ledwith, Patrick J. and Dong, Junkai and Vishwanath, Ashvin and Khalaf, Eslam},
  journal = {Phys. Rev. X},
  volume = {15},
  issue = {2},
  pages = {021087},
  numpages = {40},
  year = {2025},
  month = {Jun},
  publisher = {American Physical Society},
  doi = {10.1103/PhysRevX.15.021087},
  url = {https://link.aps.org/doi/10.1103/PhysRevX.15.021087}
}

@article{girvin2000,
    author = {Girvin, Steven M.},
    title = {Spin and Isospin: Exotic Order in Quantum Hall Ferromagnets},
    journal = {Physics Today},
    volume = {53},
    number = {6},
    pages = {39-45},
    year = {2000},
    month = {06},
    issn = {0031-9228},
    doi = {10.1063/1.1306366},
    url = {https://doi.org/10.1063/1.1306366}
}

@misc{girvin1999quantum,
      title={The Quantum Hall Effect: Novel Excitations and Broken Symmetries}, 
      author={Steven M. Girvin},
      year={1999},
      eprint={cond-mat/9907002},
      archivePrefix={arXiv},
      primaryClass={cond-mat.mes-hall}
}

@article{zhou_fuzzy2,
  title = {3D Conformal Field Theories with $\mathrm{Sp}(N)$ Global Symmetry on a Fuzzy Sphere},
  author = {Zhou, Zheng and He, Yin-Chen},
  journal = {Phys. Rev. Lett.},
  volume = {135},
  issue = {2},
  pages = {026504},
  numpages = {7},
  year = {2025},
  month = {Jul},
  publisher = {American Physical Society},
  doi = {10.1103/xstj-xvcy},
  url = {https://link.aps.org/doi/10.1103/xstj-xvcy}
}

@Article{hofmann_fuzzy,
	title={{Quantum Monte Carlo simulation of the 3D Ising transition on the fuzzy sphere}},
	author={Johannes Stephan Hofmann and Florian Goth and Wei Zhu and Yin-Chen He and Emilie Huffman},
	journal={SciPost Phys. Core},
	volume={7},
	pages={028},
	year={2024},
	publisher={SciPost},
	doi={10.21468/SciPostPhysCore.7.2.028},
	url={https://scipost.org/10.21468/SciPostPhysCore.7.2.028},
}

@article{Troyer08,
  title = {Deconfined Criticality: Generic First-Order Transition in the SU(2) Symmetry Case},
  author = {Kuklov, A. B. and Matsumoto, M. and Prokof'ev, N. V. and Svistunov, B. V. and Troyer, M.},
  journal = {Phys. Rev. Lett.},
  volume = {101},
  issue = {5},
  pages = {050405},
  numpages = {4},
  year = {2008},
  month = {Aug},
  publisher = {American Physical Society},
  doi = {10.1103/PhysRevLett.101.050405},
  url = {https://link.aps.org/doi/10.1103/PhysRevLett.101.050405}
}

@article{
sandvik_science,
author = {Hui Shao  and Wenan Guo  and Anders W. Sandvik },
title = {Quantum criticality with two length scales},
journal = {Science},
volume = {352},
number = {6282},
pages = {213-216},
year = {2016},
doi = {10.1126/science.aad5007},
URL = {https://www.science.org/doi/abs/10.1126/science.aad5007},
eprint = {https://www.science.org/doi/pdf/10.1126/science.aad5007}}

@article{SF06,
  title = {Competing orders, nonlinear sigma models, and topological terms in quantum magnets},
  author = {Senthil, T. and Fisher, Matthew P. A.},
  journal = {Phys. Rev. B},
  volume = {74},
  issue = {6},
  pages = {064405},
  numpages = {11},
  year = {2006},
  month = {Aug},
  publisher = {American Physical Society},
  doi = {10.1103/PhysRevB.74.064405},
  url = {https://link.aps.org/doi/10.1103/PhysRevB.74.064405}
}

@article{TH05,
  title = {Many-Body Spin Berry Phases Emerging from the $\ensuremath{\pi}$-Flux State: Competition between Antiferromagnetism and the Valence-Bond-Solid State},
  author = {Tanaka, Akihiro and Hu, Xiao},
  journal = {Phys. Rev. Lett.},
  volume = {95},
  issue = {3},
  pages = {036402},
  numpages = {4},
  year = {2005},
  month = {Jul},
  publisher = {American Physical Society},
  doi = {10.1103/PhysRevLett.95.036402},
  url = {https://link.aps.org/doi/10.1103/PhysRevLett.95.036402}
}

@article{dqcp_prb_04,
  title = {Quantum criticality beyond the Landau-Ginzburg-Wilson paradigm},
  author = {Senthil, T. and Balents, Leon and Sachdev, Subir and Vishwanath, Ashvin and Fisher, Matthew P. A.},
  journal = {Phys. Rev. B},
  volume = {70},
  issue = {14},
  pages = {144407},
  numpages = {33},
  year = {2004},
  month = {Oct},
  publisher = {American Physical Society},
  doi = {10.1103/PhysRevB.70.144407},
  url = {https://link.aps.org/doi/10.1103/PhysRevB.70.144407}
}

@misc{zaletel_dmrg,
      title={Superconductivity in a Topological Lattice Model with Strong Repulsion}, 
      author={Rahul Sahay and Stefan Divic and Daniel E. Parker and Tomohiro Soejima and Sajant Anand and Johannes Hauschild and Monika Aidelsburger and Ashvin Vishwanath and Shubhayu Chatterjee and Norman Y. Yao and Michael P. Zaletel},
      year={2023},
      eprint={2308.10935},
      archivePrefix={arXiv},
      primaryClass={cond-mat.str-el}
}

@article{SSRMP,
  title = {Colloquium: Order and quantum phase transitions in the cuprate superconductors},
  author = {Sachdev, Subir},
  journal = {Rev. Mod. Phys.},
  volume = {75},
  issue = {3},
  pages = {913--932},
  numpages = {0},
  year = {2003},
  month = {Jul},
  publisher = {American Physical Society},
  doi = {10.1103/RevModPhys.75.913},
  url = {https://link.aps.org/doi/10.1103/RevModPhys.75.913}
}

@article{Serlin2020Science,
  title   = {Intrinsic quantized anomalous Hall effect in a moir{\'e} heterostructure},
  author  = {Serlin, Marec and Tschirhart, C. Lealand and Polshyn, Hryhoriy and Zhang, Yuxuan and Zhu, Jiacheng and Watanabe, Kenji and Taniguchi, Takashi and Balents, Leon and Young, Andrea F.},
  journal = {Science},
  year    = {2020},
  volume  = {367},
  number  = {6480},
  pages   = {900--903},
  doi     = {10.1126/science.aay5533}
}

@article{TBG4,
	author = {Lian, Biao and Song, Zhi-Da and Regnault, Nicolas and Efetov, Dmitri K. and Yazdani, Ali and Bernevig, B. Andrei},
	doi = {10.1103/PhysRevB.103.205414},
	issue = {20},
	journal = {Phys. Rev. B},
	month = {May},
	numpages = {41},
	pages = {205414},
	publisher = {American Physical Society},
	title = {Twisted bilayer graphene. IV. Exact insulator ground states and phase diagram},
	url = {https://link.aps.org/doi/10.1103/PhysRevB.103.205414},
	volume = {103},
	year = {2021}}

@article{KangVafekPRL,
	author = {Kang, Jian and Vafek, Oskar},
	doi = {10.1103/PhysRevLett.122.246401},
	issue = {24},
	journal = {Phys. Rev. Lett.},
	month = {Jun},
	numpages = {6},
	pages = {246401},
	publisher = {American Physical Society},
	title = {Strong Coupling Phases of Partially Filled Twisted Bilayer Graphene Narrow Bands},
	url = {https://link.aps.org/doi/10.1103/PhysRevLett.122.246401},
	volume = {122},
	year = {2019}}

@article{Fertig94,
  title = {Charged spin-texture excitations and the Hartree-Fock approximation in the quantum Hall effect},
  author = {Fertig, H. A. and Brey, L. and C\^ot\'e, R. and MacDonald, A. H.},
  journal = {Phys. Rev. B},
  volume = {50},
  issue = {15},
  pages = {11018--11021},
  numpages = {0},
  year = {1994},
  month = {Oct},
  publisher = {American Physical Society},
  doi = {10.1103/PhysRevB.50.11018},
  url = {https://link.aps.org/doi/10.1103/PhysRevB.50.11018}
}

@article{Fertig97,
  title = {Hartree-Fock theory of Skyrmions in quantum Hall ferromagnets},
  author = {Fertig, H. A. and Brey, Luis and C\^ot\'e, R. and MacDonald, A. H. and Karlhede, A. and Sondhi, S. L.},
  journal = {Phys. Rev. B},
  volume = {55},
  issue = {16},
  pages = {10671--10680},
  numpages = {0},
  year = {1997},
  month = {Apr},
  publisher = {American Physical Society},
  doi = {10.1103/PhysRevB.55.10671},
  url = {https://link.aps.org/doi/10.1103/PhysRevB.55.10671}
}

@article{yc_tbg1,
	author = {Cao, Yuan and Fatemi, Valla and Fang, Shiang and Watanabe, Kenji and Taniguchi, Takashi and Kaxiras, Efthimios and Jarillo-Herrero, Pablo},
	date = {2018/04/01},
	doi = {10.1038/nature26160},
	id = {Cao2018},
	isbn = {1476-4687},
	journal = {Nature},
	number = {7699},
	pages = {43--50},
	title = {Unconventional superconductivity in magic-angle graphene superlattices},
	url = {https://doi.org/10.1038/nature26160},
	volume = {556},
	year = {2018}}

@article{yc_tbg2,
	author = {Cao, Yuan and Fatemi, Valla and Demir, Ahmet and Fang, Shiang and Tomarken, Spencer L. and Luo, Jason Y. and Sanchez-Yamagishi, Javier D. and Watanabe, Kenji and Taniguchi, Takashi and Kaxiras, Efthimios and Ashoori, Ray C. and Jarillo-Herrero, Pablo},
	date = {2018/04/01},
	doi = {10.1038/nature26154},
	id = {Cao2018},
	isbn = {1476-4687},
	journal = {Nature},
	number = {7699},
	pages = {80--84},
	title = {Correlated insulator behaviour at half-filling in magic-angle graphene superlattices},
	url = {https://doi.org/10.1038/nature26154},
	volume = {556},
	year = {2018}}

@article{yc_tbg3,
	author = {Cao, Yuan and Rodan-Legrain, Daniel and Rubies-Bigorda, Oriol and Park, Jeong Min and Watanabe, Kenji and Taniguchi, Takashi and Jarillo-Herrero, Pablo},
	date = {2020/07/01},
	doi = {10.1038/s41586-020-2260-6},
	id = {Cao2020},
	isbn = {1476-4687},
	journal = {Nature},
	number = {7815},
	pages = {215--220},
	title = {Tunable correlated states and spin-polarized phases in twisted bilayer--bilayer graphene},
	url = {https://doi.org/10.1038/s41586-020-2260-6},
	volume = {583},
	year = {2020}}

@article{Pierce21,
	author = {Pierce, Andrew T. and Xie, Yonglong and Park, Jeong Min and Khalaf, Eslam and Lee, Seung Hwan and Cao, Yuan and Parker, Daniel E. and Forrester, Patrick R. and Chen, Shaowen and Watanabe, Kenji and Taniguchi, Takashi and Vishwanath, Ashvin and Jarillo-Herrero, Pablo and Yacoby, Amir},
	journal = {Nature Physics},
	number = {11},
	pages = {1210--1215},
	title = {Unconventional sequence of correlated Chern insulators in magic-angle twisted bilayer graphene},
	volume = {17},
	year = {2021}}

@article{Jiang19,
	author = {Jiang, Yuhang and Lai, Xinyuan and Watanabe, Kenji and Taniguchi, Takashi and Haule, Kristjan and Mao, Jinhai and Andrei, Eva Y.},
	journal = {Nature},
	number = {7772},
	pages = {91--95},
	title = {Charge order and broken rotational symmetry in magic-angle twisted bilayer graphene},
	volume = {573},
	year = {2019}}

@article{Nuckolls20,
	author = {Nuckolls, Kevin P. and Oh, Myungchul and Wong, Dillon and Lian, Biao and Watanabe, Kenji and Taniguchi, Takashi and Bernevig, B. Andrei and Yazdani, Ali},
	journal = {Nature},
	number = {7839},
	pages = {610--615},
	title = {Strongly correlated Chern insulators in magic-angle twisted bilayer graphene},
	volume = {588},
	year = {2020}}

@article{fakher_scaling,
  title = {Fermionic quantum criticality in honeycomb and $\ensuremath{\pi}$-flux Hubbard models: Finite-size scaling of renormalization-group-invariant observables from quantum Monte Carlo},
  author = {Parisen Toldin, Francesco and Hohenadler, Martin and Assaad, Fakher F. and Herbut, Igor F.},
  journal = {Phys. Rev. B},
  volume = {91},
  issue = {16},
  pages = {165108},
  numpages = {18},
  year = {2015},
  month = {Apr},
  publisher = {American Physical Society},
  doi = {10.1103/PhysRevB.91.165108},
  url = {https://link.aps.org/doi/10.1103/PhysRevB.91.165108}
}

@Article{alf,
	title={{The ALF (Algorithms for Lattice Fermions) project release 2.0. Documentation for the auxiliary-field quantum Monte Carlo code}},
	author={F. F. Assaad and M. Bercx and F. Goth and A. Götz and J. S. Hofmann and E. Huffman and Z. Liu and F. Parisen Toldin and J. S. E. Portela and J. Schwab},
	journal={SciPost Phys. Codebases},
	pages={1},
	year={2022},
	publisher={SciPost},
	doi={10.21468/SciPostPhysCodeb.1},
	url={https://scipost.org/10.21468/SciPostPhysCodeb.1},
}

@Article{assaad19,
author={Liu, Yuhai
and Wang, Zhenjiu
and Sato, Toshihiro
and Hohenadler, Martin
and Wang, Chong
and Guo, Wenan
and Assaad, Fakher F.},
title={Superconductivity from the condensation of topological defects in a quantum spin-Hall insulator},
journal={Nature Communications},
year={2019},
month={Jun},
day={14},
volume={10},
number={1},
pages={2658},
issn={2041-1723},
doi={10.1038/s41467-019-10372-0},
url={https://doi.org/10.1038/s41467-019-10372-0}
}

@article{assaad21,
  title = {Doping-Induced Quantum Spin Hall Insulator to Superconductor Transition},
  author = {Wang, Zhenjiu and Liu, Yuhai and Sato, Toshihiro and Hohenadler, Martin and Wang, Chong and Guo, Wenan and Assaad, Fakher F.},
  journal = {Phys. Rev. Lett.},
  volume = {126},
  issue = {20},
  pages = {205701},
  numpages = {5},
  year = {2021},
  month = {May},
  publisher = {American Physical Society},
  doi = {10.1103/PhysRevLett.126.205701},
  url = {https://link.aps.org/doi/10.1103/PhysRevLett.126.205701}
}

@article{assaad23,
  title = {Bandwidth-controlled quantum phase transition between an easy-plane quantum spin Hall state and an $s$-wave superconductor},
  author = {Hou, Disha and Liu, Yuhai and Sato, Toshihiro and Guo, Wenan and Assaad, Fakher F. and Wang, Zhenjiu},
  journal = {Phys. Rev. B},
  volume = {107},
  issue = {15},
  pages = {155107},
  numpages = {18},
  year = {2023},
  month = {Apr},
  publisher = {American Physical Society},
  doi = {10.1103/PhysRevB.107.155107},
  url = {https://link.aps.org/doi/10.1103/PhysRevB.107.155107}
}

@article{Zhang2018,
	author = {Dan-Wei Zhang and Yan-Qing Zhu and Y. X. Zhao and Hui Yan and Shi-Liang Zhu},
	doi = {10.1080/00018732.2019.1594094},
	eprint = {https://doi.org/10.1080/00018732.2019.1594094},
	journal = {Advances in Physics},
	number = {4},
	pages = {253--402},
	publisher = {Taylor \& Francis},
	title = {Topological quantum matter with cold atoms},
	url = {https://doi.org/10.1080/00018732.2019.1594094},
	volume = {67},
	year = {2018}}

@article{Miyake13,
  title = {Realizing the Harper Hamiltonian with Laser-Assisted Tunneling in Optical Lattices},
  author = {Miyake, Hirokazu and Siviloglou, Georgios A. and Kennedy, Colin J. and Burton, William Cody and Ketterle, Wolfgang},
  journal = {Phys. Rev. Lett.},
  volume = {111},
  issue = {18},
  pages = {185302},
  numpages = {5},
  year = {2013},
  month = {Oct},
  publisher = {American Physical Society},
  doi = {10.1103/PhysRevLett.111.185302},
  url = {https://link.aps.org/doi/10.1103/PhysRevLett.111.185302}
}

@article{Aidelsburger13,
  title = {Realization of the Hofstadter Hamiltonian with Ultracold Atoms in Optical Lattices},
  author = {Aidelsburger, M. and Atala, M. and Lohse, M. and Barreiro, J. T. and Paredes, B. and Bloch, I.},
  journal = {Phys. Rev. Lett.},
  volume = {111},
  issue = {18},
  pages = {185301},
  numpages = {5},
  year = {2013},
  month = {Oct},
  publisher = {American Physical Society},
  doi = {10.1103/PhysRevLett.111.185301},
  url = {https://link.aps.org/doi/10.1103/PhysRevLett.111.185301}
}

@article{Jotzu14,
	author = {Jotzu, Gregor and Messer, Michael and Desbuquois, R{\'e}mi and Lebrat, Martin and Uehlinger, Thomas and Greif, Daniel and Esslinger, Tilman},
	journal = {Nature},
	number = {7526},
	pages = {237--240},
	title = {Experimental realization of the topological Haldane model with ultracold fermions},
	volume = {515},
	year = {2014}}

@article{Cooper_rmp19,
  title = {Topological bands for ultracold atoms},
  author = {Cooper, N. R. and Dalibard, J. and Spielman, I. B.},
  journal = {Rev. Mod. Phys.},
  volume = {91},
  issue = {1},
  pages = {015005},
  numpages = {55},
  year = {2019},
  month = {Mar},
  publisher = {American Physical Society},
  doi = {10.1103/RevModPhys.91.015005},
  url = {https://link.aps.org/doi/10.1103/RevModPhys.91.015005}
}

\end{document}